%% file: softwall.tex
\documentclass[aip,apl,reprint]{revtex4-2}

\usepackage[english]{babel}
\usepackage[T1]{fontenc}

\usepackage{amsmath}
\usepackage{amssymb}
\usepackage{graphicx}
\usepackage{float}
\usepackage[colorlinks=true, allcolors=blue]{hyperref}
\usepackage{cleveref}
\usepackage[version=4]{mhchem}
\usepackage{xcolor}
\usepackage{rotating}
\usepackage{caption}
\usepackage{subcaption}	
\usepackage{notes2bib}
\usepackage{booktabs}
\usepackage{upgreek}
\usepackage{placeins}

\bibnotesetup{
note-name = ,
use-sort-key = false
}

\newcommand*\citeit[1]{\citeauthor{#1}\cite{#1}}
\newcommand*\citeitcomma[1]{\citeauthor{#1},\cite{#1}}
\newcommand*\citeitperiod[1]{\citeauthor{#1}.\cite{#1}}

\newcommand*\HelFEM{\textsc{HelFEM}}
\newcommand*\Libxc{\textsc{Libxc}}
\newcommand*\ie{\emph{i.e.}}
\newcommand*\eg{\emph{e.g.}}

\newcommand*\citeref[1]{ref.~\citenum{#1}}
\newcommand*\citerefs[1]{refs.~\citenum{#1}}

\newcommand*\Eh{\ensuremath{\mathrm{E}_\mathrm{h}}}

\input{tables/shifttabmg}

\input{tables/shifttabca}
\begin{document}

\title{Atomic Confinement Potentials and the Generation of Numerical Atomic Orbitals}

\author{Hugo {\AA}str{\"o}m}
\affiliation{University of Helsinki, Department of Chemistry, 
P.O. Box
  55 (A.I. Virtanens plats 1), FI-00014 University of Helsinki, Finland}
\author{Susi Lehtola}
\email{susi.lehtola@alumni.helsinki.fi}
\affiliation{University of Helsinki, Department of Chemistry, 
P.O. Box
  55 (A.I. Virtanens plats 1), FI-00014 University of Helsinki, Finland}

\begin{abstract}
  We aim to develop novel reusable open source infrastructure [Lehtola, J. Chem. Phys. 159, 180901 (2023)] for numerical atomic orbitals (NAOs). Soft confinement potentials are typically used to force the NAO radial basis functions ${\psi}_{nl}(r)$ to vanish smoothly in increasing $r$ and to generate localized unoccupied states; we review such potentials and other commonly-used techniques in NAO generation as a follow-up to our recent study on atoms in hard-wall confinement [Åström and Lehtola, J. Phys. Chem. A 129, 2791 (2025)]. In addition to NAO generation, confinement potentials are also employed to simulate environmental effects in other research areas, such as studies of (i) atoms in solids, (ii) quantum dots, and (iii) high-pressure chemistry.
As in our earlier work, we perform fully numerical density functional calculations with spherically averaged densities, as is usual in NAO studies. Our calculations employ the the finite element method (FEM) implemented in the HelFEM program, yielding variational energies and enabling the use of various boundary conditions.
We consider four families of potentials to study the Mg and Ca atoms, which are textbook examples of extended electronic structures. We show that the resulting ground-state orbitals are surprisingly insensitive to the employed form of the confinement potential, and that the orbitals decay quickly under confinement. We study increasingly steep potentials and examine how they approach the hard-wall limit. Finally, we assess NAO basis set truncation errors for types of singular potentials that are now broadly used in the NAO literature. 

\end{abstract}

\maketitle

\section{Introduction \label{sec:intro}}

Thanks to advances in the theory and computational algorithms of quantum chemistry, as well as in computer and software engineering in the last several decades, density functional theory\cite{Hohenberg1964_PR_864, Kohn1965_PR_1133} (DFT) has become the standard tool for studying the structure and properties of molecules and solids across disciplines.\cite{Orio2009_PR_443, Neugebauer2013_WIRCMS_438, Jones2015_RMP_897, Mardirossian2017_MP_2315}
The key step in the computer implementation of DFT is the choice of the discretization of the single-particle states $\psi_{i}$, commonly known as molecular orbitals (MOs).
The MOs are typically expressed as a linear combination of basis functions $\chi_\alpha$\cite{Lehtola2020_M_1218}
\begin{equation}
  \label{eqn:lcao}
  \psi_{i}({\bf r}) = \sum_\alpha C_{\alpha i} \chi_\alpha({\bf r}),
\end{equation}
where ${\bf C}$ is the matrix of expansion coefficients.
Many kinds of basis sets have been proposed in the literature,\cite{Lehtola2019_IJQC_25968} but atomic orbitals (AOs)
\begin{equation}
  \label{eqn:ao}
  \chi_{nlm}({\bf r}) = R_{nl}(r) Y_{lm}(\hat{\bf r})
\end{equation}
are the most widely used ones in chemistry, yielding the linear combination of atomic orbitals (LCAO) approach in \cref{eqn:lcao,eqn:ao}.

The angular functions of \cref{eqn:ao} are the standard spherical harmonics $Y_{lm}(\hat{\bf r})$, which are typically used in the real form, but the radial basis functions $R_{nl}(r)$ can be chosen in many ways.\cite{Lehtola2019_IJQC_25968}
This work focuses on numerical atomic orbitals (NAOs),\cite{Averill1973_JCP_6412} which have been shown to enable linear-scaling DFT calculations on large systems. \cite{Sankey1989_PRB_3979, Demkov1995_PRB_1618, Ordejon1996_PRB_10441, Windl1998_PRB_2431, Gan2001_PRB_205109, Junquera2001_PRB_235111, Anglada2002_PRB_205101, Ozaki2003_PRB_155108, Ozaki2004_PRB_195113, Louwerse2012_PRB_35108, Lin2024_WIRCMS_1687, Kokott2024_JCP_24112}
Importantly, already small NAO basis sets afford a precision similar to that of much larger plane-wave basis sets in applications on solid state systems,\cite{Sankey1989_PRB_3979, SanchezPortal1997_IJQC_453, Kenny2000_PRB_4899, Gan2001_PRB_205109} yielding results in good agreement with experiment.\cite{Demkov1995_PRB_1618, Lewis1997_PRB_6880, Windl1998_PRB_2431}
NAO basis sets can also reach a high level of precision in molecular DFT calculations, as demonstrated by a recent benchmark against fully numerical results.\cite{Jensen2017_JPCL_1449}

The key reason for the excellent computational performance of NAO basis sets lies in the extreme sparsity of operator matrix elements that arises when the NAO basis functions have finite support, that is, when they are non-zero only within a given distance from their center, that is typically chosen to be in the order of 5 Å, as will be discussed later in this work.
This locality is crucial in polyatomic calculations where one has to calculate matrix elements between all overlapping basis functions located at different centers: by going from orbitals with global support to orbitals with local support one can screen out integrals exactly with the cutoff radius, and this method is used in various solid state programs, such as {\sc BAND},\cite{Velde1991_PRB_7888} {\sc FHI-aims},\cite{Blum2009_CPC_2175} {\sc SIESTA},\cite{Artacho2008_JPCM_64208} and {\sc GPAW}.\cite{Mortensen2024_JCP_92503}

The strict localization of the basis functions is typically achieved in practice with the help of a confinement potential $V_\mathrm{c}(r)$ in the atomic calculations used to generate the NAO basis functions.
The confinement potential ensures that the orbitals and their derivatives go smoothly to zero, vanishing altogether beyond the employed cutoff radius $r_c$.\cite{Eschrig1978_PSSB_621, Sankey1989_PRB_3979, Porezag1995_PRB_12947, Horsfield1997_PRB_6594, Delley2000_JCP_7756, Junquera2001_PRB_235111, Ozaki2003_PRB_155108}

This localization of the orbitals also has a physical meaning, as orbitals are known to contract when atoms form chemical bonds.\cite{Craig1956_JCS_4895, Kutzelnigg1982_PRA_2361, Levine2017_JPCL_1967, Bacskay2017_JPCA_9330, Bacskay2018_JPCA_7880, Bacskay2022_JCP_204122, Sterling2024_JACS_9532}
In fact, confinement potentials are also often used in computational studies of atoms confined in materials,\cite{Watson1958_PR_1108, Boeyens1994_JCSFT_3377, Connerade1998_JPBAMOP_3557, Garza1998_PRE_3949, Connerade2000_JPBAMOP_251, Connerade2000_JPBAMOP_869, Connerade2000_JPBAMOP_3467, Garza2000_JMST_183, Sen2000_CPL_29, Eschrig2004, Dolmatov2004_RPC_417, Garza2005_JCS_379, Witek2007_JPCA_5712, Guerra2009_AQC_1, Wahiduzzaman2013_JCTC_4006, LozanoEspinosa2017_PM_284, Saha2020_PRA_52824, Deshmukh2021_EPJD_166, Cui2024_JCTC_5276, Chou2016_JCTC_53}, quantum dots,\cite{Takagahara1992_PRB_15578, Gerchikov1998_APL_532, Bednarek1999_PRB_13036, Capelle2007_PRL_010402} and high-pressure chemistry\cite{Chattaraj2003_CPL_805, Chattaraj2003_JPCA_4877, Cruz2004_CPL_138, Sarkar2009_JPCA_10759, Rahm2019_JACS_10253, Pasteka2020_MP_1730989, Connerade2020_EPJD_211, Scheurer2020_JCTC_583, Stauch2020_JCP_134503, Rahm2021_JACS_10804, Gale2021_JCP_244103, Zeller2024_WIRCMS_1708} to simulate effects of the environment.
Even though the generation of localized basis functions and the simulation of atoms and molecules in confined environments employ similar techniques, the connection between these dissimilar fields does not appear to be widely appreciated in the literature, as the form of the employed confinement potential does depend on the task at hand.
Analytically solvable models of confined systems can play an important role in testing approximate theories, or as a first step in simulations of more involved models of confined systems.\cite{Franklin1999_MPLA_2409, Micu2003_MPLA_2895, Karwowski2005_JMST_1}
The literature on the simplest possible systems---the confined hydrogen and helium atoms---is especially broad.\cite{Jaskolski1996_PR_1}
We recently reported a systematic study of the effects of hard-wall confinement on the electronic structure of many-electron atoms in \citeref{Aastroem2025_JPCA_2791}.

Various confinement potentials for NAO generation have been suggested in the literature.\cite{Eschrig1978_PSSB_621, Porezag1995_PRB_12947, Horsfield1997_PRB_6594, Junquera2001_PRB_235111, Ozaki2003_PRB_155108, Ozaki2004_PRB_195113, Witek2007_JPCA_5712}
The NAO confinement potential $V_\mathrm{c}(r)$ typically has several adjustable parameters, which are used to fine-tune the form of the NAOs, the typical aims being (i) the accuracy of the NAO basis and (ii) the facility of the quadrature of the molecular integrals in the NAO basis.
To achieve strict locality, NAO confinement potentials often diverge at a finite radius $r_c>0$, which is a cut-off parameter.
For $r>r_c$, the NAO basis function strictly vanishes.


In contrast, physical models of confinement typically use regular confinement potentials with finite width and depth/height (which may still diverge in the limit $r \to \infty$).
For example, a quadratic confinement potential arises naturally in an external magnetic field: the field confines the electrons' motion in the orthogonal directions, while also coupling to the electrons' spin and angular momentum around the field.
These effects introduce significant numerical challenges to the reliable modeling of molecular electronic structure in strong magnetic fields with standard techniques,\cite{Lehtola2020_MP_1597989, Aastroem2023_JPCA_10872} which is why we expected in \citeref{Aastroem2025_JPCA_2791} other situations where the Hamiltonian is modified (\ie{}, the addition of a confinement potential) to be affected by similar challenges, as well.
However, as will be discussed later on in this work, typical NAO setups employ relatively weak confinement: the confinement potential is only turned on relatively far from the nucleus in the outer valence region, so that the atoms' essential chemistry is not changed.

Our present interest in the study of confinement effects is motivated by the aim to develop reusable libraries\cite{Lehtola2023_JCP_180901} for electronic structure calculations with NAO basis sets, following up on a series of studies by the senior author that employ a modern high-order finite element method (FEM).\cite{Lehtola2019_IJQC_25945, Lehtola2019_IJQC_25944, Lehtola2020_JCP_144105, Lehtola2020_MP_1597989, Lehtola2020_PRA_12516, Lehtola2020_PRA_32504, Lehtola2021_JCTC_943, Lehtola2023_JCTC_2502, Lehtola2023_JCTC_4033, Lehtola2023_JPCA_4180, Lehtola2024_ES_15015}
Compared to the finite difference method (FDM) commonly used in the literature for NAO applications, the FEM approach is variational, and enables easy control over the boundary conditions of the solution.\cite{Lehtola2019_IJQC_25968}
Yet, the biggest benefit in employing modern FEM methodologies is the ability to employ numerical basis functions of very high order, which enables extremely compact numerical representations.
A mere $\mathcal{O}(100)$ radial basis functions are sufficient to reach n\Eh{} precision in non-relativistic Hartree--Fock and DFT calculations on atoms.\cite{Lehtola2019_IJQC_25945, Lehtola2023_JPCA_4180}

The reduction in the necessary number of numerical basis functions by orders of magnitude from previous methods opens the door to novel avenues of NAO basis set design, as the full set of unoccupied, \emph{i.e.}, virtual orbitals is now also accessible.
The construction of novel NAO basis sets is key to the effort of establishing a new open source framework for all-electron molecular calculations with NAOs, and as a first step towards this goal, we revisit calculations of atoms in soft confinement as a follow-up to our study of atoms in hard-wall confinement in \citeref{Aastroem2025_JPCA_2791}.

In this earlier study, we examined the behavior of ground and low-lying excited states of the H--Xe atoms in hard-wall confinement.\cite{Aastroem2025_JPCA_2791}
The hard-wall potential is trivial to implement in FEM and is controlled by a single parameter, which allows for easy analysis of the behavior of the various electron configurations as a function of the strength of the confinement.
Furthermore, the potential ensures strict localization of the orbitals, which is key for NAOs;\cite{Lewis2001_PRB_195103, Lewis2011_PSSB_1989} however, hard-wall confinement leads to a first-derivative discontinuity that makes hard-wall NAOs unattractive for polyatomic calculations, as will be discussed later in this work.

In this work, we discuss various soft confinement potentials employed in the literature for atomic calculations.
As opposed to the hard-wall potential, soft confinement potentials ensure smooth decay of the radial functions, which is important for guaranteeing facile numerical integrability of NAO matrix elements in polyatomic calculations.
We will also discuss other techniques commonly used in the NAO context, specifically for generating breathing and polarization functions.

The layout of the paper is the following.
We review various soft confinement potentials used in the NAO literature in \cref{sec:theory}, introducing a novel exponential confinement potential, and then review the above-mentioned further techniques used for NAO generation in \cref{sec:other}.
We discuss the computational details and present the numerical approach of this work in \cref{sec:computational-details}.
We present the results for four families of confinement potentials in \cref{sec:results} for the orbitals of the Mg and Ca atoms: the finite-barrier potential in \cref{sec:finite-barrier}, the polynomial and exponential potentials in \cref{sec:poly-exp}, and the singular potentials in \cref{sec:singular}.
We conclude the article with a summary and an outlook in \cref{sec:summary}.
Hartree atomic units are employed throughout unless otherwise specified.

\section{Theory \label{sec:theory}}

We consider Hamiltonians of the form
\begin{equation} \label{eqn:confinement}
H=H_0+V_\mathrm{c}(r),
\end{equation}
where $H_0$ is the standard electronic Hamiltonian for an atom and $V_\mathrm{c}(r)$ is the confinement potential.
From the Schrödinger equation
\begin{equation}
[-\nabla^2/2 + V({\bf r})]\psi({\bf r}) = E\psi({\bf r})
\end{equation}
we see that a global shift to the potential $V({\bf r}) \to V({\bf r})+V_0$ simply translates to a global shift of the orbital energies, $E \to E + V_0$.
Therefore any finite potential can be dressed in either attractive or repulsive form: since orbital energies are only determined up to an additive constant, attraction at small $r$ or repulsion at large $r$ can be thought to be two sides of the same coin.
A finite confinement potential can thus be written either as attractive at small $r$, or repulsive at large $r$.
Importantly, even a globally repulsive confinement potential, $V_\mathrm{c}(r)>0$, has exactly the same solutions as a potential that has been shifted down by a constant offset to achieve $N$ states with negative orbital energies; this formally justifies the extraction of low-lying unoccupied orbitals for use as polarization/correlation functions in the NAO basis even if they have positive orbital energies.

As already mentioned in \cref{sec:intro}, various confinement potentials have been suggested in the literature.
The use of a hard wall potential was proposed by \citeit{Sankey1989_PRB_3979}
\begin{equation} \label{eqn:hard-wall}
V_\mathrm{c}(r) =
\begin{cases}
0,\quad r < r_c, \\
\infty,\quad r \geq r_c.
\end{cases}
\end{equation}
An analogous procedure was also employed by \citeitcomma{SanchezPortal1997_IJQC_453} \citeitcomma{Basanta2007_CMS_759} and \citeitcomma{Nakata2020_JCP_164112} for instance.
A related technique based on spherical Bessel functions due to \citeit{Haynes1997_CPC_17} is discussed in \cref{sec:other}.
However, it is easy to see from the radial Schr\"odinger equation that the derivative of the resulting radial function is not continuous at $r=r_c$, where the potential of \cref{eqn:hard-wall} has an infinite jump.
This leads to a jump discontinuity in the first derivative of the radial wave function, which makes the hard-wall potential unattractive for NAO generation, as matrix elements of NAOs situated on different centers will be hard to compute accurately by quadrature.

As it is not necessary for the potential to diverge to make the wave function negligible beyond a given cutoff, a soft wall can be used, instead.
The simplest model of soft confinement is to use a finite barrier, and this was already done in the pioneering work of \citeit{Averill1973_JCP_6412} for determining breathing and polarization functions:
\begin{equation} \label{eqn:soft-wall}
V_\mathrm{c}(r) =
\begin{cases}
0,\quad r < r_0\\
V_0,\quad r \geq r_0.
\end{cases}
\end{equation}
In the context of atoms in confinement, \citeit{Connerade2000_JPBAMOP_251} carried out calculations on 3d and 4d atoms and ions with a finite barrier of height $V_0=10$ \Eh, which they claimed to suffice to make penetration effects ``very small''.\cite{Connerade2000_JPBAMOP_251}
Finite-barrier potentials of the form of \cref{eqn:soft-wall} have also been used in the literature to simulate pressure effects on atoms with the extreme pressure polarized continuum model (XP-PCM).\cite{Cammi2012_JCP_154112, Cammi2015_JCC_2246, Cammi2018_JCC_2243, Cammi2018__273, Rahm2019_JACS_10253, Rahm2020_C_2441, Rahm2021_CS_2397, Eeckhoudt2022_CS_9329, Cammi2022_JCC_1176}

Again, from the Schrödinger equation, while the first derivative is now continuous as long as $V_0$ is finite,\cite{Branson1979_AJP_1000, Home1982_AJP_552} it appears that the discontinuity in the potential of \cref{eqn:soft-wall} is reflected by a finite jump in the second derivative of the resulting orbital.

Following a suggestion by \citeitcomma{Ozaki2003_PRB_155108} the transition can be smoothened by polynomial interpolation
\begin{equation}\label{eqn:pot3}
V_\mathrm{c}(r) =
\begin{cases}
0,\quad r \leq r_i, \\
\displaystyle \sum_{n=0}^3 b_n r^n,\quad r_i < r \leq r_0, \\
V_0,\quad r> r_0,
\end{cases}
\end{equation}
where $r_i$ is a new parameter that controls the initiation of the confinement, and the four expansion coefficients $\{b_{n}\}_{n=0}^{3}$ are solved by demanding continuity of $V_\mathrm{c}(r)$ and $V_\mathrm{c}'(r)$ at $r=r_i$ and $r=r_0$.

We note here that the approach of \citeit{Ozaki2003_PRB_155108} did not actually employ a confinement potential of the form of \cref{eqn:pot3}.
Instead, \citeauthor{Ozaki2003_PRB_155108} modified the classical Coulomb attraction potential of the nucleus with atomic number $Z$ such that it switches from $-Z/r$ at $r=r_i$ to a polynomial that attains the constant value $V_0$ at $r=r_0$.
Such an approach should still yield quite similar results to the use of \cref{eqn:pot3}, as the difference of the two at $r \ge r_i$ is $Z/r$ which is likely small compared to the value of $V_0=3\times10^4$ \Eh{} provided by \citeitperiod{Ozaki2003_PRB_155108}
Also the original formalism of \citeit{Averill1973_JCP_6412} is slightly different to what was discussed above in the context of \cref{eqn:soft-wall}.

%
%

\begin{figure}
  \centering
  \includegraphics[width=.7\columnwidth]{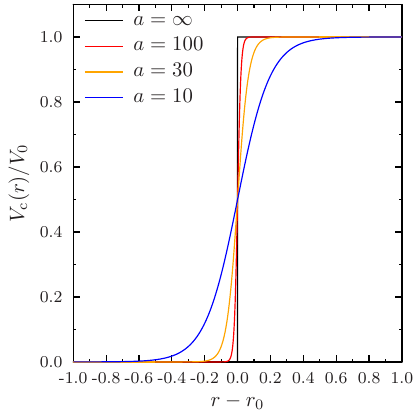}
  \caption{The Woods--Saxon potential of \cref{eqn:woodssaxon} with various values of $a$. The case $a=\infty$ coincides with \cref{eqn:soft-wall}.}
  \label{fig:woodssaxon}
\end{figure}

Another smooth alternative to the finite-barrier potential is afforded by the Woods--Saxon potential\cite{Woods1954_PR_577}
\begin{equation}
  \label{eqn:woodssaxon}
  V_\mathrm{c}(r)=\frac{V_0}{1+e^{-a(r-r_0)}},
\end{equation}
which appears to have found use in tight-binding DFT.\cite{Witek2007_JPCA_5712, Chou2016_JCTC_53, Huran2018_JCTC_2947, Huran2019_PSSB_1900634, Panosetti2021_JPCA_691, Hutama2021_JPCA_2184}
The parameter $a$ can be adjusted to control the rapidity of the onset of the potential, as illustrated in \cref{fig:woodssaxon}; \cref{eqn:woodssaxon} approaches \cref{eqn:soft-wall} at the limit $a \to \infty$.

The NAO study of \citeit{Eschrig1978_PSSB_621} introduced a general polynomial confinement potential of the form
\begin{equation}\label{eqn:rn-pot}
V_\mathrm{c}(r) = \left(\frac{r}{r_0} \right)^N,
\end{equation}$
r_0>0$ being a parameter that describes the strength of the potential, and $N > 0$ controls its form.
\citeit{Eschrig1978_PSSB_621} employed $N=2$, which was later used also by \citeitperiod{Porezag1995_PRB_12947}
In turn, \citeauthor{Koepernik1997_PRB_5717}\cite{Koepernik1997_PRB_5717,Koepernik1999_PRB_1743} employed $N=4$, while \citeit{Horsfield1997_PRB_6594} proposed using $N=6$ for NAO generation, instead, to make the resulting orbitals more strongly localized.
\citeit{Wahiduzzaman2013_JCTC_4006} introduced the use of fractional values of $N$.
The study of atoms and molecules in soft and hard confinement potentials of \citeit{Pasteka2020_MP_1730989} considered values of $N$ up to $N=20$, but these calculations were limited to Gaussian basis sets, which, as discussed by \citeit{Pasteka2020_MP_1730989} and in \citeref{Aastroem2025_JPCA_2791} are likely unreliable as Gaussian basis functions have the same asymptotic form as the solutions for $N=2$, only.

As the orbitals resulting from a confinement potential of the form of \cref{eqn:rn-pot} are not formally strictly zero anywhere, \citeit{Junquera2001_PRB_235111} suggested a confinement potential of the form
\begin{equation} \label{eqn:junq-pot}
V_\mathrm{c}(r) =
\begin{cases}
0,\quad r \le r_i\\
V_0 \displaystyle \frac{\exp\left(-\displaystyle \frac{r_c-r_i}{r-r_i}\right)}{r_c-r},\quad r_i < r < r_c, \\
\infty,\quad r \ge r_c
\end{cases}
\end{equation}
where $V_0$ controls the strength of the potential and $r_i$ is again an adjustable parameter that controls the onset of the potential, thus avoiding its application in the core region that may later be represented by a pseudopotential.
The SIESTA\cite{Artacho2008_JPCM_64208} manual gives defaults $r_i = 0.9 r_c$ and $V_0 = 20$ \Eh.
The same function has also been used by \citeit{Larsen2009_PRB_195112} with default values $r_i = 0.6 r_c$ and $V_0 = 12$ \Eh{} in the GPAW program.\cite{Mortensen2024_JCP_92503}

In order to spread the damping of the radial function more evenly across the width of the confinement potential, \citeit{Blum2009_CPC_2175} proposed using a higher power in the denominator in \cref{eqn:junq-pot}, that is,
\begin{equation} \label{eqn:blum-pot}
V_\mathrm{c}(r) =
\begin{cases}
0,\quad r \le r_i, \\
V_0 \displaystyle \frac{\exp\left(\displaystyle -\frac{r_c-r_i}{r-r_i}\right)}{(r_c-r)^2},\quad r_i < r < r_c, \\
\infty,\quad r \ge r_c.
\end{cases}
\end{equation}
\citeit{Blum2009_CPC_2175} provide ``safe default values'' of $V_0=200$ \Eh{} and $r_c=r_i+2.0$ Å; however, the \textsc{FHI-aims} program appears to soon thereafter have switched to a higher default value $V_0 = 250$ \Eh.
Note that the critical minus sign is missing in the expression of the exponential term of \cref{eqn:blum-pot} in \citeref{Blum2009_CPC_2175}; yet, the implementation that was employed in \citeref{Blum2009_CPC_2175} was the correct one.\bibnote{Volker Blum, private communication, 2024.}

As a trivial generalization of \cref{eqn:junq-pot,eqn:blum-pot} we obtain
\begin{equation} \label{eqn:genblum-pot}
V_\mathrm{c}(r) =
\begin{cases}
0,\quad r \le r_i, \\
V_0 \displaystyle \frac{\exp\left(\displaystyle -\frac{r_c-r_i}{r-r_i}\right)}{(r_c-r)^n},\quad r_i < r < r_c, \\
\infty,\quad r \ge r_c.
\end{cases}
\end{equation}
where $n=1$ corresponds to \cref{eqn:junq-pot}, $n=2$ corresponds to \cref{eqn:blum-pot}, and the choice $n=3$ will be explored in detail later on in this work.

We study the asymptotic behaviors arising from the different choices of $n$ in the Appendix, and the resulting functional forms turn out to be surprisingly different.
However, the observed decay behaviors turn out similar for various $n$ (see \cref{sec:singular}).

A strict localization of the orbitals is achieved with \cref{eqn:junq-pot,eqn:blum-pot,eqn:genblum-pot} exactly as with the hard-wall potential of \cref{eqn:hard-wall}, since the potentials are singular at $r=r_c$.
The correct solution that vanishes for $r\ge r_c$ is readily captured in FEM calculations by truncating the radial grid at $r=r_c$; this is discussed in more detail in \cref{sec:computational-details}.

As discussed by \citeitcomma{Delley2000_JCP_7756} an alternative to attain the same strict localization is to use a soft confinement potential in combination with a hard-wall boundary
\begin{equation}\label{eqn:hybrid-pot}
V_\mathrm{c}(r) = \begin{cases}
\left( \displaystyle \frac {r} {r_0} \right)^N,\quad r < r_c,\\
\infty,\quad r \geq r_c.
\end{cases}
\end{equation}
While the hard-wall boundary alone can lead to a significant derivative discontinuity at $r=r_c$, and a soft confinement potential alone in principle leads to basis functions with global support, their combination offers a good compromise: the localization is mostly achieved by the soft confinement part, but the basis function is also made to strictly vanish beyond the cutoff.
The hybrid potential is also trivial to implement: the soft part of the potential is smooth, while the hard-wall potential is again implemented in practice by truncating the wave function expansion at $r=r_c$; such clipped orbitals were also employed by \citeitcomma{Horsfield1997_PRB_6594} for example.
The hybrid approach is exact for a sufficiently large value of $r_c$, as the choice of this parameter corresponds to the choice of the discretization domain in FEM, as we have recently discussed in \citeref{Aastroem2025_JPCA_2791}; this connection has also been previously used by \citeit{Pasteka2020_MP_1730989}, for example.

In the numerical approach employed in this work, we can also force the first derivative to vanish at the employed value of $r_c$ by using a Hermite interpolating polynomial basis set.\cite{Lehtola2023_JPCA_4180}
We show in \cref{sec:barrier-trunc,sec:poly-exp-trunc} that this approach indeed affords a practically parameter-free approach for choosing $r_c$, since the location of the hard wall can be found by bracketing the value of $r_c$ that leads to a negligible change in the total energy.

Finally, as a hitherto unexplored alternative, we introduce a family of smooth, soft potentials that lead to exponential localization of the radial orbitals: the exponential potential defined by
\begin{equation}\label{eqn:exp-pot}
V_\mathrm{c}(r) = N! \left[ \exp \left( \frac r {r_0} \right) - \sum_{k=0}^{N-1} \frac 1 {k!} \left( \frac r {r_0} \right)^k \right].
\end{equation}
The idea of the potential in \cref{eqn:exp-pot} is that like \cref{eqn:rn-pot}, it is controlled by a single parameter $r_0$, assuming the form which is controlled by $N$ has been fixed.
Moreover, like \cref{eqn:rn-pot}, the small-$r$ behavior of \cref{eqn:exp-pot} is $V(r) \propto (r/r_0)^{N}$; this is achieved by removing the lower-order Taylor series terms from $\exp(r/r_0)$, and then renormalizing the remainder by the prefactor of the lowest surviving Taylor series term.

In contrast to \cref{eqn:rn-pot}, \cref{eqn:exp-pot} grows exponentially quickly at large $r$ and should therefore lead to improved localization.
We show in \cref{sec:poly-exp-hw} that the potential of \cref{eqn:exp-pot} can be combined with a hard wall at a value of $r_c$ that has a negligible effect on the total energy, similarly to \cref{eqn:hybrid-pot} above.
The value of $r_c$ for the exponential potential is systematically smaller than that for the polynomial potential with the same $N$, confirming improved locality of the orbitals.

As discussed by \citeit{Junquera2001_PRB_235111} and \citeitcomma{Blum2009_CPC_2175} in the context of NAO basis set generation it is generally desirable for the confinement potential to only affect the valence region, as the potential for the core electrons should not be modified.
We note that the polynomial and exponential confinement potentials of \cref{eqn:rn-pot,eqn:exp-pot}, respectively, can readily be modified by shifting the potential to only turn on at $r=\delta$
\begin{equation} \label{eqn:piecewise}
V_\mathrm{c}(r) \mapsto V_\mathrm{c}^\mathrm{shifted}(r) =
\begin{cases}
0,\quad r < \delta, \\
V_\mathrm{c}(r-\delta),\quad r \geq \delta.
\end{cases}
\end{equation}
It is easy to see that in combination with \cref{eqn:rn-pot,eqn:exp-pot} \cref{eqn:piecewise} leads to a $C^{N-1}$ continuous potential, and a $C^{N+1}$ continuous wave function.
We expect that turning on such a confinement potential at a radial coordinate $r=\delta$ in the valence region should yield similar results as the rigid singular potentials of \citeit{Junquera2001_PRB_235111} and \citeitcomma{Blum2009_CPC_2175} provided that a sufficiently small $r_0$ and large $N$ values are employed, even though the rigid potentials of \cref{eqn:junq-pot,eqn:blum-pot,eqn:genblum-pot} are $C^\infty$ at the switch-on point $r=r_i$.
We note that shifted polynomial confinement potentials are already used in calculations employing complex absorbing potentials,\cite{Riss1993_JPBAMOP_4503} and have also been used by \citeit{Zubiaga2014_PRA_52707} for calculations on the unbound \ce{e+ H} and \ce{e+ He} systems, for instance.

\subsection{Techniques for NAO generation \label{sec:other}}

At this point we will briefly diverge from the main topic of the paper (atoms with confinement potentials), to discuss related approaches to generate NAO basis sets, continuing the discussion in our earlier paper.\cite{Aastroem2025_JPCA_2791}
The goal in the generation of NAO basis sets is to enable the robust generation of sets of radial basis functions that enable rapid and reliable materials modeling, which we hope to pursue in future work.

While the atomic occupied orbitals are well-defined and easy to solve from the equations of \citeit{Kohn1965_PR_1133} for a given density functional approximation (DFA) and the atom's ground state, the form of the optimal breathing and polarization functions that describe the atom's behavior in a polyatomic system is unknown.\cite{Lehtola2024_ES_15015}
Yet, NAO radial basis functions are extremely flexible: in principle, their form can be chosen freely.
Traditional implementations of NAO basis sets employ $\mathcal{O}(10^4)$--$\mathcal{O}(10^6)$ parameters: the values of the radial functions on the employed radial grid.
As explicit optimization of such a large number of parameters arising from the use of low-order numerical approximations is unattractive, some standard strategies appear to have emerged in the literature for more facile generation of NAO basis sets.

As already mentioned in \cref{sec:theory}, the pioneering study on NAO calculations by \citeit{Averill1973_JCP_6412} proposed extracting additional functions from the unoccupied orbitals of an atom in \emph{``a spherical well or barrier which is of sufficient depth and width to induce the appropriate number of localized eigenfunctions''}.
We note here that it is well known that while the exact Kohn--Sham potential decays like $V(r) \propto -1/r$ at large $r$,\cite{Levy1984_PRA_2745, Almbladh1985_PRB_3231} yielding an infinite spectrum of bound but extended Rydberg states, presently-available DFAs exhibit a much faster, exponential decay, leading to qualitatively incorrect form of the unoccupied orbitals.
Confinement potentials are therefore traditionally employed in NAO generation as a means to circumvent this incorrect, exponentially decaying asymptotic behavior of the Kohn--Sham potential.
The unbound unoccupied orbitals obtained with presently-available DFAs and a confinement potential may be useful for generating diffuse basis functions,\cite{GarciaGil2009_PRB_75441} however.

Following the suggestion of \citeitcomma{Averill1973_JCP_6412} the parameters of the confinement potential can be used to (roughly) optimize the form of the NAO radial functions in calculations on polyatomic systems, reducing the number of optimized parameters significantly: in the scheme of \citeitcomma{Averill1973_JCP_6412} only the location and height of the finite barrier need to be determined.
It appears to since have become standard practice to employ a different confinement potential for each angular momentum channel $l$, thus allowing the generation of custom polarization functions for each angular momentum.
Further freedom can be introduced by employing separate confinement potentials for the unoccupied orbitals that are included in the NAO basis as breathing and polarization functions, see \citeitcomma{Corsetti2013_JPCM_435504} for example.

Also avenues more similar to those prevalent in quantum chemistry have been employed: \citeit{Zunger1977_PRB_4716} and \citeit{Blum2009_CPC_2175} employ hydrogenic functions for describing polarization effects, while \citeit{Larsen2009_PRB_195112} use Gaussian radial functions; we have described a technique based on completeness optimization\cite{Manninen2006_JCC_434, Lehtola2015_JCC_335} for systematical formation of such basis sets.\cite{Rossi2015_JCP_94114}

The similarities do not stop there.
\citeit{Roos1985_CP_43} pointed out in the quantum chemistry literature that polarization functions could be generated from the first-order response of an atomic wave function to an electric field; 14 years later, apparently unaware of \citeref{Roos1985_CP_43}, \citeit{Artacho1999_PSSB_809} reported the analogous procedure for NAOs (see \citeit{Soler2002_JPCM_2745} for details on the implementation).
\citeit{Artacho1999_PSSB_809} also proposed a scheme to generate breathing functions by radially splitting the NAO radial orbitals into core and valence regions, again following established practice in the quantum chemistry literature.
This can be achieved, \eg{}, with a suitable smooth interpolation function $0 \leq \phi(r) \leq 1$: multiplying $R_{nl}(r)$ with $\phi(r)$ and $1-\phi(r)$ yields two radial functions that describe the head and the tail of the orbital.

Increasing the charge states of atoms can be used to generate increasingly confined radial functions.
For example, \citeit{Delley1990_JCP_508} discussed numerical double-$\zeta$ basis sets obtained by augmenting the NAO basis for an atom with radial functions for the cation with charge +2, +1.5, or +1, as well as the generation of additional functions for hydrogen by a fractional increase of the nuclear charge; the use of cationic and anionic orbitals had previously been discussed by \citeitperiod{Zunger1977_PRB_4716}

Changes to the charge state of atoms in NAO generation has also been employed by \citeit{Junquera2001_PRB_235111} and \citeitcomma{Bennett2025_PRB_125123} for instance.
Also other types of schemes can be investigated; for example, \citeit{Watson1958_PR_1108} employed an attractive potential corresponding to a uniformly charged sphere to stabilize the \ce{O^{2-}} anion.
The analogous stabilization of the \ce{H-} anion with a hard-wall potential has been discussed by \citeitperiod{Shore1977_PRB_2858}
\citeit{Corsetti2013_JPCM_435504} suggested employing a Yukawa screened Coulomb confinement potential to generate NAO basis functions
\begin{equation}\label{eqn:yukawa-pot}
V_\mathrm{c}(r) = -Q_0 \frac {\exp(-\lambda r)} {\sqrt{r^2+\delta^2}},
\end{equation}
where $Q_0$ is a parameter that controls the strength of the potential, $\delta$ is a parameter introduced to avoid numerical difficulties at $r=0$, and $\lambda$ is a parameter used to fine-tune the orbital tail.
\citeit{Corsetti2013_JPCM_435504} gave the value $\delta = 0.01\ a_0$ and $\lambda=0$, reverting \cref{eqn:yukawa-pot} into a soft Coulomb potential, $V(r) \propto - Q_0/\sqrt{r^2 + \delta^2}$.
Thus, the method of \citeit{Corsetti2013_JPCM_435504} is likewise similar in spirit to increasing the charge state of the atom; here, instead, the nuclear charge is changed.
Scalings of the nuclear charge have been used in the quantum chemistry literature to calculate discrete components of resonance states,\cite{Nestmann1985_JPBAMP_615} for instance.

In summary of this subsection, the flexibility of NAO basis sets has traditionally demanded the use of confinement potentials for the generation of polarization functions in polyatomic calculations, and while a face-to-face assessment of the various schemes discussed above would be interesting, it is outside the scope of this work.

Finally, we note a closely related approach to the fireball orbitals arising from hard-wall confinement of \cref{eqn:hard-wall}: the use of spherical Bessel functions, which was proposed by \citeit{Haynes1997_CPC_17} and benchmarked by \citeit{Gan2001_PRB_205109}.
In this approach, the radial basis functions are chosen as
\begin{equation}
\label{eqn:bessel-basis}
R_{nl}=\begin{cases}
j_{l}(q_{nl}r), & r<a,\\
0, & r\ge a,
\end{cases}
\end{equation}
where the parameter $q_{nl}$ is chosen to be the $n$:th zero of $j_l(qa)=0$.\cite{Gan2001_PRB_205109}
The functions in \cref{eqn:bessel-basis} are an analogy of plane waves: they correspond to free-electron solutions $(\nabla^2 + k^2) \psi = 0$ in a spherical cavity, and the technique of \citeit{Haynes1997_CPC_17} relies on Fourier transforms for polyatomic integral evaluation.
Since the kinetic energy of \cref{eqn:bessel-basis} is $q_{nl}^2/2$, a single kinetic energy cutoff suffices to determine the basis set for each $l$.\cite{Haynes1997_CPC_17, Gan2001_PRB_205109}
All the applications of this basis in the literature appear to rely on the use of pseudopotentials.
\citeit{Chen2010_JPCM_445501} and \citeit{Li2016_CMS_503} examined valence-only NAO basis sets expanded in terms of these spherical Bessel functions.
\citeit{Papior2018_JPCM_1} suggested using the (uncontracted) Bessel functions as diffuse functions.
However, since the Bessel functions have non-zero derivatives $j_l'(qa)$ when $j_l(qa)=0$, all of the aforementioned approaches have issues with derivative discontinuities at the boundary.
The technique was revisited by \citeit{Monserrat2010_JPMT_465205} to remove the contributions from the first derivative discontinuity on the boundary.

\section{Computational Details \label{sec:computational-details}}

As in our previous work,\cite{Aastroem2025_JPCA_2791} we carry out all calculations with the free and open-source\cite{Lehtola2022_WIRCMS_1610} \HelFEM{} program,\cite{Lehtola2019_IJQC_25945, Lehtola2020_PRA_12516, Lehtola2023_JCTC_2502, Lehtola2023_JPCA_4180} which is publicly available on GitHub in its present form.\cite{HelFEM}
The FEM approach has been described in detail in \citerefs{Lehtola2019_IJQC_25945} and \citenum{Lehtola2023_JPCA_4180}, while the details of FEM calculations with hard-wall confinement have been recently discussed in \citeref{Aastroem2025_JPCA_2791}.
We also refer here to \citeref{GarciaMiranda2023_PRE_35302} for an alternative discussion of FEM for a limited form of \cref{eqn:soft-wall,eqn:rn-pot}.

For completeness, we provide a brief outline of the approach.
We divide the radial domain into $N_\mathrm{elem}$ segments $r \in [r_{i}^\mathrm{start},r_{i}^\mathrm{end}]$ called elements.
A basis of piecewise polynomials, called shape functions, $B_n(r)$, is then constructed within each element.\cite{Lehtola2019_IJQC_25945, Lehtola2023_JPCA_4180}
The numerical radial basis functions in \cref{eqn:ao} are set up from these shape functions as
\begin{equation}
  \label{eqn:radbas}
    R_{n}(r)=r^{-1}B_n(r).
\end{equation}
As implied by \cref{eqn:radbas}, the same radial basis set is used for all angular momenta $l$.
By default the radial shape functions are 15-node Lagrange interpolating polynomials (LIPs),\cite{Lehtola2019_IJQC_25945} but we will also employ 8-node Hermite interpolating polynomials (HIPs),\cite{Lehtola2023_JPCA_4180} resulting in a basis with the same accuracy as the LIP basis.

The radial domain is truncated at the point $r=r_\infty$ which is called the practical infinity, \ie{}, the end point of the last element.
All basis functions are built to vanish at this point;\cite{Lehtola2019_IJQC_25945} in the HIP basis calculations, also the derivative can be forced to vanish at this point.\cite{Lehtola2023_JPCA_4180}
In calculations on unconfined atoms, $r_\infty$ is a parameter that needs to be converged such that the solution does not change even if a larger value is employed (possibly in combination with more radial elements), and $r_\infty=40\ a_0$ by default; this value is sufficient for neutral atoms.
However, we will see that (much) smaller values can often be employed in confinement with insignificant changes in the solution.
For the singular potentials of \cref{eqn:hard-wall,eqn:junq-pot,eqn:blum-pot,eqn:genblum-pot}, the correct discretization is obtained with $r_\infty=r_c$.

The finite element discretization requires some further thought in the context of the confinement potentials studied in this work.
The radial Schr\"odinger equation
\begin{equation}
  \label{eqn:radschrodinger}
  \begin{split}
  &\left[ \frac {\partial^2} {\partial r^2} + 2 \frac 1 r \frac {\partial} {\partial r} + \frac {l(l+1)} {r^2} + V(r) \right] R_{nl} (r) \\ &= E_{nl} R_{nl}(r)
  \end{split}
\end{equation}
shows that a finite discontinuity in $V(r)$ leads to a finite discontinuity in $R_{nl}''(r)$; more generally, finite discontinuities in $V^{(k)}(r)$ lead to a finite discontinuity in $R_{nl}^{(k+2)}(r)$.
Because the finite element shape functions are infinitely differentiable ($C^\infty$) polynomials, such a discontinuity can only be achieved when an element boundary is placed at the discontinuity of the potential: the LIP basis is only $C^0$ continuous across element boundaries, while the HIP basis is $C^1$ continuous.
For this reason, we add an extra element boundary at the discontinuity: $r=r_0$ for \cref{eqn:soft-wall}, $r=r_i$ for \cref{eqn:junq-pot,eqn:blum-pot,eqn:genblum-pot}, and $r=\delta$ for \cref{eqn:piecewise}.

The calculations are performed with DFT and the employed electronic structure approach is the same as in \citeref{Aastroem2025_JPCA_2791}; in short, we employ the fractional occupation formalism\cite{Lehtola2020_PRA_12516, Lehtola2023_JCTC_2502} and perform calculations within the generalized-gradient approximation (GGA), employing the Perdew--Burke--Ernzerhof (PBE) exchange-correlation functional\cite{Perdew1996_PRL_3865, Perdew1997_PRL_1396} as implemented in \Libxc{}\cite{Lehtola2018_S_1} using the \verb|gga_x_pbe-gga_c_pbe| keywords.
In \cref{sec:bste} we will additionally perform calculations within the local density approximation\cite{Bloch1929_ZP_545, Dirac1930_MPCPS_376} employing the Perdew--Wang (PW92) correlation functional,\cite{Perdew1992_PRB_13244} and within the meta-GGA approximation employing the $r^2$SCAN exchange-correlation functional,\cite{Furness2020_JPCL_8208, Furness2020_JPCL_9248} also as implemented in \Libxc{} using the \texttt{lda\_x-lda\_c\_pw}, and \texttt{mgga\_x\_r2scan-mgga\_c\_r2scan} keywords, respectively.
The employed level of theory gave reliable results in the case of hard-wall confinement\cite{Aastroem2025_JPCA_2791} and we expect this to hold in this work as well.
As in our previous work, all calculations considered in this work are converged with respect to the number of radial elements.
Unless stated otherwise, the calculations were converged such that the energy changes less than 1 $\upmu \Eh{}$ upon the addition of further elements.

\section{Results \label{sec:results}}

We recently studied confinement effects in the H--Xe atoms in \citeref{Aastroem2025_JPCA_2791}, and found the behavior to be more or less systematic for most atoms.
In this study, we are mostly motivated by NAO generation where the confinement is usually weak enough so that it does not result in changes to the ground state configuration.
However, in order to illustrate potential derivative discontinuities, we will examine stronger confinement than what is typically used in practice.

In the following, we will study the Mg and Ca atoms in soft confinement in their 1s$^2$2s$^2$2p$^6$3s$^2$ and 1s$^2$2s$^2$2p$^6$3s$^2$3p$^6$4s$^2$ configurations, respectively.
Since Mg and Ca have extended 3s and 4s orbitals, respectively, they are textbook examples of cases where confining potentials are beneficial for NAO generation, as the sparsity resulting from the finite support of NAO basis functions leads to large savings in large systems.
Due to the similarity of the results for Mg and Ca, we focus exclusively on the Mg atom in the main text, and present the analogous results for the Ca atom in the Supporting Information (SI).

As confinement potentials, we will consider the finite-barrier potential of \cref{eqn:soft-wall} in \cref{sec:finite-barrier}, the polynomial and exponential potentials of \cref{eqn:rn-pot,eqn:exp-pot} in \cref{sec:poly-exp}, and the singular potentials of  \cref{sec:singular} with various exponents $n \in \{1,2,3\}$ in \cref{eqn:genblum-pot}.
The exponential potential of \cref{eqn:exp-pot} and the singular potential of \cref{eqn:genblum-pot} with $n=3$ have not been considered in the literature so far to the best of our knowledge.

We perform the same analysis for the various potentials, using the same confinement potential for all orbitals.
For each potential, we study the contraction of the occupied orbitals of the ground state of Mg and Ca, and demonstrate how the various confinement potentials lead to similarly localized orbitals (\cref{sec:barrier-contract,sec:poly-exp-contract,sec:singular-contract} for finite barriers, polynomial and exponential potentials, and singular potentials, respectively).
For the regular finite-barrier potential, as well as the polynomial and exponential potentials we also demonstrate the strict localization of the solution, which is easily demonstrated with the facile control over boundary conditions in FEM calculations (\cref{sec:barrier-trunc,sec:poly-exp-trunc}).
This procedure, which amounts to combining the soft-wall confinement with a hard-wall boundary, results in the best of both worlds: the hard wall ensures strict finite support of the resulting NAO basis, while the soft potential ensures smooth radial decay.
We study how all potentials approach the hard-wall limit, and point out that the approach is smooth and systematic (\cref{sec:barrier-hw,sec:poly-exp-hw,sec:singular-hw}).

We finalise the analysis in \cref{sec:bste} with an assessment of the basis set truncation errors (BSTEs) obtained with NAOs generated with the singular potentials of \cref{eqn:genblum-pot}: we study how the BSTEs depend on the potentials' parameters and discuss how this relates to the computational performance in polyatomic calculations.

\subsection{Finite barrier} \label{sec:finite-barrier}

\subsubsection{Contracting the orbitals} \label{sec:barrier-contract}

We start our analysis by studying the localization of the orbitals of Mg in finite-barrier confinement (\cref{eqn:soft-wall}).
To study the orbitals' dependence on the form of the confinement potential, we will consider finite barriers with $V_0$ that range from weak confinement to confinement that approaches the hard-wall limit (see \cref{sec:barrier-hw}).
We consider $r_0\in\{2.0,3.0,4.0\}\ a_0$ in this section; $r_0 \gtrsim 2\ a_0$ is large enough that the 1s, 2s, and 2p core orbitals no longer feel the effects of the confinement, and only the valence 3s orbital is affected (see SI).

As discussed in \cref{sec:computational-details}, the exact solution is found in FEM by converging the radial grid: truncating the grid at $r=r_\infty$ is analogous to placing a hard-wall potential at $r=r_\infty$.
It was observed in \citeref{Lehtola2023_JPCA_4180} that the 3s orbital goes to zero linearly in hard-wall confinement, exhibiting the expected derivative discontinuity at $r=r_\infty$.
Under soft confinement, the orbital and its derivatives should go to zero more smoothly, and as discussed in \cref{sec:theory}, the values of the occupied radial orbitals should then quickly become negligible in increasing $r$.

\begin{figure}
\centering
\begin{subfigure}[b]{.49\textwidth}
\includegraphics[width=\textwidth]{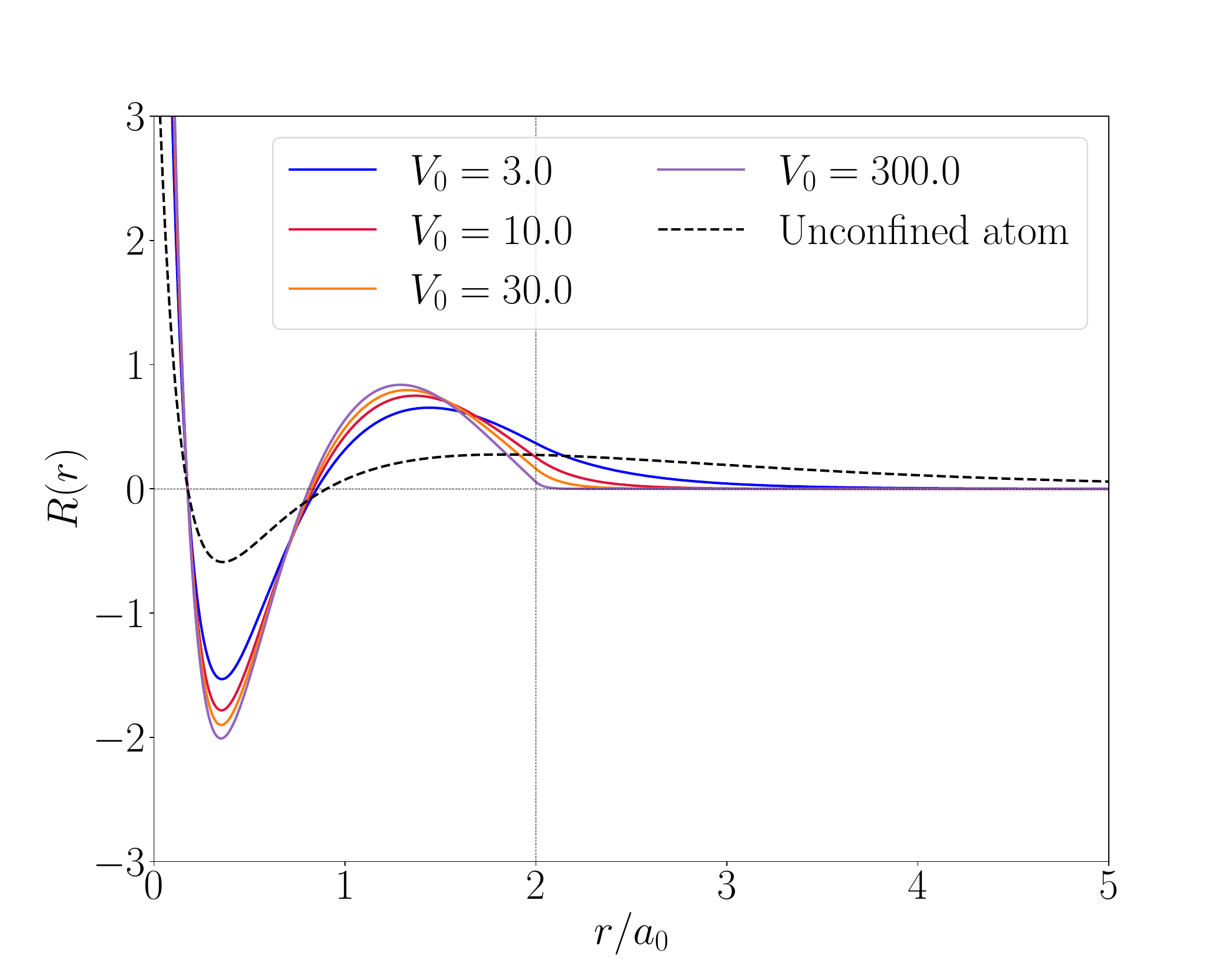}
\caption{$r_0=2\ a_0$.}
\label{fig:Mg_3s_2.0_lin_barrier}
\end{subfigure}
\begin{subfigure}[b]{.49\textwidth}
\includegraphics[width=\textwidth]{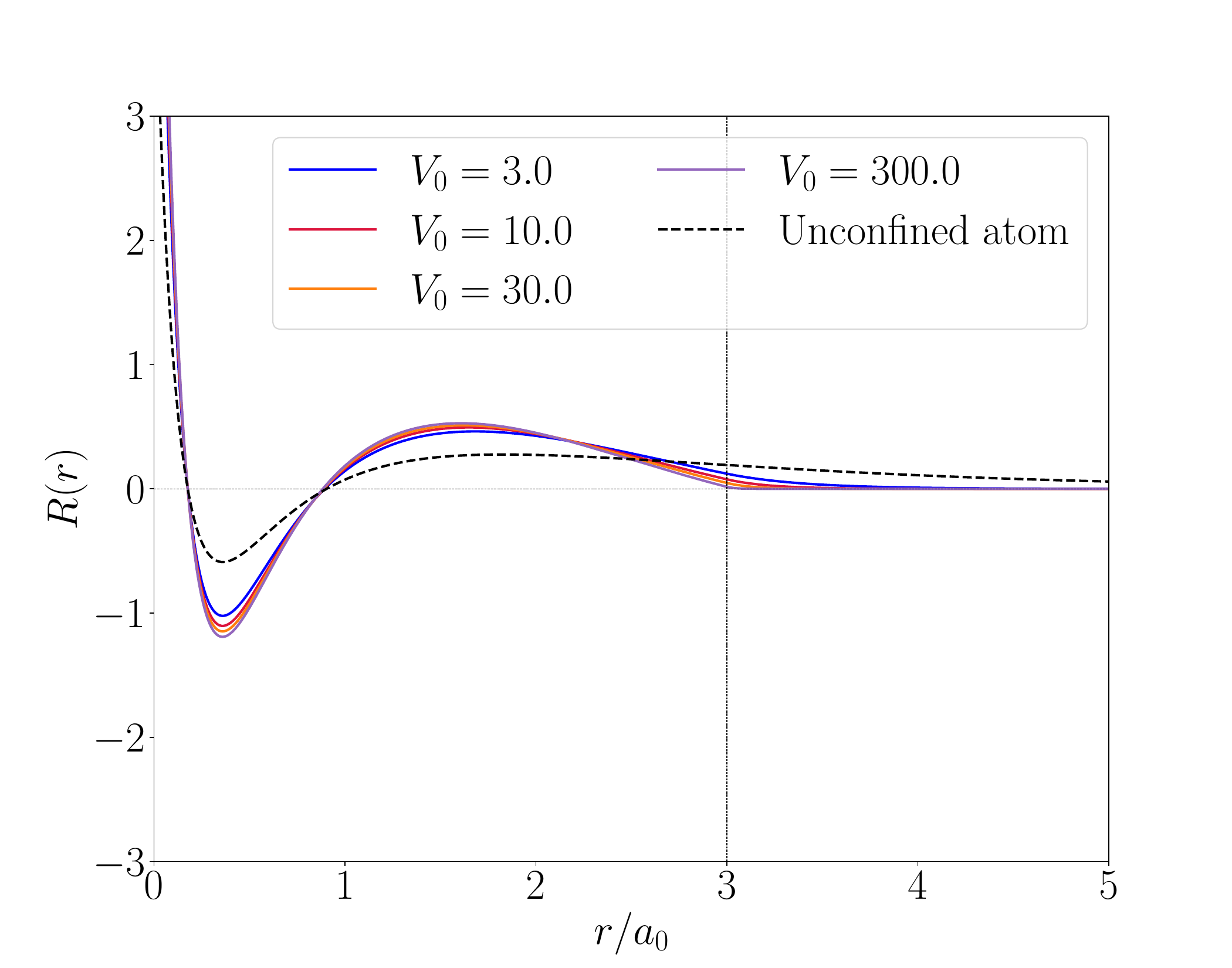}
\caption{$r_0=3\ a_0$.}
\label{fig:Mg_3s_3.0_lin_barrier}
\end{subfigure}
\begin{subfigure}[b]{.49\textwidth}
\includegraphics[width=\textwidth]{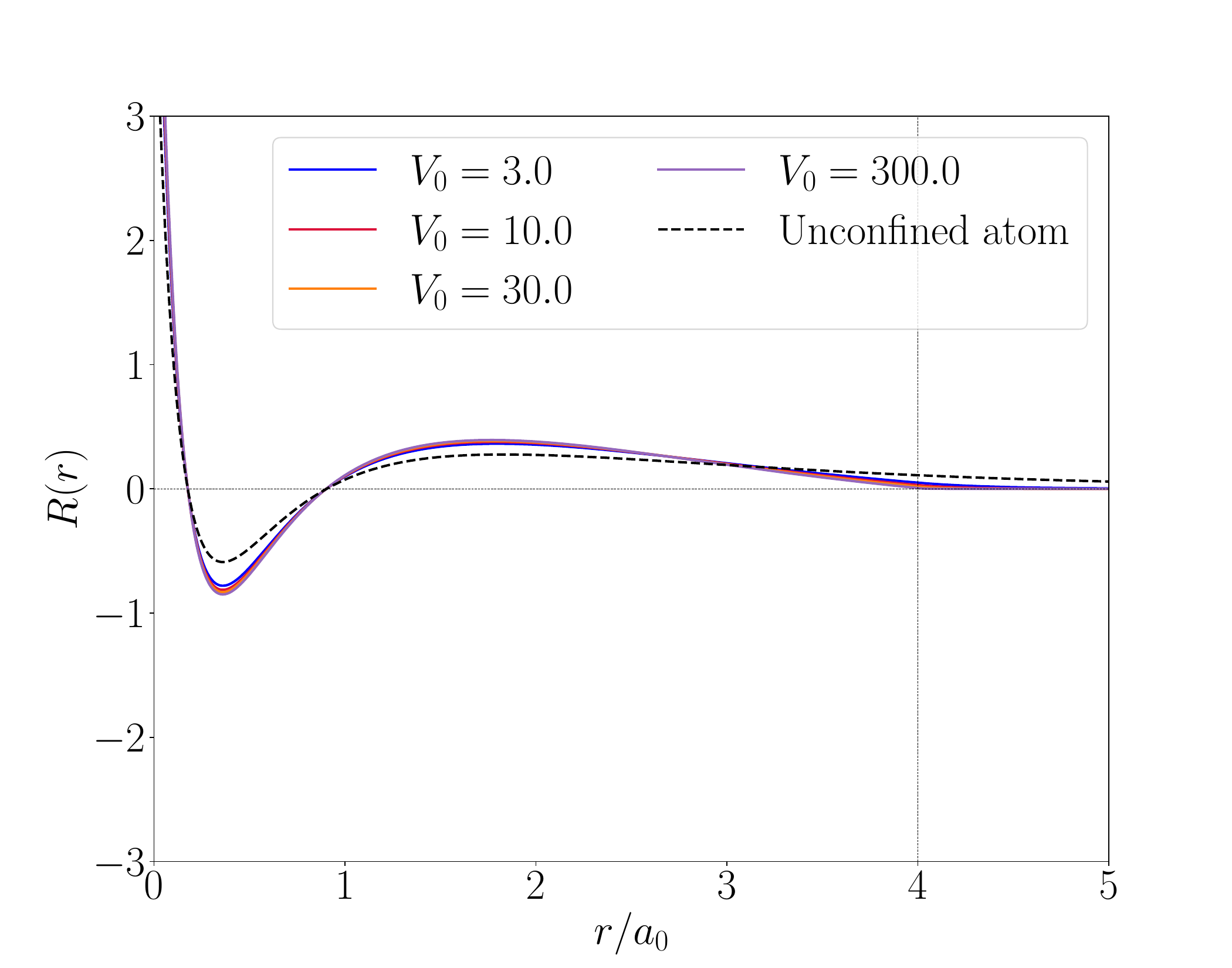}
\caption{$r_0=4\ a_0$.}
\label{fig:Mg_3s_4.0_lin_barrier}
\end{subfigure}
\caption{Radial part of the 3s orbital of Mg in finite-barrier confinement with varying $V_0$ and $r_0$.}
\label{fig:mg-barrier}
\end{figure}

The radial part of the Mg 3s orbital in finite-barrier confinement is depicted in \cref{fig:mg-barrier}.
The differences between the orbitals for various barrier heights are significant for $r_0=2\ a_0$, as can be seen in \cref{fig:Mg_3s_2.0_lin_barrier}.
Increasing the size of the cavity to $r_0 = 3\ a_0$, we see that the orbitals now behave more similarly, but also that differences in the orbital amplitudes can still be observed, especially for the lower barriers.
Increasing the size of the cavity further to $r_0=4\ a_0$, the form of the orbital becomes less dependent of the barrier height, while the orbitals still go clearly more quickly to zero than in the unconfined atom.
Analogous observations can be made for the 4s orbital of Ca, except that small differences can still be observed for different $V_0$ at $r_0=4\ a_0$, as can be expected from the Ca 4s orbital's larger extent in the unconfined atom.

\begin{figure}
\centering
\begin{subfigure}[b]{.49\textwidth}
\includegraphics[width=\textwidth]{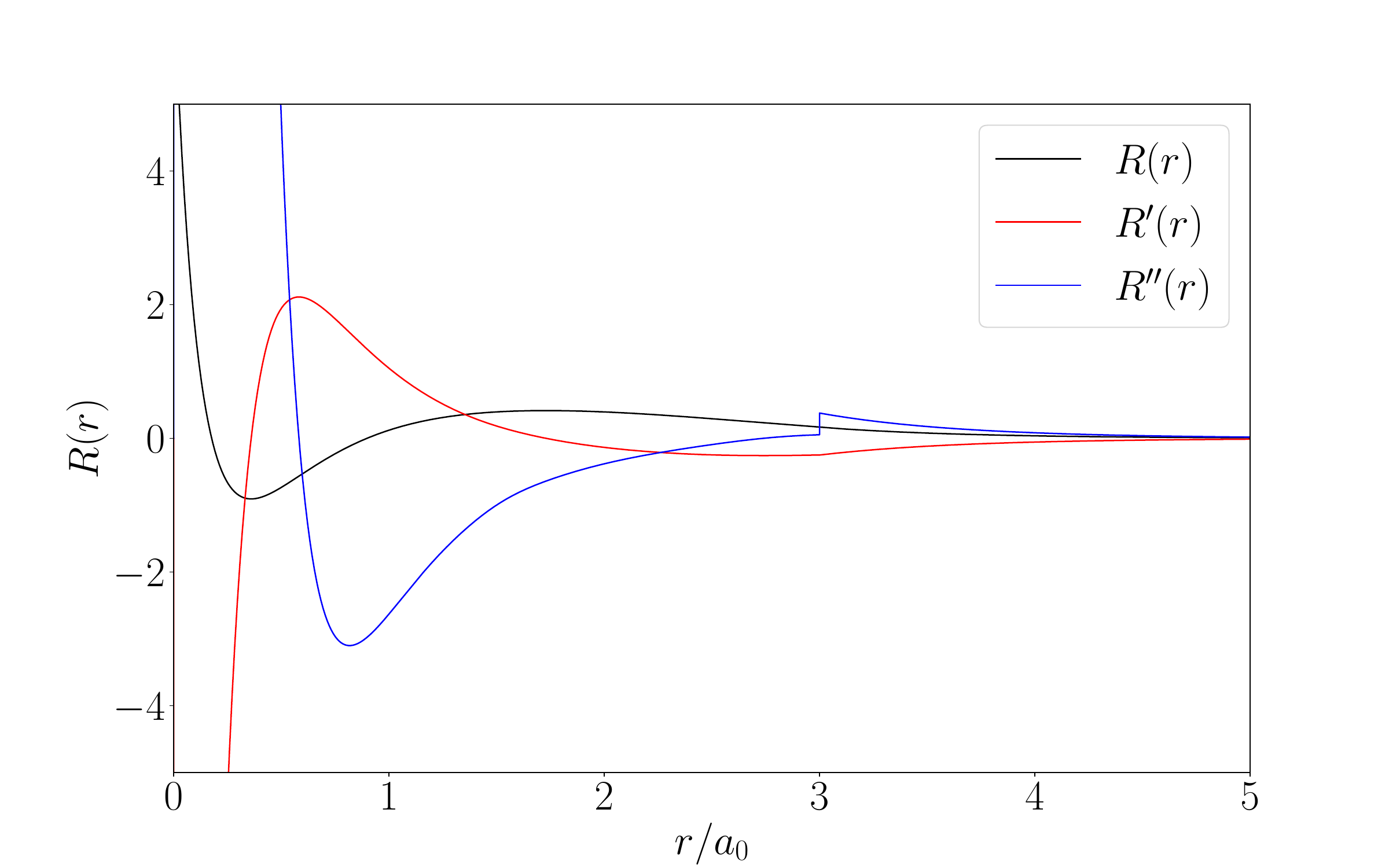}
\caption{$V_0=1.0$ \Eh.}
\label{fig:der_1.0}
\end{subfigure}
\begin{subfigure}[b]{.49\textwidth}
\includegraphics[width=\textwidth]{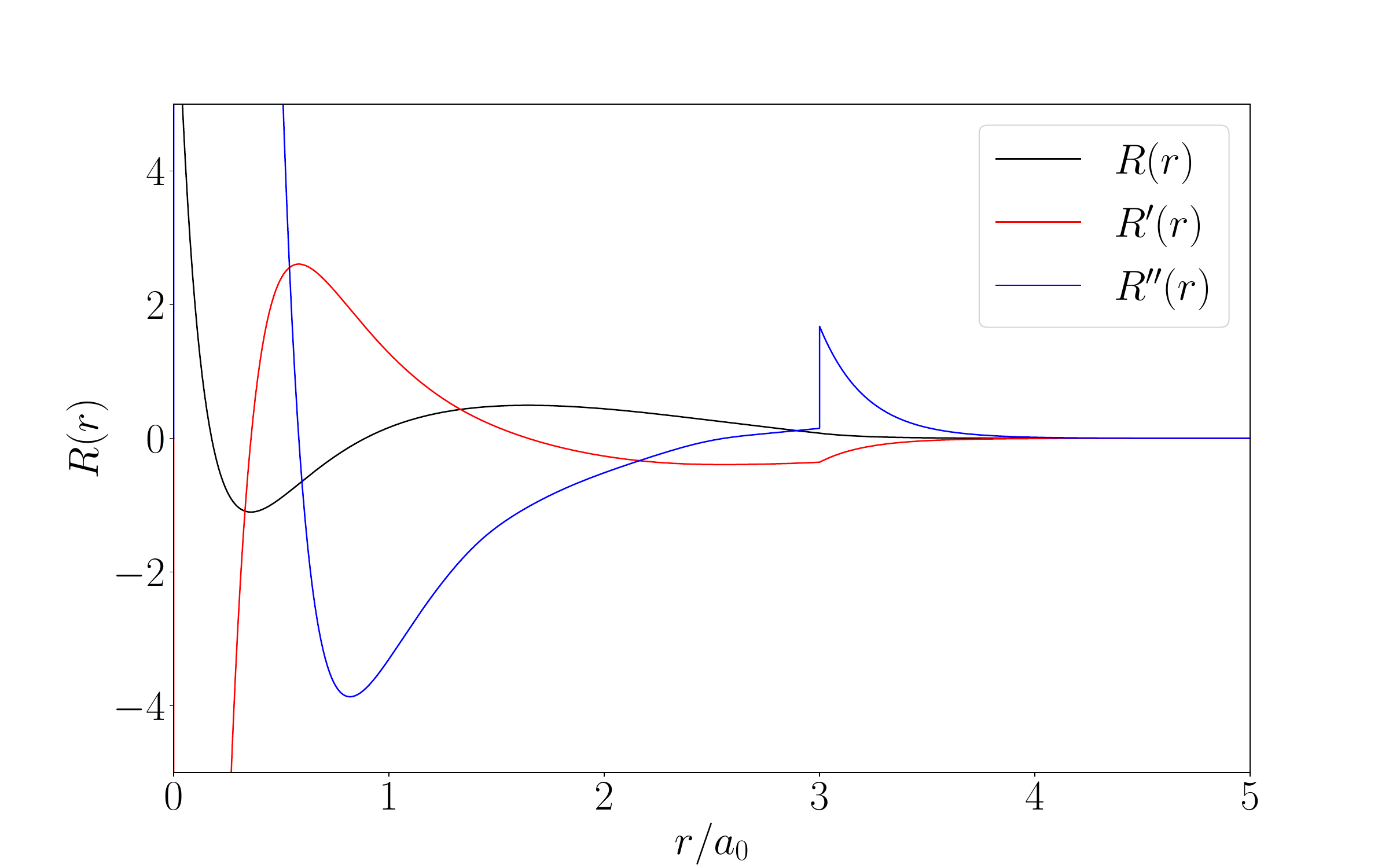}
\caption{$V_0=10.0$ \Eh.}
\label{fig:der_10.0}
\end{subfigure}
\begin{subfigure}[b]{.49\textwidth}
\includegraphics[width=\textwidth]{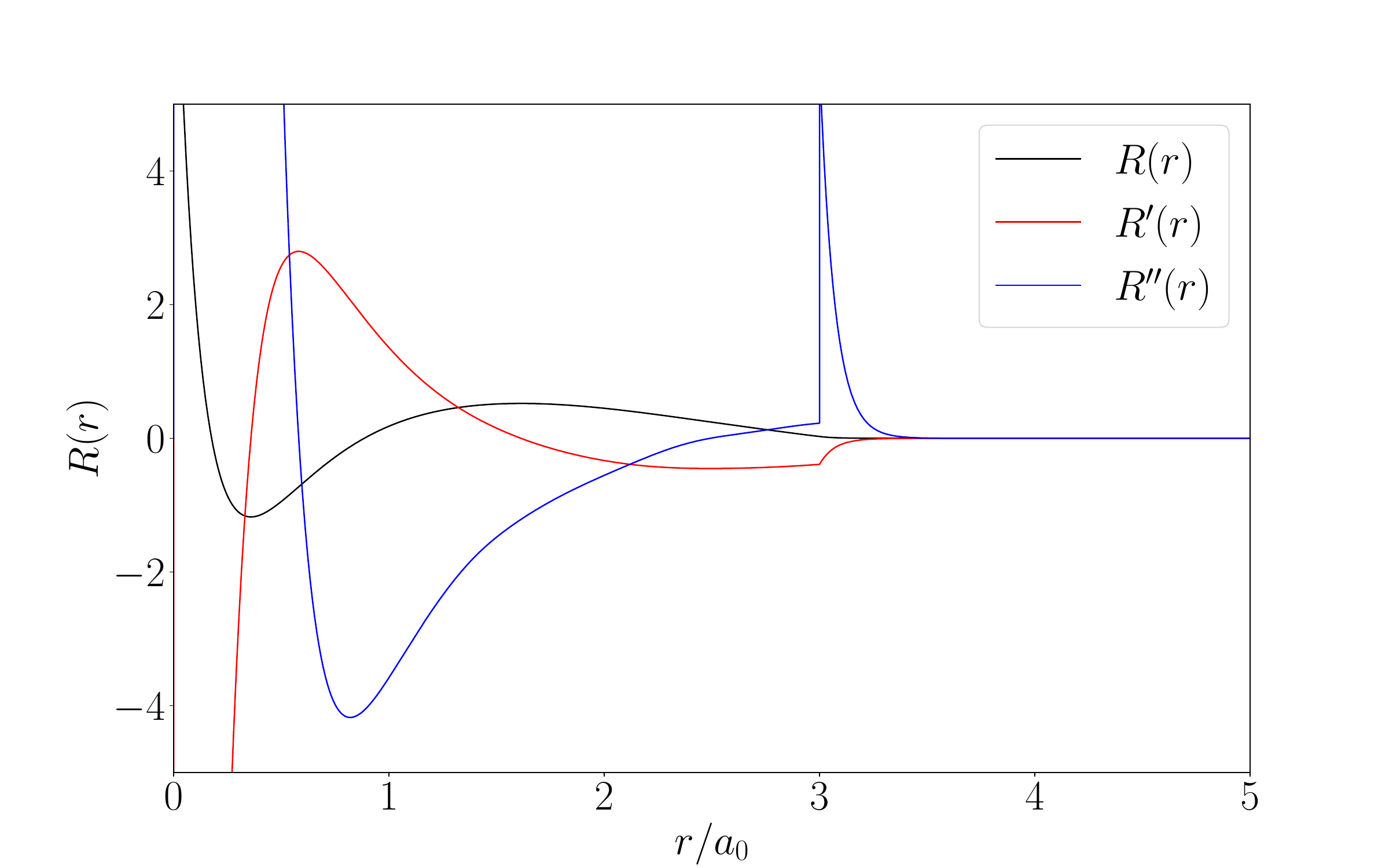}
\caption{$V_0=100.0$ \Eh.}
\label{fig:der_100.0}
\end{subfigure}
\caption{The radial part of the 3s orbital of the Mg atom as well as its first and second derivatives in the finite-barrier potential at $r_0=3\ a_0$ with various barrier heights.}
\label{fig:derivatives_barrier}
\end{figure}

However, as discussed in \cref{sec:theory}, even though the decay of the orbital is smooth, the second derivative of the radial wavefunction is discontinuous at $r=r_c$ due to the discontinuity of the potential, and a kink is observed in the first derivative.
This non-smooth behavior is demonstrated in \cref{fig:derivatives_barrier}.

\subsubsection{Truncating the radial grid} \label{sec:barrier-trunc}

When the radial orbitals go quickly to zero, the radial expansion can be truncated to finite support with the introduction of a hard-wall boundary.
We find the suitable truncation radius with a binary search algorithm: taking the calculation with a converged value of $r_\infty$ around $15\ a_0$ as reference ($r_\infty = 40\ a_0$ without confinement), we find the value for $r_\infty$ that leads to an energy that is exactly 1 $\upmu \Eh{}$ higher than the energy converged to the complete basis set (CBS) limit.
Correspondingly, the calculations for each choice of $r_\infty$ were converged to higher precision with respect to the number of employed radial elements.
A 1 $\upmu \Eh{}$ change in the total energy is minimal, and we can interpret that the orbital has already become negligible at the corresponding truncation radius just due to the soft confinement potential as discussed in \cref{sec:theory}.

For this part of the study, in addition to the LIP shape functions, we also consider Hermite interpolating polynomials (HIPs) recently introduced to atomic calculations in \citeref{Lehtola2023_JPCA_4180}.
In addition to controlling the boundary values of the orbitals, the HIP basis enables control of the values of the first derivative at the boundaries.
We therefore augment our calculations with the 15-node LIP basis with ones performed in an 8-node HIP basis, which has the same accuracy as the 15-node LIP basis.\cite{Lehtola2023_JPCA_4180}
Moreover, we consider two types of calculations with the HIP basis: one where a finite value for the derivative is allowed at $r_\infty$, and another where the derivative is forced to go to zero at $r_\infty$ (denoted as HIP').\cite{Lehtola2023_JPCA_4180}
The LIP and HIP calculations are mathematically equivalent, while HIP' provides an upper bound for this energy.

\input{tables/rmax_table-barrier-Mg.tex}

The obtained values for $r_\infty$ with these three methods are tabulated in \cref{tab:rmax-table-barrier-Mg}.
The truncation obviously depends strongly on the employed form of the potential, that is, the values of $V_0$ and $r_0$.

The expected behavior is that $r_\infty$ approaches $r_0$ at the limit of strong confinement, \ie{}, large $V_0$.
For the lowest barrier $V_0=3.0$ \Eh{}, $r_\infty$ is larger when $r_0=2\ a_0$ than when $r_0=3\ a_0$.
This appears paradoxical, but as the calculations have been converged to the CBS limit, we tentatively explain this by the weakest barrier being unable to overcome the energy increase that would be associated with contracting the orbital for an $r_0$ that is too small.

\citeit{Connerade2000_JPBAMOP_251} used $V_0=10$ \Eh{} in a study of the 3d and 4d atoms, but we see here that such a barrier is not high enough to strongly localize the Mg 3s orbital, as the value of $r_\infty$ is 1.3--1.8 $a_0$ larger than $r_0$.
The weakest barriers are even less effective at localizing the Ca 4s orbital, as can be seen in the analogous table in the SI.
\citeit{Ozaki2003_PRB_155108} on the other hand provided the value $V_0=3\times 10^4$ \Eh{} in an example figure; this value is practically at the hard-wall limit, as we will also see below in \cref{sec:barrier-hw}.

Finalising the analysis of the data in \cref{tab:rmax-table-barrier-Mg}, we also see that forcing the derivative to vanish at $r_\infty$ only changes $r_\infty$ by a small amount from the LIP value, suggesting that in addition to the wave function, also the derivative is well behaved and smooth, becoming negligible along with the wave function.
We note, however, that this is due to the use of a small energy threshold in choosing the value of $r_\infty$.

\begin{figure}
	\centering
	\begin{subfigure}[b]{.45\textwidth}
		\includegraphics[width=\textwidth]{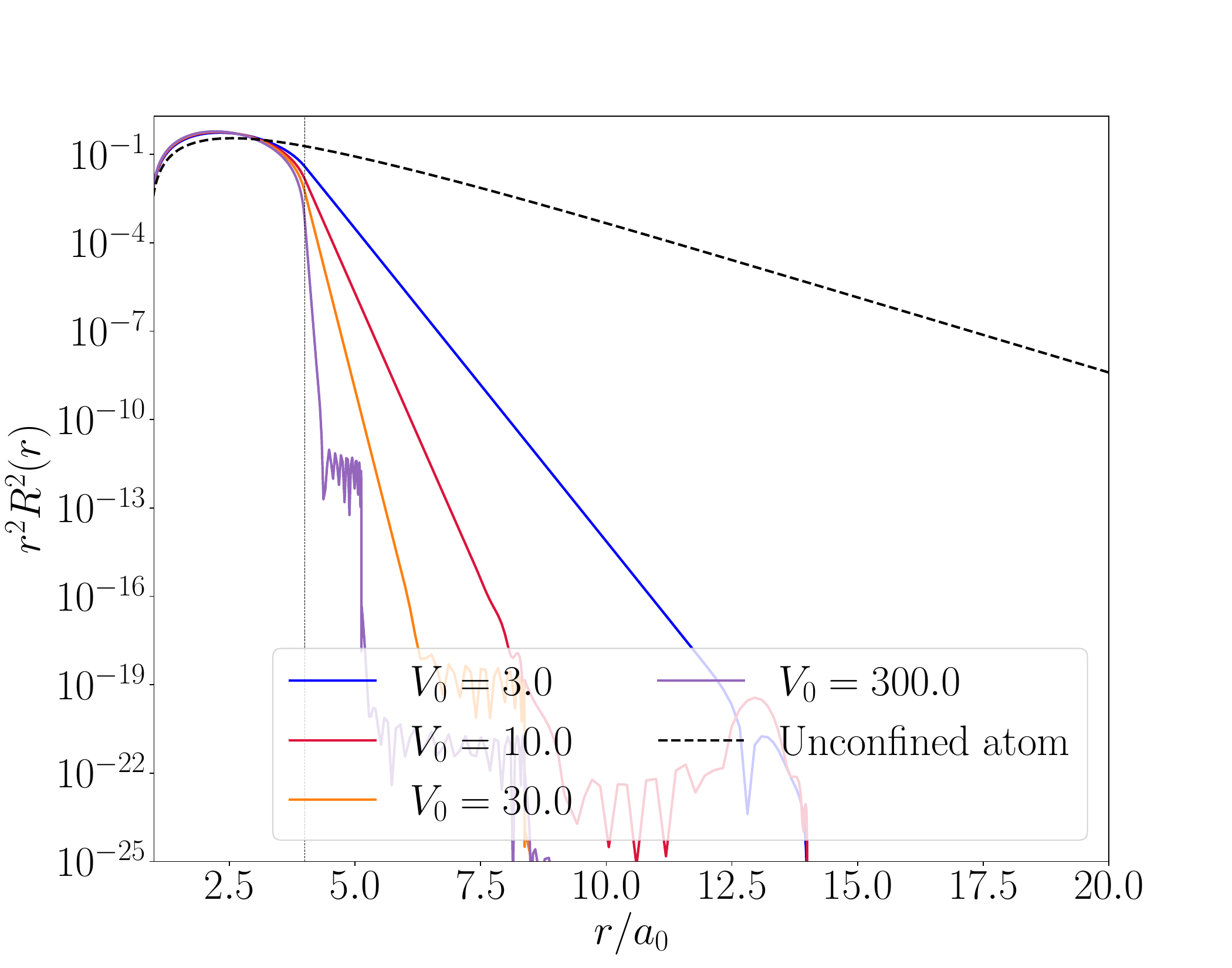}
		\caption{Polynomial confinement without cutoff}
		\label{fig:barrier_log_no-cutoff}
	\end{subfigure}
	\begin{subfigure}[b]{.45\textwidth}
		\includegraphics[width=\textwidth]{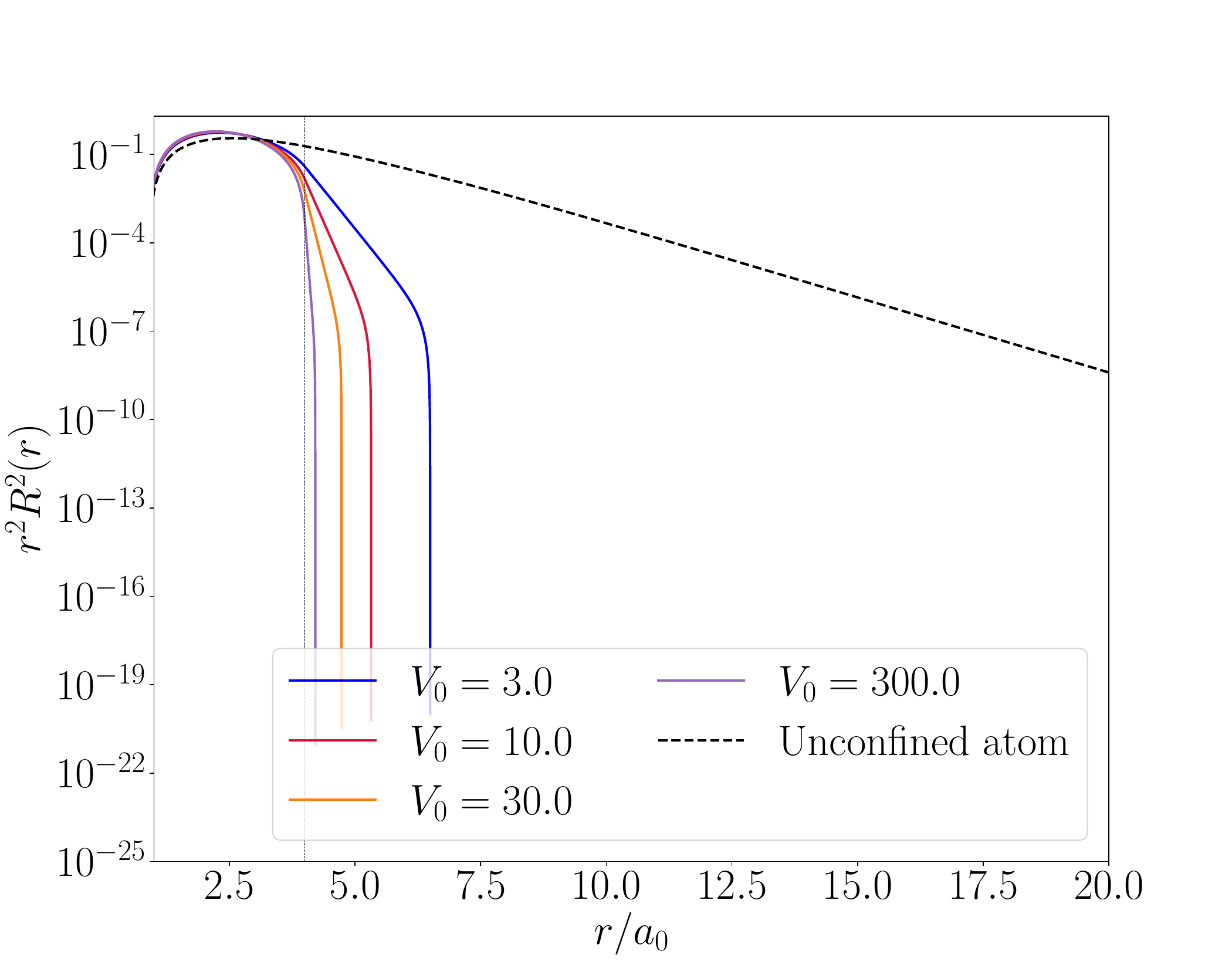}
		\caption{Polynomial confinement with cutoff}
		\label{fig:barrier_log_cutoff}
	\end{subfigure}
\caption{Radial density of the 3s orbital of Mg without (\cref{fig:barrier_log_no-cutoff}) and with (\cref{fig:barrier_log_cutoff}) adaptive cutoff in finite-barrier confinement with varying $V_0$ and $r_0=4\ a_0$, according to the procedure of \cref{tab:rmax-table-barrier-Mg}. Note semilogarithmic scale.}
\label{fig:barrier_log}
\end{figure}

The decay of the 3s orbital is more clearly demonstrated by the density plot in \cref{fig:barrier_log} for the case $r_0=4\ a_0$.
When $r_\infty=15\ a_0$ in \cref{fig:barrier_log_no-cutoff}, we see that the orbitals decay exponentially to $r^2 R(r)^2 \approx 10^{-13}$, after which only numerical noise resides.

The corresponding calculations employing the hard wall truncation with the 1 $\upmu \Eh{}$ energy criterion are depicted in \cref{fig:barrier_log_cutoff}.
Now the orbitals decay quickly to around $r^2 R(r)^2 \approx 10^{-7}$, after which they are truncated and we observe no noise.

\subsubsection{Approaching the hard wall limit} \label{sec:barrier-hw}

\begin{figure}
\centering
\includegraphics[width=0.5\textwidth]{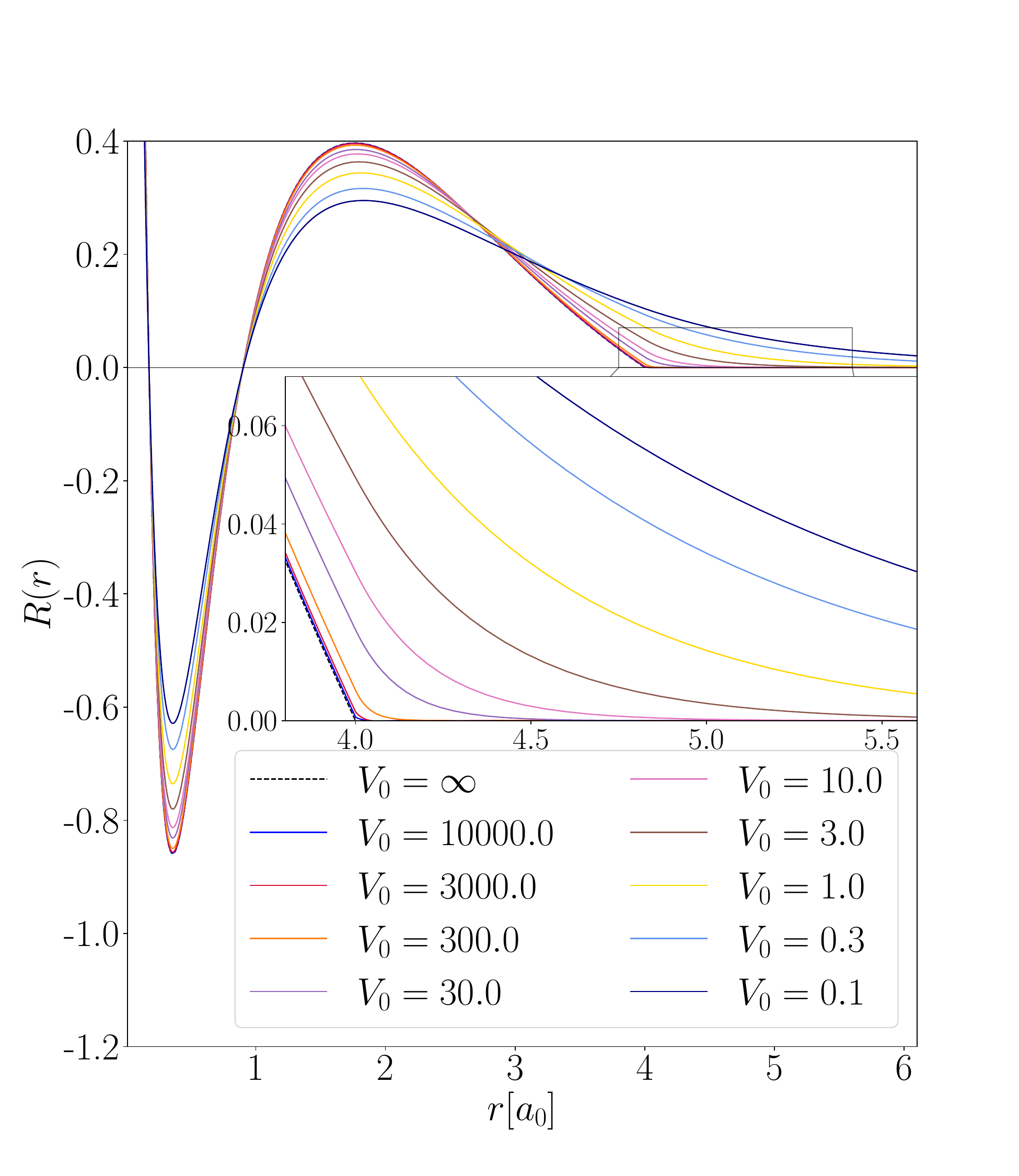}
\caption{The radial part of the 3s orbital of the Mg atom confined by the finite barrier potential of \cref{eqn:soft-wall} at $r_0=4.0$ with various values of $V_0$, and by the hard-wall potential at $r_\infty=4.0$.}
\label{fig:barrier-hw}
\end{figure}

\begin{table}
\centering
\shifttabmgbarrier
\caption{Values of $r_\infty$ in $a_0$ that minimize the norm in \cref{eqn:radial-norm} between (i) the 3s orbital of the Mg atom in the finite-barrier potential at $r=4\ a_0$ for various values of $V_0$ and (ii) the hard-wall potential at given $r_\infty$.}
\label{tab:barrier_shift}
\end{table}

We conclude the analysis of the finite-barrier potential by investigating how the orbitals in the potential approach the hard-wall limit.
The 3s orbital of Mg in the finite-barrier potential at $r=4\ a_0$ with various $V_0$ are depicted in \cref{fig:barrier-hw} together with the orbitals of the hard-wall potential.
By increasing $V_0$, the orbital approaches hard-wall confinement systematically, and when $V_0=3000$ \Eh, the form of the orbital is no longer quantitatively changed when increasing the barrier height.
A closer examination of the figure shows that a small tail still penetrates the barrier even at $V_0 =10\ 000$ \Eh.

To every orbital obtained with the specific values of $V_0$, we fit a hard-wall confined orbital in the following way.
For each $V_0$ studied in \cref{fig:barrier-hw} we find the location of the hard-wall boundary $r_\infty$ that minimizes the norm with respect to $r_\infty$
\begin{equation} \label{eqn:radial-norm}
\begin{split}
&\Vert R_\mathrm{soft}(r)-R_\mathrm{hw}(r;r_\infty)\Vert \\
&= \int_0^{r_\infty}r^2\left[R_\text{soft}(r)-R_\text{hw}(r;r_\infty)\right]^2\text{d}r \\
&= \sum_{\alpha=1}^nw_\alpha r_\alpha^2\left[R_\mathrm{soft}(r_\alpha)-R_\mathrm{hw}(r_\alpha;r_\infty)\right]^2 \\
&:=\Vert\Delta\Vert
\end{split}
\end{equation}
where we interpolate the values of $R_\mathrm{soft}(r)$ to the integration points $\{r_\alpha\}_{\alpha=1}^n$ of the hard-wall (hw) orbital $R_\mathrm{hw}(r_\alpha;r_\infty)$.
By minimizing the norm of \cref{eqn:radial-norm} with respect to $r_\infty$, we get the values of $r_\infty$ and $||\Delta||$ shown in \cref{tab:barrier_shift}.
This data confirms our analysis; we see that the value of the optimized values of $r_\infty$ and $||\Delta||$ decrease systematically and smoothly.
Furthermore, we see that when we go from $V_0=3\ 000$ \Eh{} to $V_0 =10\ 000$ \Eh, the value of $r_\infty$ does not change and the value of $||\Delta||$ decreases only by a factor of 3.
These results confirm that the hard-wall solution can indeed be approached by a sufficiently high finite barrier.

\subsection{Polynomial and exponential potentials} \label{sec:poly-exp}

\begin{figure*}
  \centering
  \begin{subfigure}[b]{.45\textwidth}
    \centering
    \includegraphics[width=\textwidth]{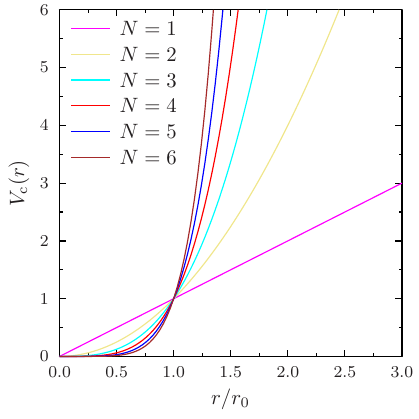}
    \caption{Polynomial confinement potential of \cref{eqn:rn-pot} as a function of $r/r_0$ with $N\in[1,6]$.}
    \label{fig:polpot}
  \end{subfigure}
  \hfill
 \begin{subfigure}[b]{.45\textwidth}
    \centering
    \includegraphics[width=\textwidth]{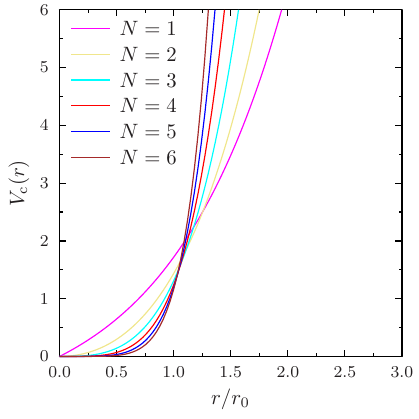}
    \caption{Exponential confinement potential of \cref{eqn:exp-pot} as a function of $r/r_0$ with $N\in[1,6]$.}
    \label{fig:exppot}
 \end{subfigure}
 \caption{The polynomial (\cref{fig:polpot}) and exponential (\cref{fig:exppot}) confinement potentials considered in this work.}
 \label{fig:pots}
\end{figure*}

\begin{figure*}
  \centering
  \begin{subfigure}[b]{.45\textwidth}
    \centering
    \includegraphics[width=\textwidth]{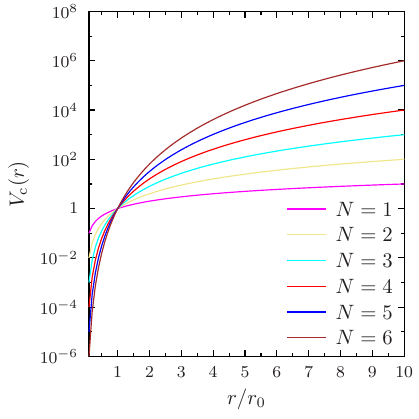}
    \caption{Polynomial confinement potential of \cref{eqn:rn-pot} as a function of $r/r_0$ with $N\in[1,6]$. Note logarithmic scale.}
    \label{fig:polpot-log}
  \end{subfigure}
  \hfill
 \begin{subfigure}[b]{.45\textwidth}
    \centering
    \includegraphics[width=\textwidth]{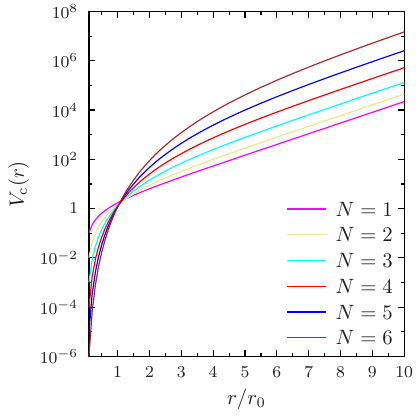}
    \caption{Exponential confinement potential of \cref{eqn:exp-pot} as a function of $r/r_0$ with $N\in[1,6]$. Note logarithmic scale.}
    \label{fig:exppot-log}
 \end{subfigure}
 \caption{Plots of the polynomial (\cref{fig:polpot-log}) and exponential (\cref{fig:exppot-log}) confinement potentials considered in this work, now in semilogarithmic scale instead to the linear scale used in \cref{fig:pots}.}
 \label{fig:logpots}
\end{figure*}

We continue the analysis with the polynomial and exponential potentials of \cref{eqn:rn-pot,eqn:exp-pot}.
We illustrate the comparison between them in \cref{fig:pots}.
The polynomial confinement potential (\cref{fig:polpot}) has the important behavior discussed by \citeit{Pasteka2020_MP_1730989} that it approaches a hard-wall potential at $r_c=r_0$ as $N\to\infty$.
Although the $N=1$ potential is significant already at small $r$, the potentials for higher $N$ are damped at $r<r_0$ and grow more rapidly at $r>r_0$ than the $N=1$ curve.
For this reason, even though the form of the polynomial potential is simple, calculations with various values for $N$ and $r_0$ allow exploration of vastly different confinement situations.

The exponential confinement potential (\cref{fig:exppot}) behaves similarly to the polynomial potential for $r<r_0$, becoming more flat as $N$ grows.
However, the exponential potential grows much more rapidly than the polynomial potential for large $r$.
This is much clearer in the semilogarithmic plots shown in \cref{fig:logpots} that fit a wider range of $x$ and $y$ values than the analogous plots in linear scale in \cref{fig:pots}.
Because of the similarity of the small-$r$ Taylor series of \cref{eqn:rn-pot,eqn:exp-pot}, also the exponential potential approaches a hard-wall potential for $N\to\infty$.

\subsubsection{Contracting the orbitals} \label{sec:poly-exp-contract}

In analogy to the analysis on finite-barrier potentials in \cref{sec:barrier-contract}, we study the orbitals of the ground state of Mg in the polynomial and exponential confinement potentials.
We consider $N\in\{1,2,4,6,8,10\}$ and again notice that radial part of the core orbitals 1s, 2s and 2p are independent of $N$ for $r_0 \gtrsim 2\ a_0$ (see SI).
We therefore consider $r_0\in\{2.0,3.0,4.0\}\ a_0$, as for the finite barrier.

\begin{figure}
    \centering
  \begin{subfigure}[b]{.45\textwidth}
    \centering
    \includegraphics[width=\textwidth]{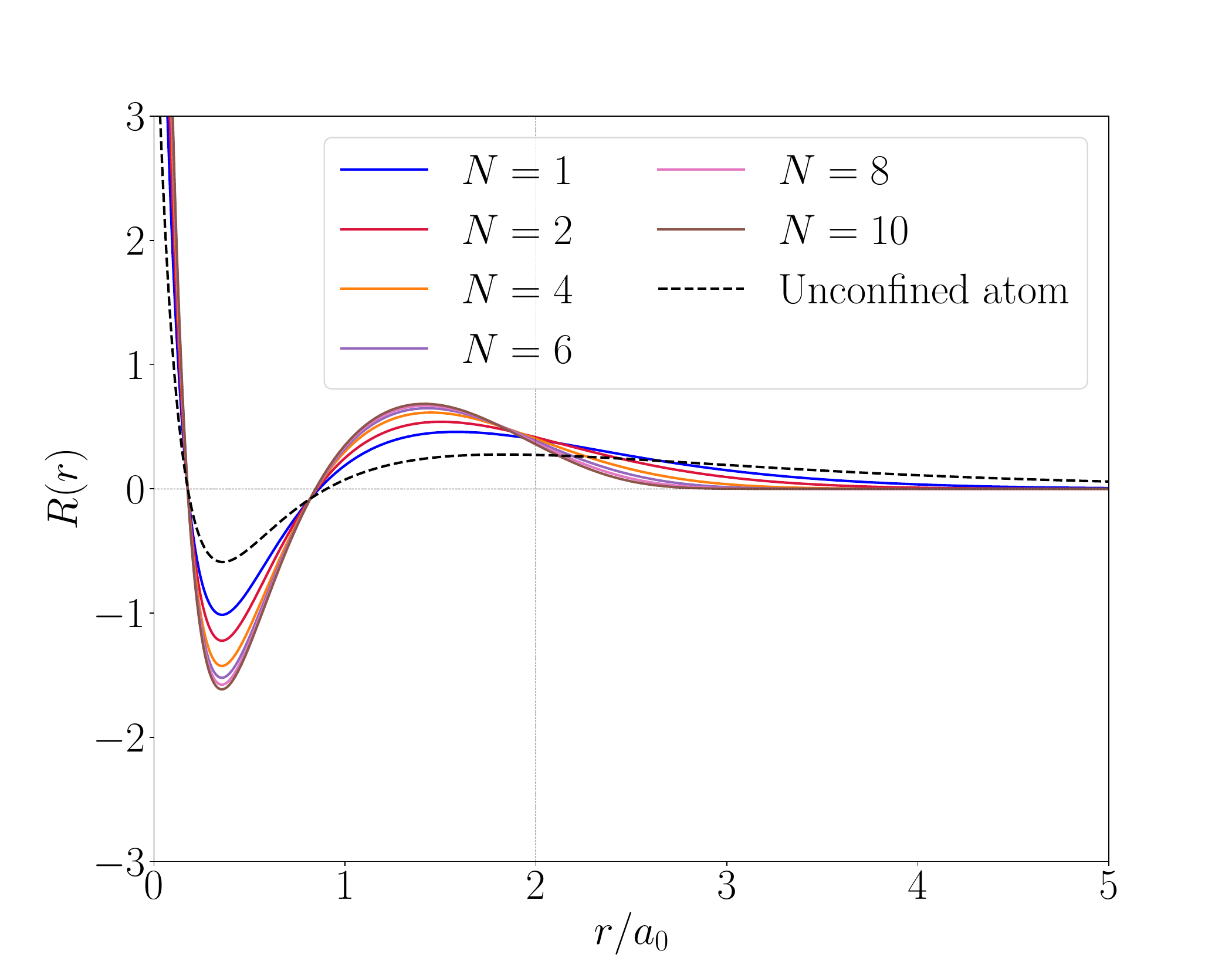}
    \caption{$r_0=2\ a_0$}
    \label{fig:Mg-poly-2}
  \end{subfigure}
  \hfill
  \begin{subfigure}[b]{.45\textwidth}
    \centering
    \includegraphics[width=\textwidth]{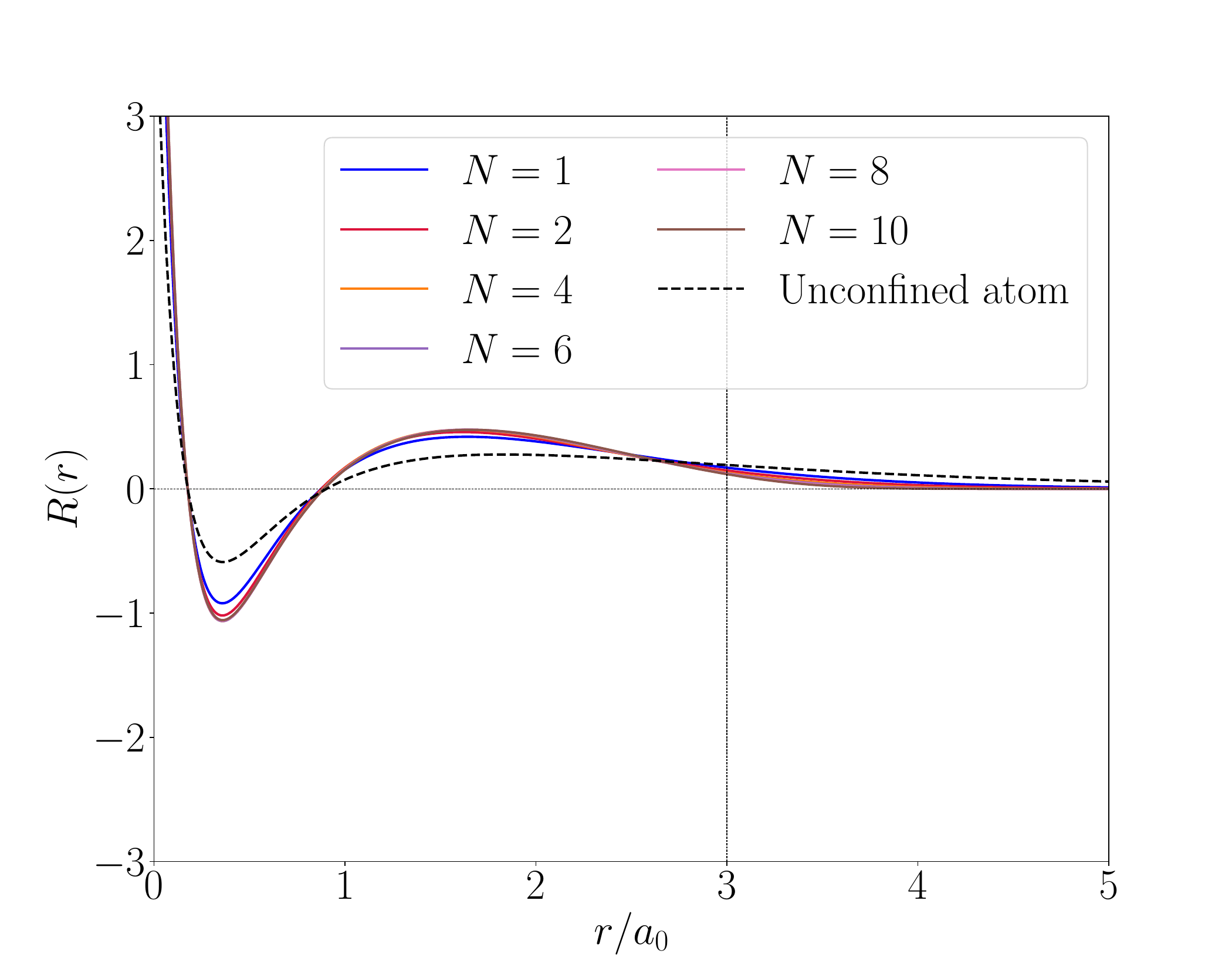}
    \caption{$r_0=3\ a_0$}
    \label{fig:Mg-poly-3}
  \end{subfigure}
  \hfill
  \begin{subfigure}[b]{.45\textwidth}
    \centering
    \includegraphics[width=\textwidth]{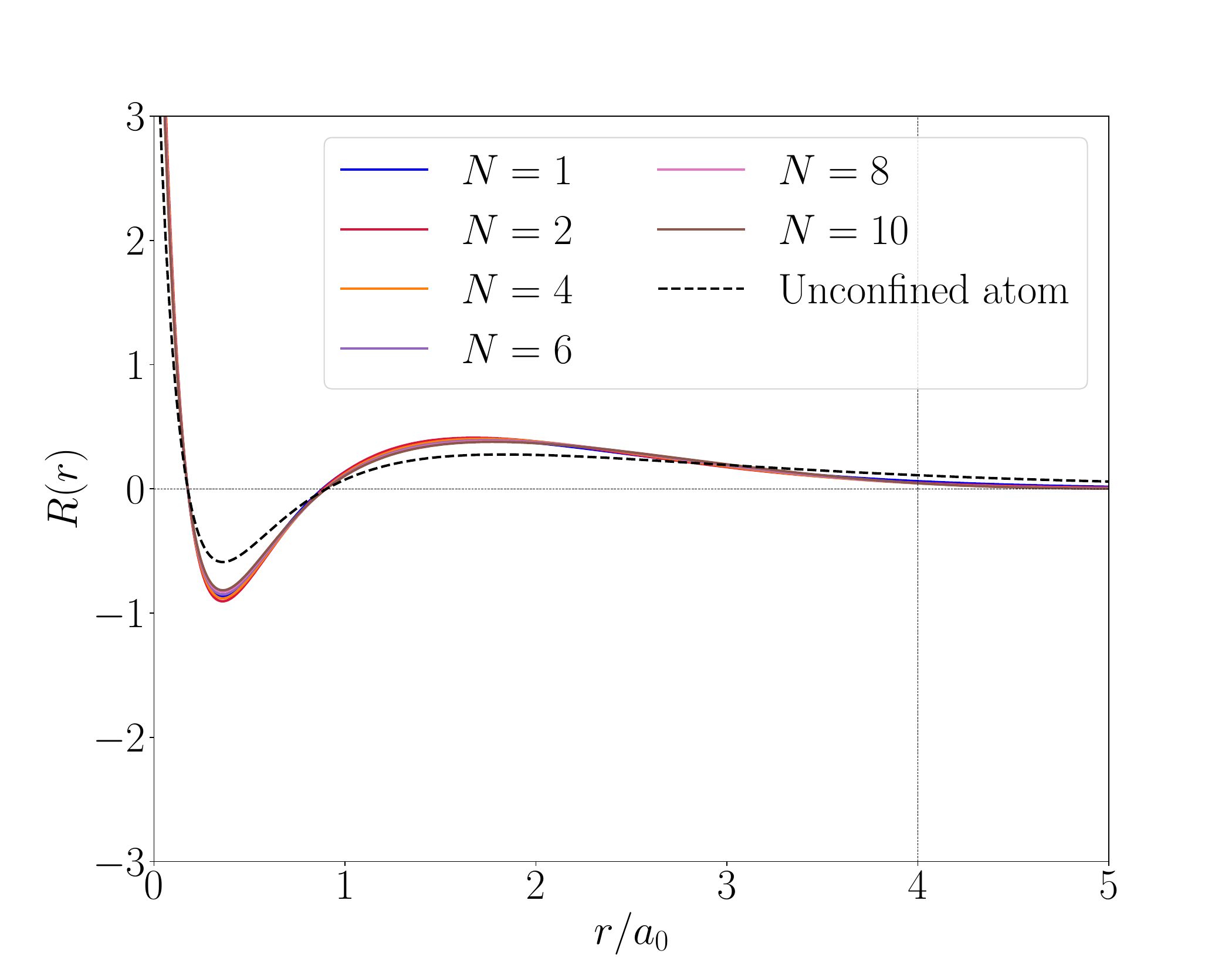}
    \caption{$r_0=4\ a_0$}
    \label{fig:Mg-poly-4}
  \end{subfigure}
  \caption{The Mg 3s orbital in polynomial confinement with various $N$ and $r_0=2\ a_0$ (\cref{fig:Mg-poly-2}), $r_0=3\ a_0$ (\cref{fig:Mg-poly-3}), and $r_0=4\ a_0$ (\cref{fig:Mg-poly-4}).}
  \label{fig:mg_valence_polynomial}
\end{figure}

\begin{figure}
    \centering
  \begin{subfigure}[b]{.45\textwidth}
    \centering
    \includegraphics[width=\textwidth]{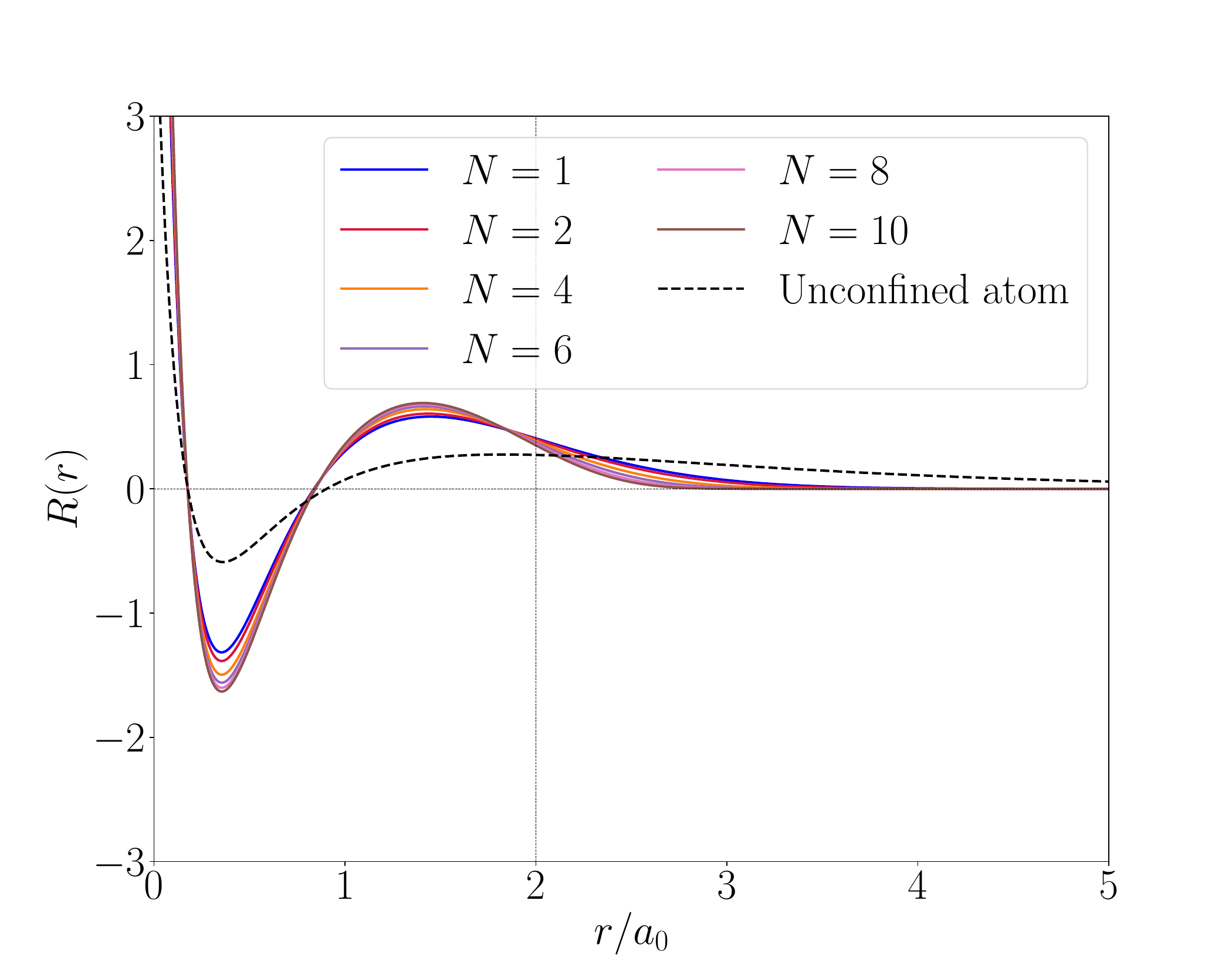}
    \caption{$r_0=2\ a_0$}
    \label{fig:Mg-expn-2}
  \end{subfigure}
  \hfill
  \begin{subfigure}[b]{.45\textwidth}
    \centering
    \includegraphics[width=\textwidth]{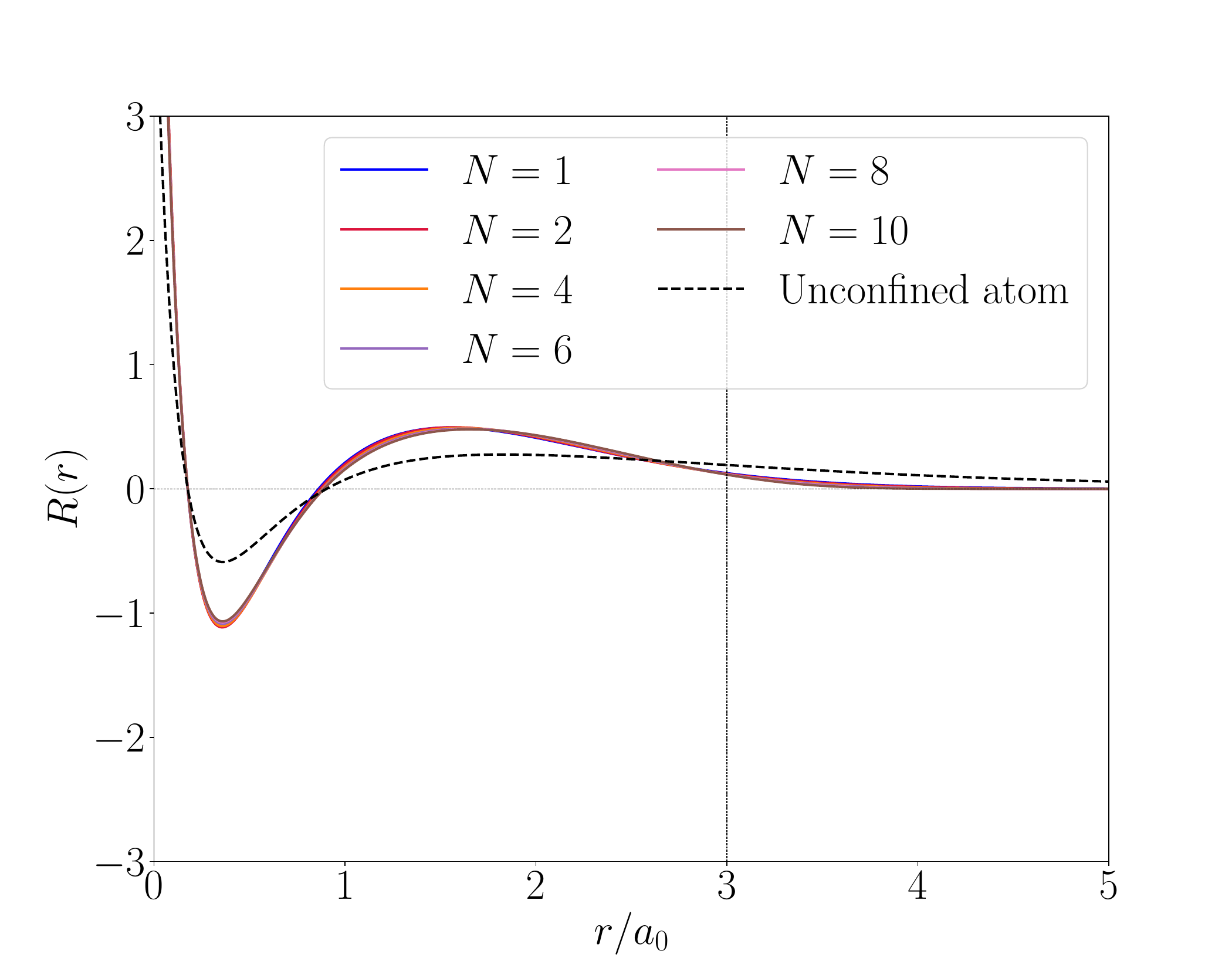}
    \caption{$r_0=3\ a_0$}
    \label{fig:Mg-expn-3}
  \end{subfigure}
  \hfill
  \begin{subfigure}[b]{.45\textwidth}
    \centering
    \includegraphics[width=\textwidth]{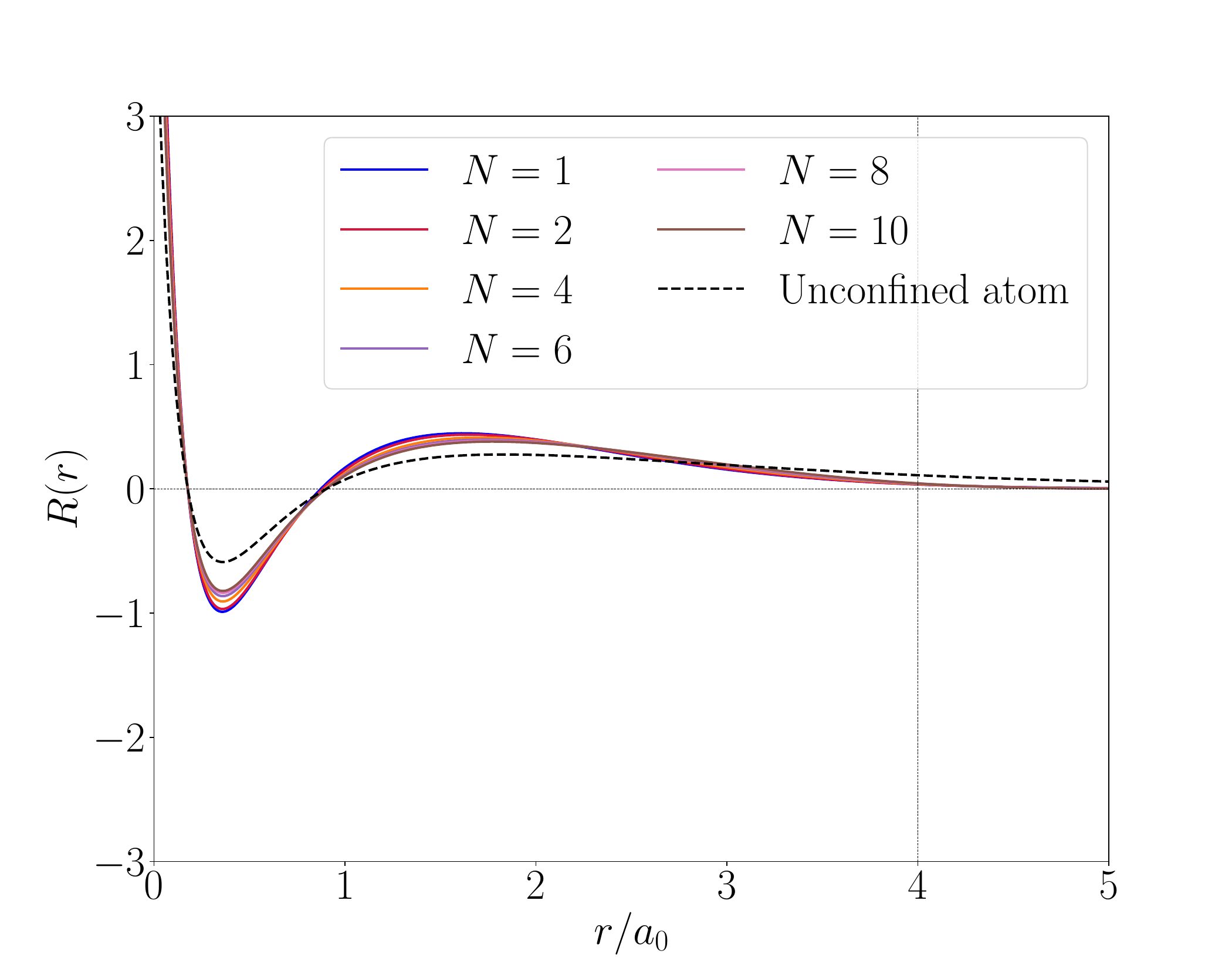}
    \caption{$r_0=4\ a_0$}
    \label{fig:Mg-expn-4}
  \end{subfigure}
  \caption{The Mg 3s orbital in exponential confinement with various $N$ and $r_0=2\ a_0$ (\cref{fig:Mg-expn-2}), $r_0=3\ a_0$ (\cref{fig:Mg-expn-3}), and $r_0=4\ a_0$ (\cref{fig:Mg-expn-4}).}
  \label{fig:mg_valence_exponential}
\end{figure}

The plots of the radial part of the 3s orbital confined by the polynomial potential are shown in \cref{fig:mg_valence_polynomial}.
The differences between the orbitals obtained with various $N$ values are significant for $r_0=2\ a_0$, but increasing the confinement radius to $r_0 = 3\ a_0$, only $N=1$ and $N=2$ differ from the others.
For $r_0=4\ a_0$, the form of the orbital appears qualitatively independent of $N$, but again the orbitals clearly vanish more rapidly than the unconfined orbital.

Analogous plots for the exponential potential are shown in \cref{fig:mg_valence_exponential}.
The differences at $r_0=2\ a_0$ between the orbitals for various $N$ now appear smaller than in polynomial confinement; a similar observation is also made for $r=3\ a_0$.
For $r_0=4\ a_0$, somewhat more dependence on $N$ in the orbital form is observed than in polynomial confinement.
In all cases, the orbitals clearly vanish faster than in polynomial confinement.

Comparison of \cref{fig:mg_valence_polynomial,fig:mg_valence_exponential} with \cref{fig:mg-barrier} shows that the polynomial and exponential potentials result in similar behavior of the Mg 3s orbital as in weak finite-barrier confinement.
The main qualitative difference is that the polynomial and exponential potentials result in a smoother decay of the orbital than that observed with the finite-barrier potential.
Furthermore, as the strength of the polynomial and exponential confinement potentials increase with radius, the outer parts of the orbital are more strongly damped by these potentials.

\begin{figure}
\centering
\begin{subfigure}[b]{.49\textwidth}
\includegraphics[width=\textwidth]{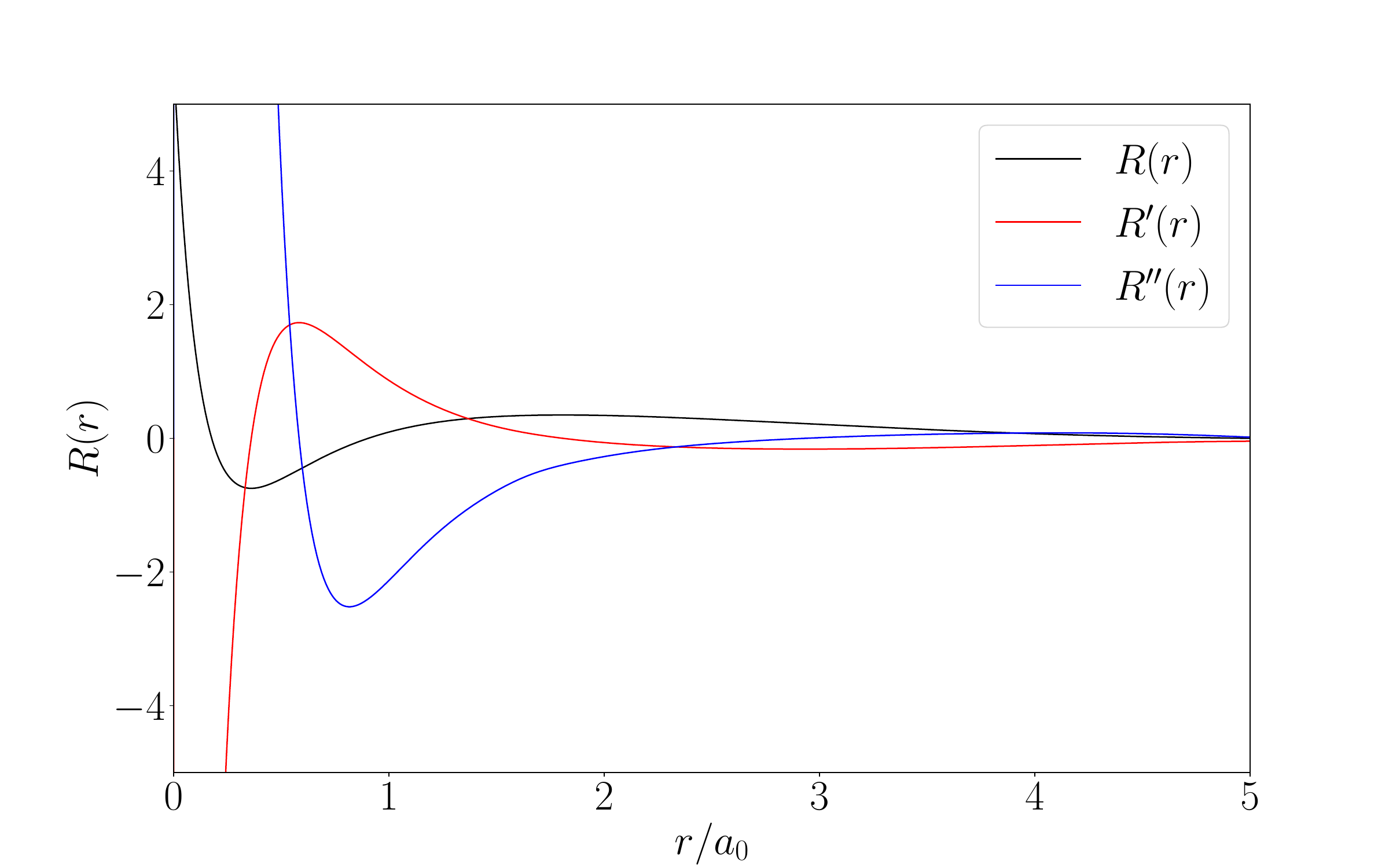}
\caption{$N=1$.}
\label{fig:der_1}
\end{subfigure}
\begin{subfigure}[b]{.49\textwidth}
\includegraphics[width=\textwidth]{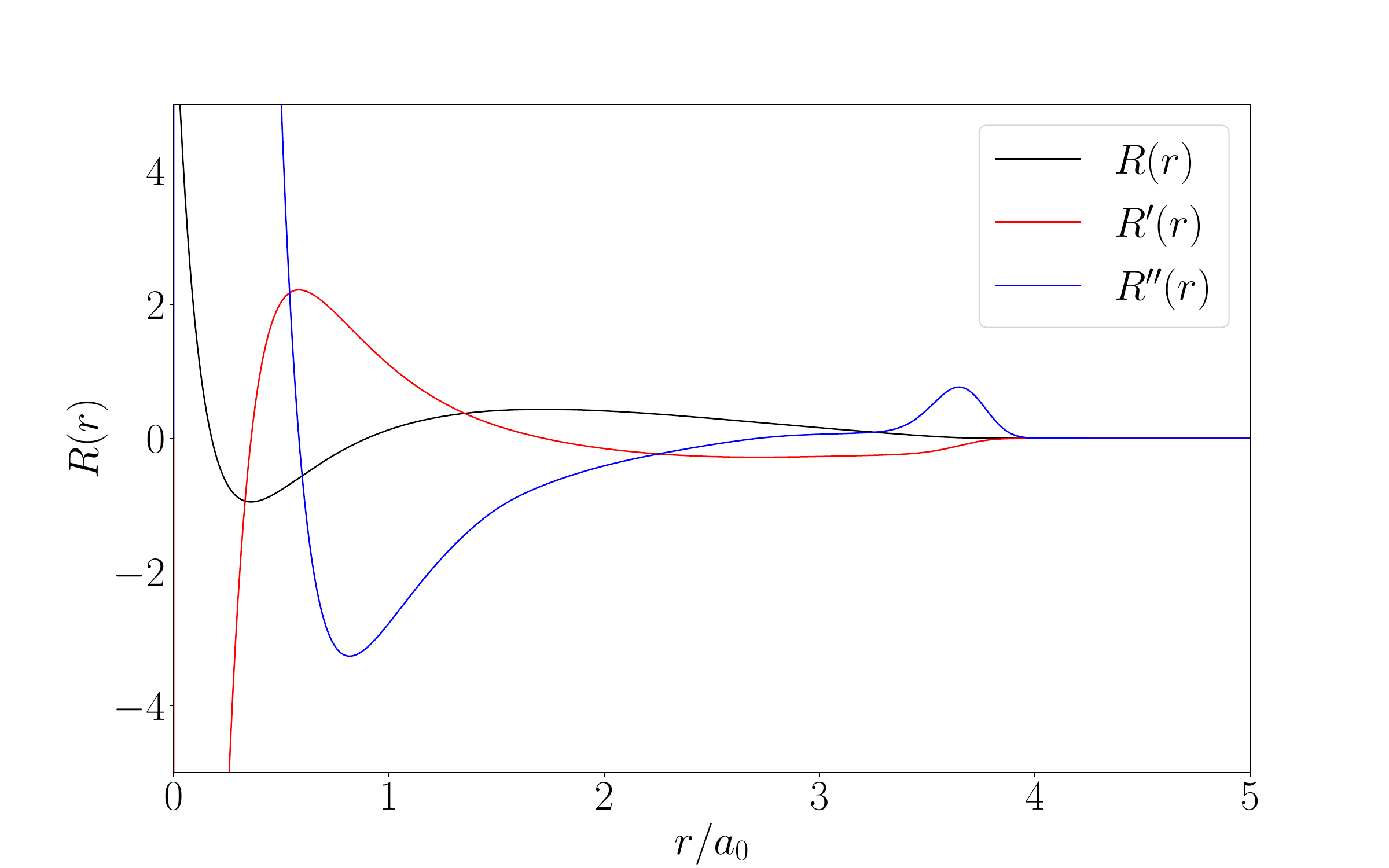}
\caption{$N=6$.}
\label{fig:der_2}
\end{subfigure}
\begin{subfigure}[b]{.49\textwidth}
\includegraphics[width=\textwidth]{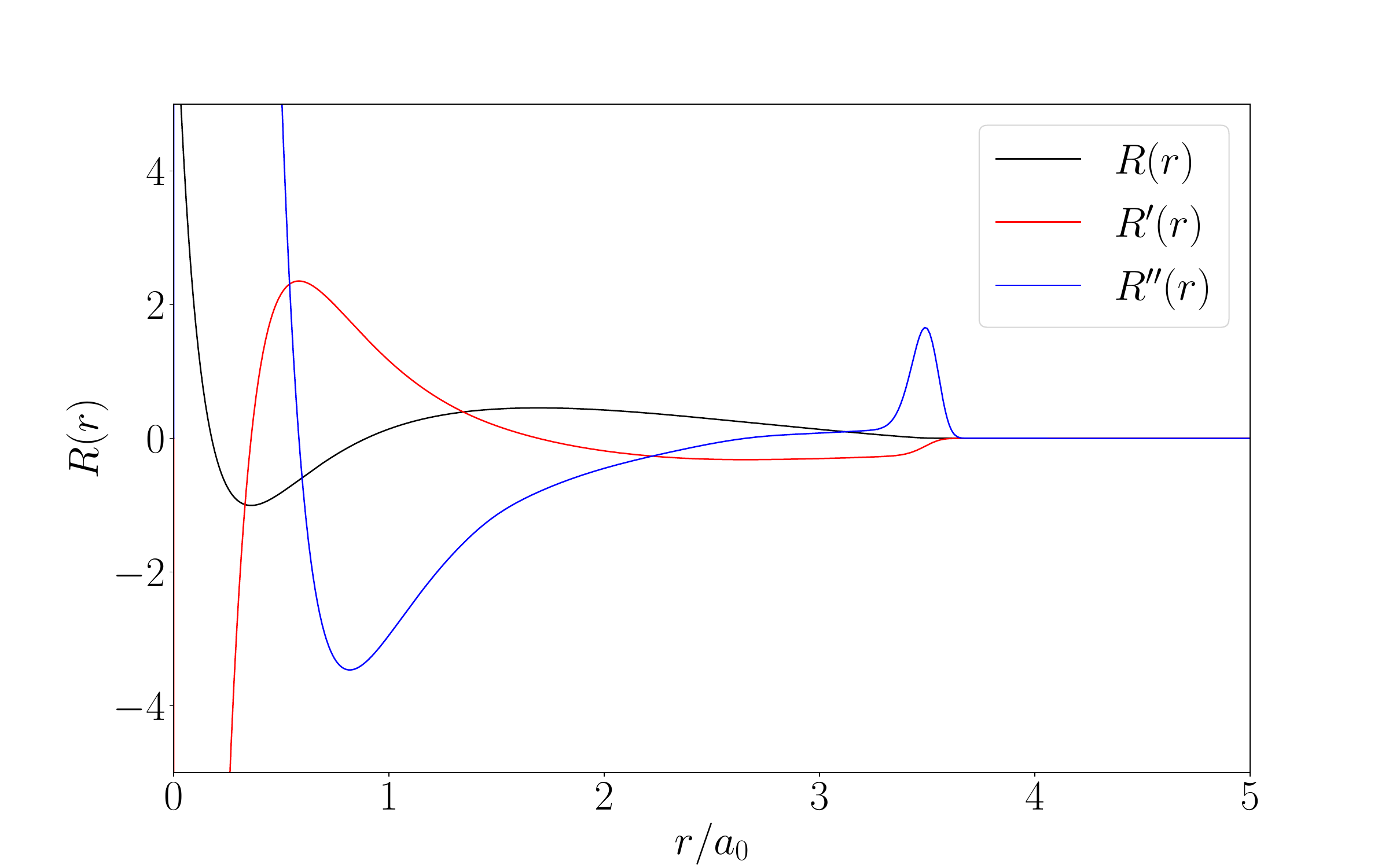}
\caption{$N=10$.}
\label{fig:der_3}
\end{subfigure}
\caption{The radial part of the 3s orbital of the Mg atom as well as its first and second derivatives in the shifted polynomial potential with various $N$ and $\delta=3\ a_0$ and $r_0=0.2\ a_0$.}
\label{fig:derivatives_poly}
\end{figure}

Further demonstration of the smoother decay compared to the finite-barrier potential can be done by studying the first and second derivatives of the radial wavefunction in \cref{fig:derivatives_poly}: we see that we now avoid the second derivative discontinuities, however, with increasing $N$ we still observe kinks in the second derivative as the potential becomes more and more steep.

\subsubsection{Truncating the radial grid} \label{sec:poly-exp-trunc}

Next, we study the truncation of the radial grid in the case of the polynomial and exponential confinement potentials.
We follow the same logic as in \cref{sec:barrier-hw}.

\input{tables/rmax_table-polynomial-Mg.tex}

\input{tables/rmax_table-exponential-Mg.tex}

The obtained values for $r_\infty$ are tabulated in \cref{tab:rmax-table-polynomial-Mg,tab:rmax-table-exponential-Mg}.
The large variation in $r_\infty$ demonstrates the vastly different confinement situations we achieve when varying $N$.
For the linear potential, even though the confinement potential is already significant for small $r$, the orbital does not vanish quickly as demonstrated by the large values of $r_\infty$ in \cref{tab:rmax-table-polynomial-Mg}.
Interestingly, large values of $r_\infty$ are also observed for the $N=1$ exponential potential in \cref{tab:rmax-table-exponential-Mg}.
When increasing $N$, we notice that $r_\infty$ decreases in line with our expectations; still, the orbital does not appear to vanish as quickly as it did for the range of finite-barrier potentials studied in \cref{tab:rmax-table-barrier-Mg}.

We again note that also the first derivative decays smoothly, as the truncation with the HIP' basis only changes $r_\infty$ by a small amount from the LIP value.

Finally, we note by comparing the data in \cref{tab:rmax-table-polynomial-Mg,tab:rmax-table-exponential-Mg} that the exponential soft confinement potential introduced in this work leads to more localized radial functions than those produced by a polynomial potential with the same $N$.
The differences between the potentials are largest at small $N$ and decrease in increasing $N$, as also the polynomial potential becomes steeper and steeper.
However, as was demonstrated in \cref{fig:logpots}, the exponential potential grows faster regardless of $N$.

\begin{figure*}
\centering
\begin{subfigure}[b]{0.49\textwidth}
\centering
\includegraphics[width=\textwidth]{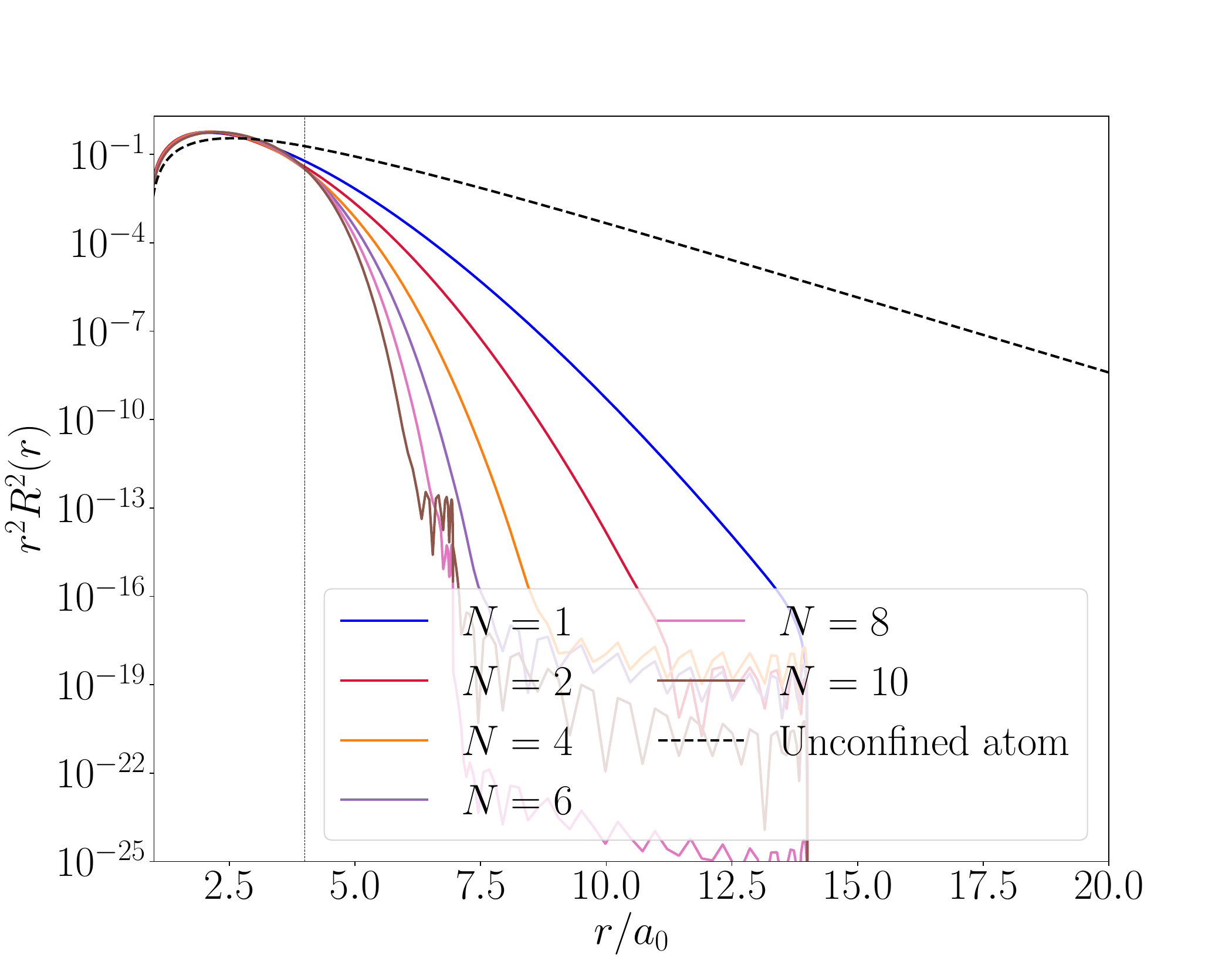}
\caption{Polynomial confinement without cutoff}
\label{fig:Mg_poly_log}
\end{subfigure}
\begin{subfigure}[b]{0.49\textwidth}
\centering
\includegraphics[width=\textwidth]{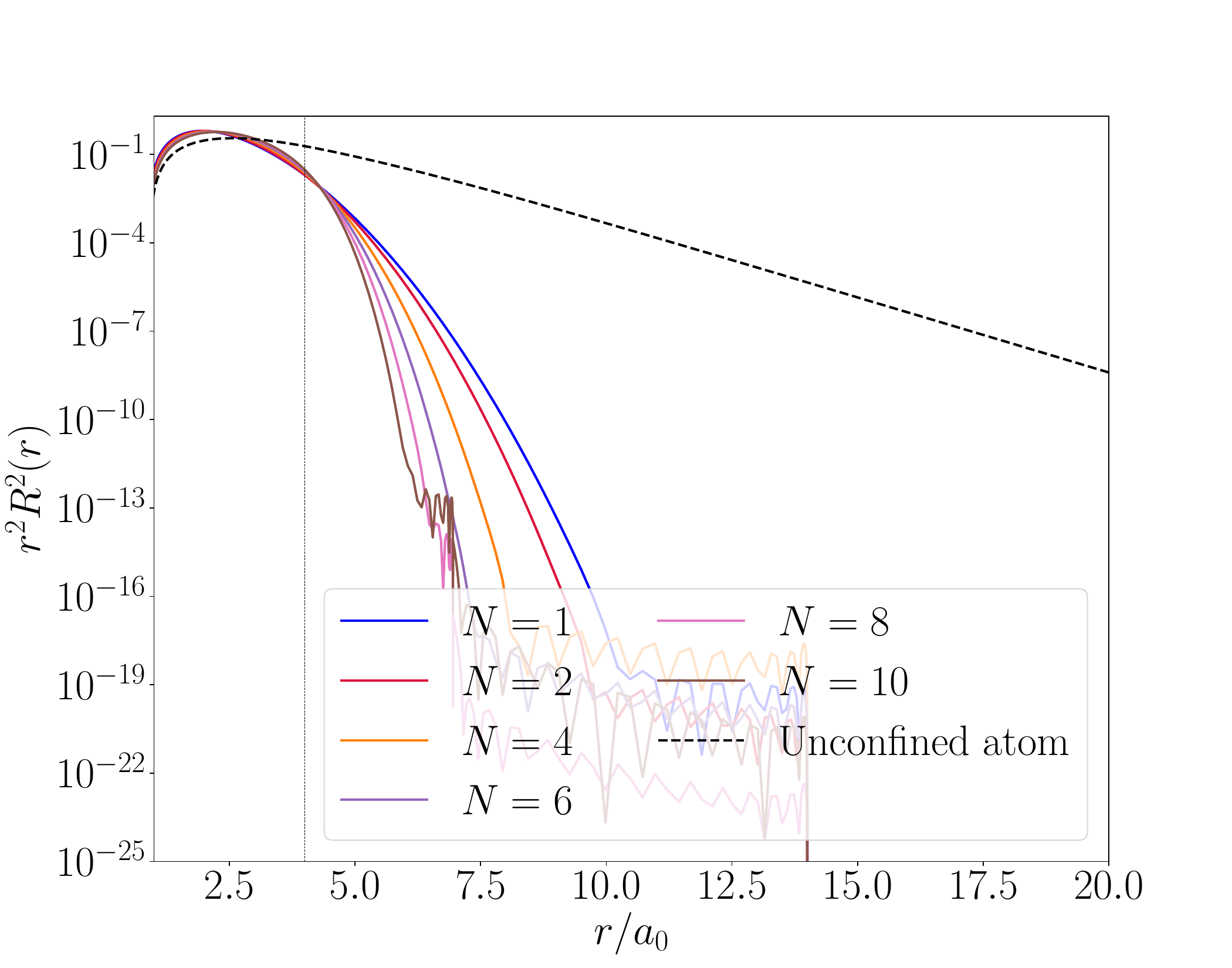}
\caption{Exponential confinement without cutoff}
\label{fig:Mg_exp_log}
\end{subfigure}
\begin{subfigure}[b]{0.49\textwidth}
\centering
\includegraphics[width=\textwidth]{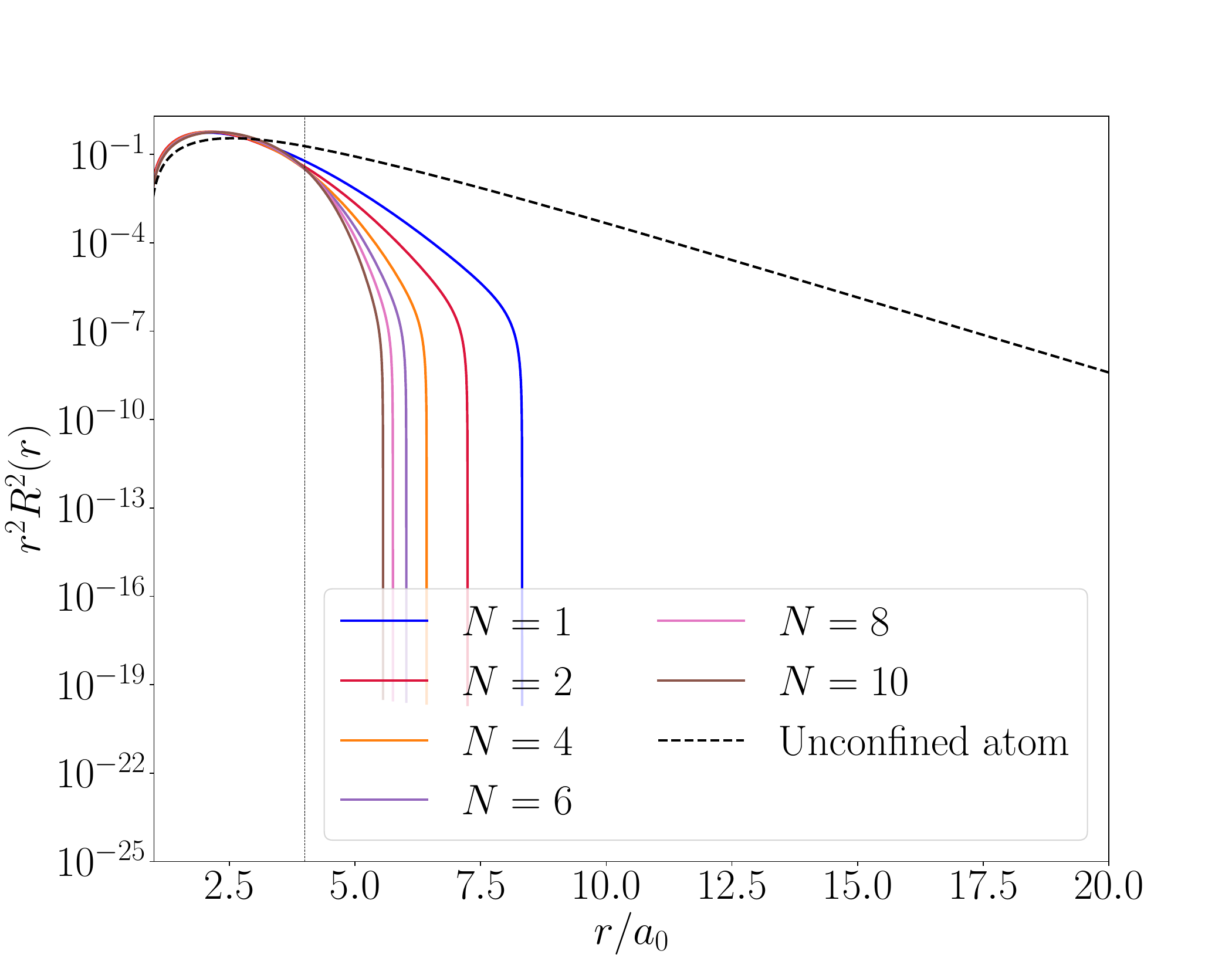}
\caption{Polynomial confinement with cutoff}
\label{fig:Mg_poly_trunc_log}
\end{subfigure}
\begin{subfigure}[b]{0.49\textwidth}
\centering
\includegraphics[width=\textwidth]{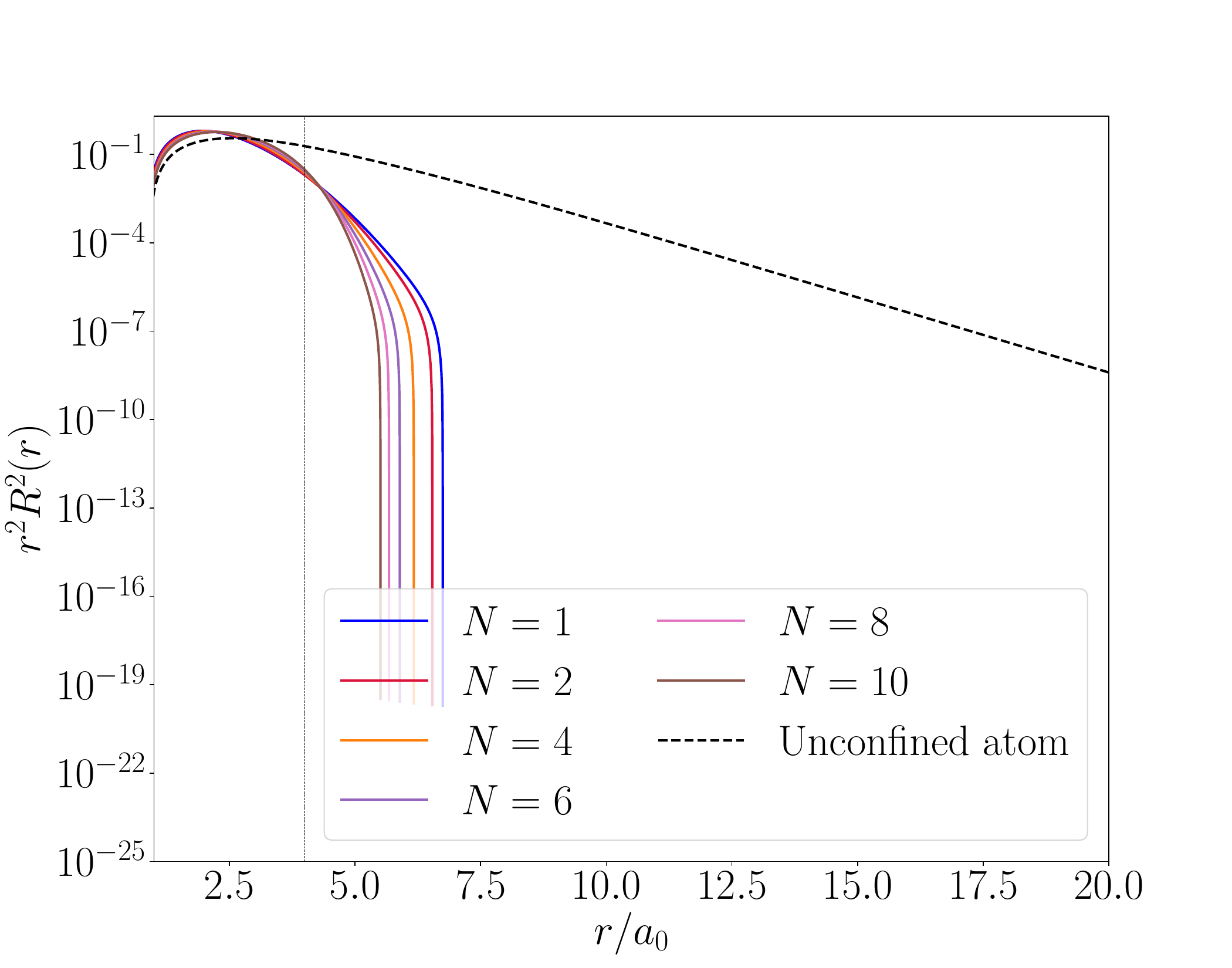}
\caption{Exponential confinement with cutoff}
\label{fig:Mg_exp_trunc_log}
\end{subfigure}
\caption{Radial density of the 3s orbital of Mg without cutoff (upper) and with cutoff (lower) in polynomial (\cref{fig:Mg_poly_log} and \cref{fig:Mg_poly_trunc_log})  and exponential (\cref{fig:Mg_exp_log} and \cref{fig:Mg_exp_trunc_log}) confinement with varying $N$ and $r_0=4\ a_0$. Note semilogarithmic scale.}
\label{fig:poly-exp_log}
\end{figure*}

We demonstrate the decay of the 3s orbital in the polynomial and exponential confinement potentials in \cref{fig:poly-exp_log}.
We observe how both the polynomial and exponential confinement potentials (\cref{fig:Mg_poly_log,fig:Mg_exp_log}, respectively) make the orbital negligible in a rapid and smooth manner, after which only numerical noise is left in a calculation employing $r_\infty=15\ a_0$.

The corresponding calculations employing the hard-wall truncation with the 1 $\upmu  \Eh{}$ energy criterion are depicted in \cref{fig:Mg_poly_trunc_log,fig:Mg_exp_trunc_log}, respectively.
We observe that the truncated versions do not exhibit visible numerical noise, and go rapidly to zero as soon as the radial density has decreased to $r^2 R(r)^2 \approx 10^{-7}$.

\subsubsection{Approaching the hard wall limit} \label{sec:poly-exp-hw}

Finalising the analysis of the polynomial and exponential confinement potentials, we study how they approach the hard-wall limit.
We now consider the shifted potential of \cref{eqn:piecewise} where $V_c(r-\delta)$ is either the polynomial potential of \cref{eqn:rn-pot}, or the exponential potential of \cref{eqn:exp-pot}.
The shifted potentials have several advantages.
First, they leave the core orbitals explicitly unaffected, and only affect the valence orbitals.
Second, the shift allows us to employ smaller values of $r_0$, thus making the potential grow more rapidly and to more strongly localize the orbitals for chosen values of $N$ and $\delta$.

\begin{figure}
\centering
\begin{subfigure}[b]{.49\textwidth}
\includegraphics[width=\textwidth]{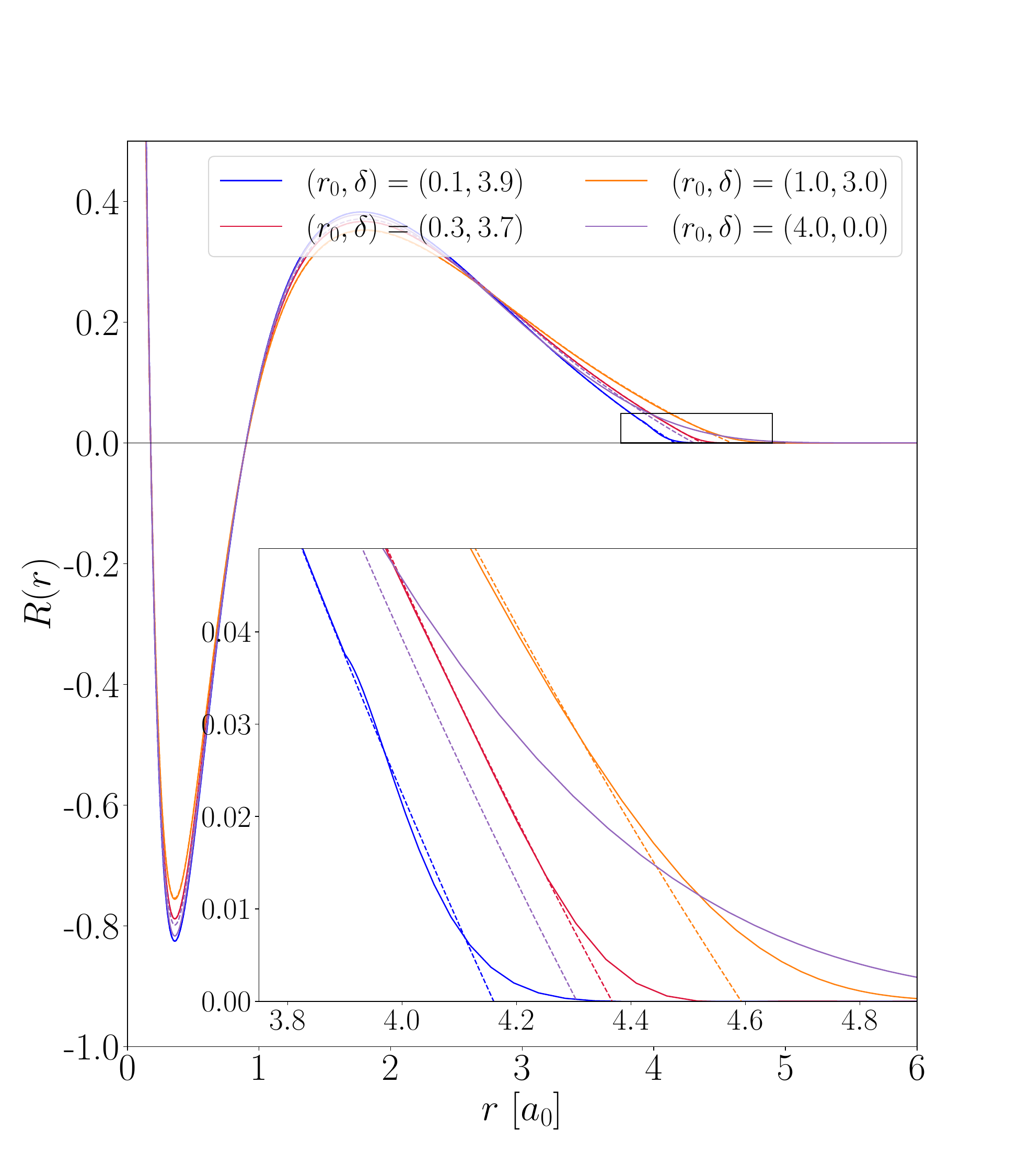}
\caption{Polynomial.}
\label{fig:Mg_shift_poly}
\end{subfigure}
\begin{subfigure}[b]{.49\textwidth}
\includegraphics[width=\textwidth]{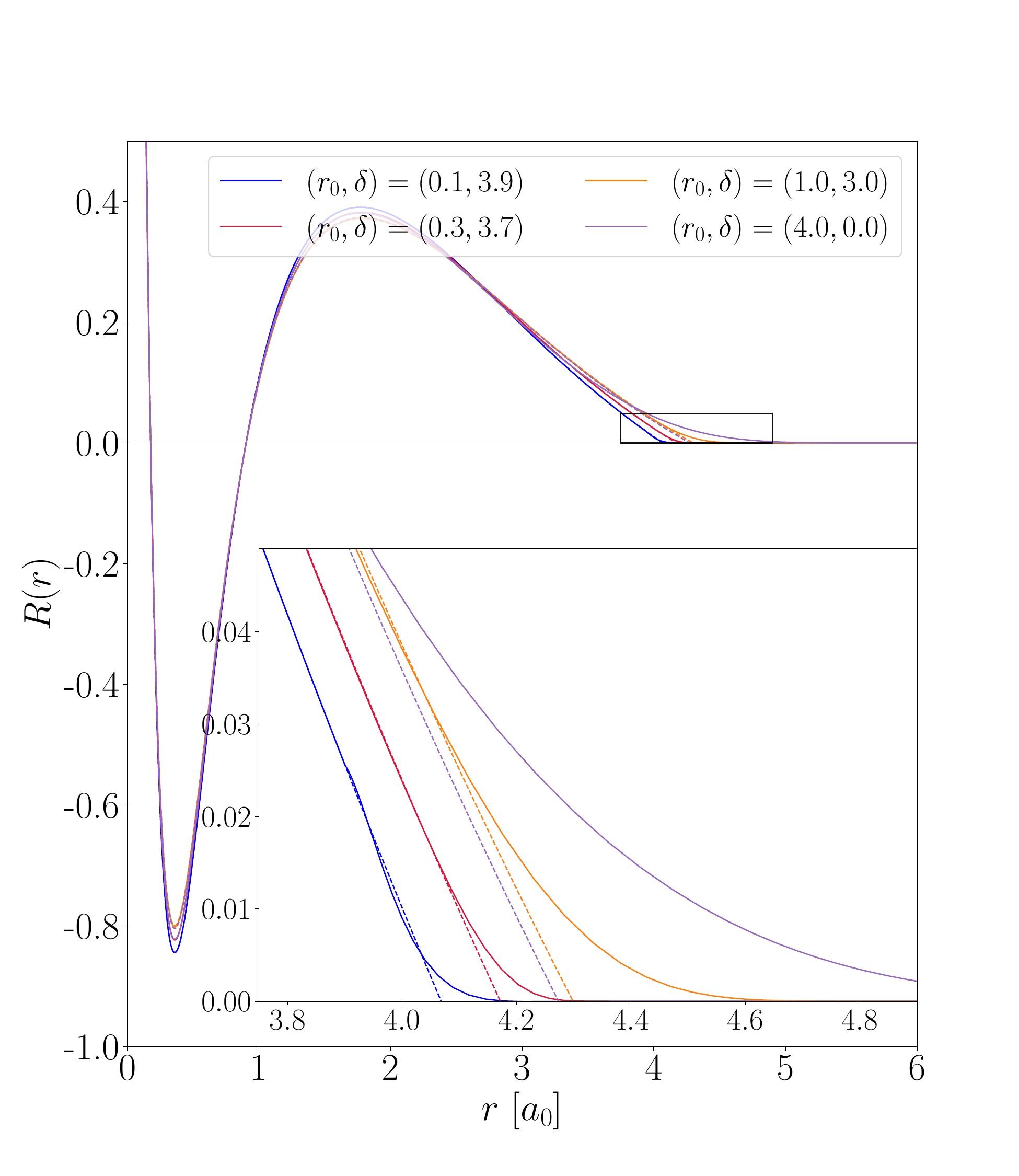}
\caption{Exponential.}
\label{fig:Mg_shift_exp}
\end{subfigure}
\caption{The radial part of the 3s orbital of the Mg atom confined by the shifted polynomial (\cref{fig:Mg_shift_poly}) and exponential (\cref{fig:Mg_shift_exp}) potential with $N=10$ and varying $r_0$ and $\delta$ (values in parentheses as $(r_0,\delta)$) in solid lines, as well as by the hard-wall potential with locations of $r_\infty$ given in \cref{tab:shift} in dashed lines.}
\label{fig:Mg_shift}
\end{figure}

The radial part of the 3s orbital of the Mg atom confined by the shifted polynomial potential with $N=10$ and various values for $r_0$ and $\delta$ is plotted in \cref{fig:Mg_shift_poly}.
The analogous plot for the exponential potential is in \cref{fig:Mg_shift_exp}.

We again fit a hard-wall confined orbital to  every orbital obtained with the specific values of $r_0$ and $\delta$, as we did in \cref{sec:barrier-hw}.
By minimizing the norm of \cref{eqn:radial-norm} with respect to the $r_\infty$ parameter used in the corresponding hard-wall calculation, we get the values of $r_\infty$ and difference norms $||\Delta||$ shown in \cref{tab:shift}.
The 3s orbitals of the Mg atom with the hard-wall locations $r_\infty$ from \cref{tab:shift} are shown along with the soft-confined orbitals in \cref{fig:Mg_shift_poly,fig:Mg_shift_exp}, respectively.

\begin{table*}
\centering
	\begin{subtable}[h]{\textwidth}
		\centering
		\shifttabmgpolynomial
		\caption{Polynomial}
		\label{tab:poly_norm}
	\end{subtable}
	\begin{subtable}[h]{\textwidth}
		\centering
		\shifttabmgexponential
		\caption{Exponential}
		\label{tab:exp_norm}
	\end{subtable}
\caption{Values of $r_\infty$ in $a_0$ that minimize the norm in \cref{eqn:radial-norm} between the 3s orbital of the Mg atom in the shifted potentials for various values of $N$, $r_0$ and $\delta$ in $a_0$ and the hard-wall potential.}
\label{tab:shift}
\end{table*}

From \cref{fig:Mg_shift_poly,fig:Mg_shift_exp} we see that the soft-confined orbitals appear practically identical to the hard-wall confined orbitals; the only difference is that the soft-confined orbitals' tails go smoothly to zero, while the hard-wall confined orbitals go linearly to zero.
The similarity between the orbitals is further confirmed by the small values of $||\Delta||$ in \cref{tab:shift}.
Compared to the unshifted potentials, $(r_0,\delta)=(4.0,0.0)$, we see that the orbital in the shifted potentials has a significantly larger overlap with the orbital in the hard-wall potential.
This demonstrates that with the shifted potentials, we are able to smoothly approach the hard-wall limit at arbitrary locations of $r_\infty$.
We can thus leave the core orbitals strictly unaffected but force the valence orbital to go to zero significantly more quickly than with the unshifted potentials.

An interesting feature in \cref{fig:Mg_shift_poly,fig:Mg_shift_exp} is that close examination of the $r_0=0.1\ a_0$ curves shows how the orbital starts to deviate from the hard-wall solution exactly at the point the confining potential is turned on, $r=\delta$.
Because the potential increases in $r$, slightly more density is placed in the region $r>\delta$.
As some density penetrates to $r>\delta+r_0$, this density is removed from the intermediate region.
As a result, one observes the complicated shape of the orbital.
Obviously, $r_0=0.1\ a_0$ is likely too small for practical use, and the other combinations show smoother orbitals at the cost of slightly less locality.

\subsection{Singular potentials \label{sec:singular}}

Finally, in this section, we assess the singular potentials of \cref{eqn:genblum-pot} with the choices $n=1$ as in \cref{eqn:junq-pot}, $n=2$ as in \cref{eqn:blum-pot}, as well as $n=3$, whose asymptotics are discussed in the Appendix.
We note that we also attempted calculations for $n=4$, but they failed to reach convergence.

\subsubsection{Contracting the orbitals \label{sec:singular-contract}}

\begin{figure}
\centering
	\begin{subfigure}[b]{.45\textwidth}
		\includegraphics[width=\textwidth]{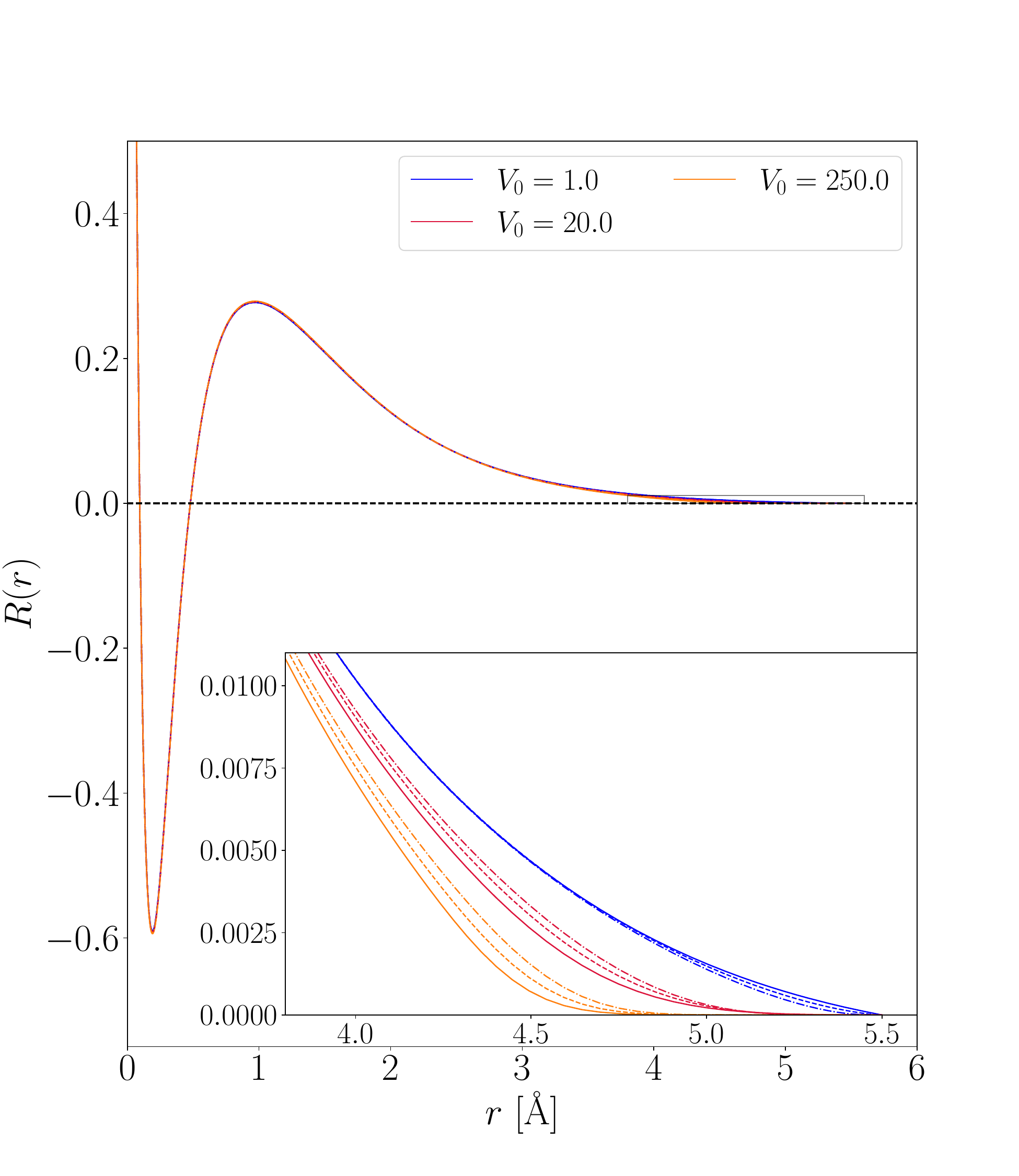}
		\caption{``light'' setting: $r_i=4.0$ \AA{} and $r_c=5.5$ \AA{}}
		\label{fig:singular_light}
	\end{subfigure}
	\begin{subfigure}[b]{.45\textwidth}
		\includegraphics[width=\textwidth]{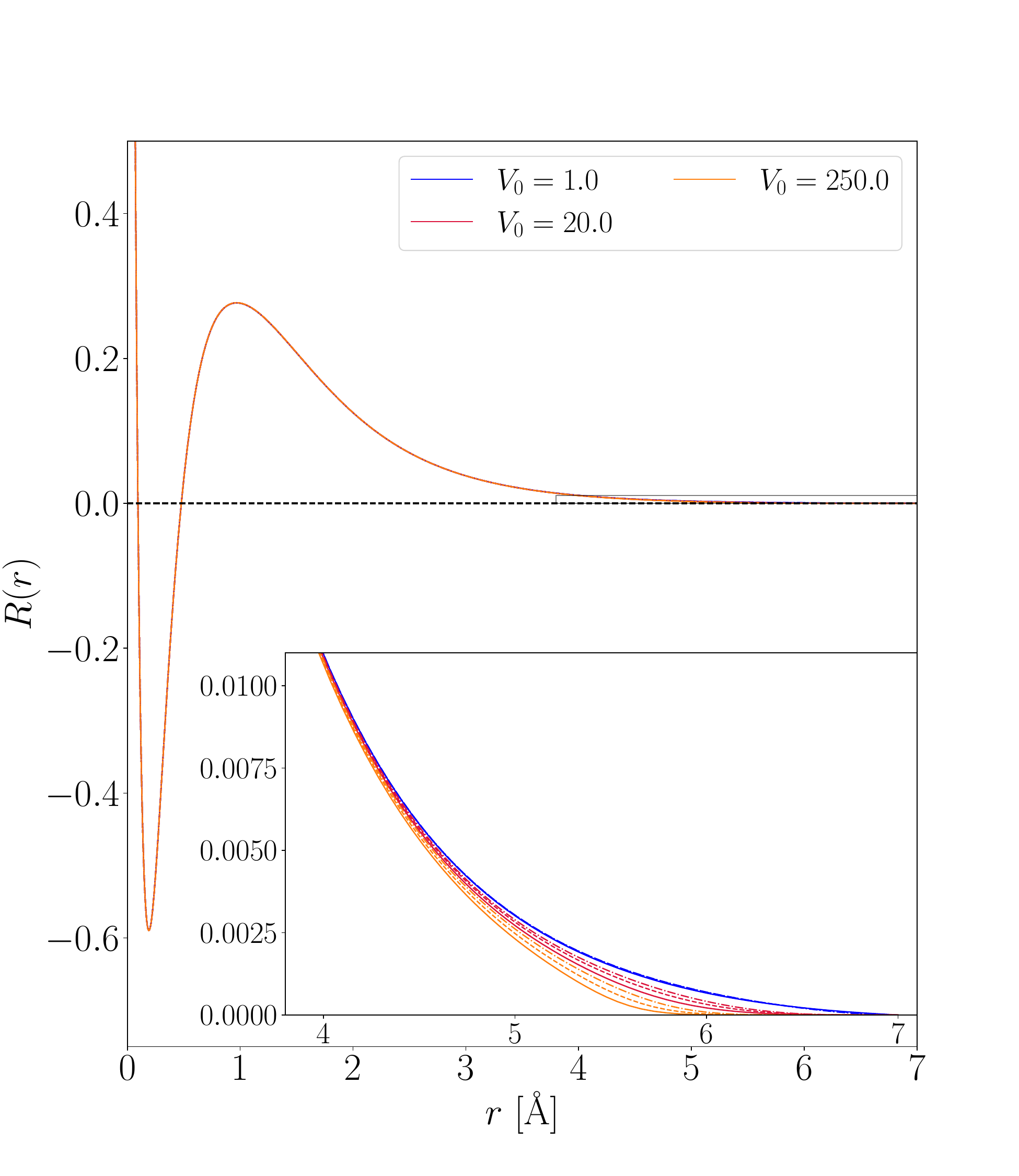}
		\caption{``tight'' setting: $r_i=5.0$ \AA{} and $r_c=7.0$ \AA{}}
		\label{fig:singular_tight}
	\end{subfigure}
\caption{The radial part of the 3s orbital of Mg confined by the singular potential for $n=1$ resulting in \cref{eqn:junq-pot} (solid lines), $n=2$ resulting in \cref{eqn:blum-pot} (dashed lines), and $n=3$ (dash-dotted lines) with various $V_0$ with the 2020 ``light'' (\cref{fig:singular_light}) and ``tight'' (\cref{fig:singular_tight}) defaults for $r_i$ and $r_c$ in \textsc{FHI-aims}.
Note that the unit of $r$ is \AA{} and not $a_0$ as in the other figures.}
\label{fig:junq}
\end{figure}

We start by studying the radial part of the Mg 3s orbital in \cref{fig:junq} for various values of $V_0$ used in \textsc{FHI-aims}, \textsc{SIESTA}, and \textsc{GPAW}, using the \textsc{FHI-aims} 2020 default values for $r_i$ and $r_c$, which are classified into ``light'', ``intermediate'', and ``tight'' settings.
For Mg the ``intermediate'' and ``tight'' settings are the same ($r_i=5.0$ \AA{} and $r_c=7.0$ \AA{}), so it suffices to study only ``light'' ($r_i=4.0$ \AA{} and $r_c=5.5$ \AA{}) and ``tight'' in \cref{fig:junq}.
We see that the orbital goes smoothly to zero in both cases and it behaves qualitatively the same regardless of the value of $n$.
However, the orbital decays faster for $n=1$ than for $n=2$ or $n=3$; similarly, the orbital for $n=2$ decays faster than that for $n=3$.

\begin{figure}
\centering
\begin{subfigure}[b]{.49\textwidth}
\includegraphics[width=\textwidth]{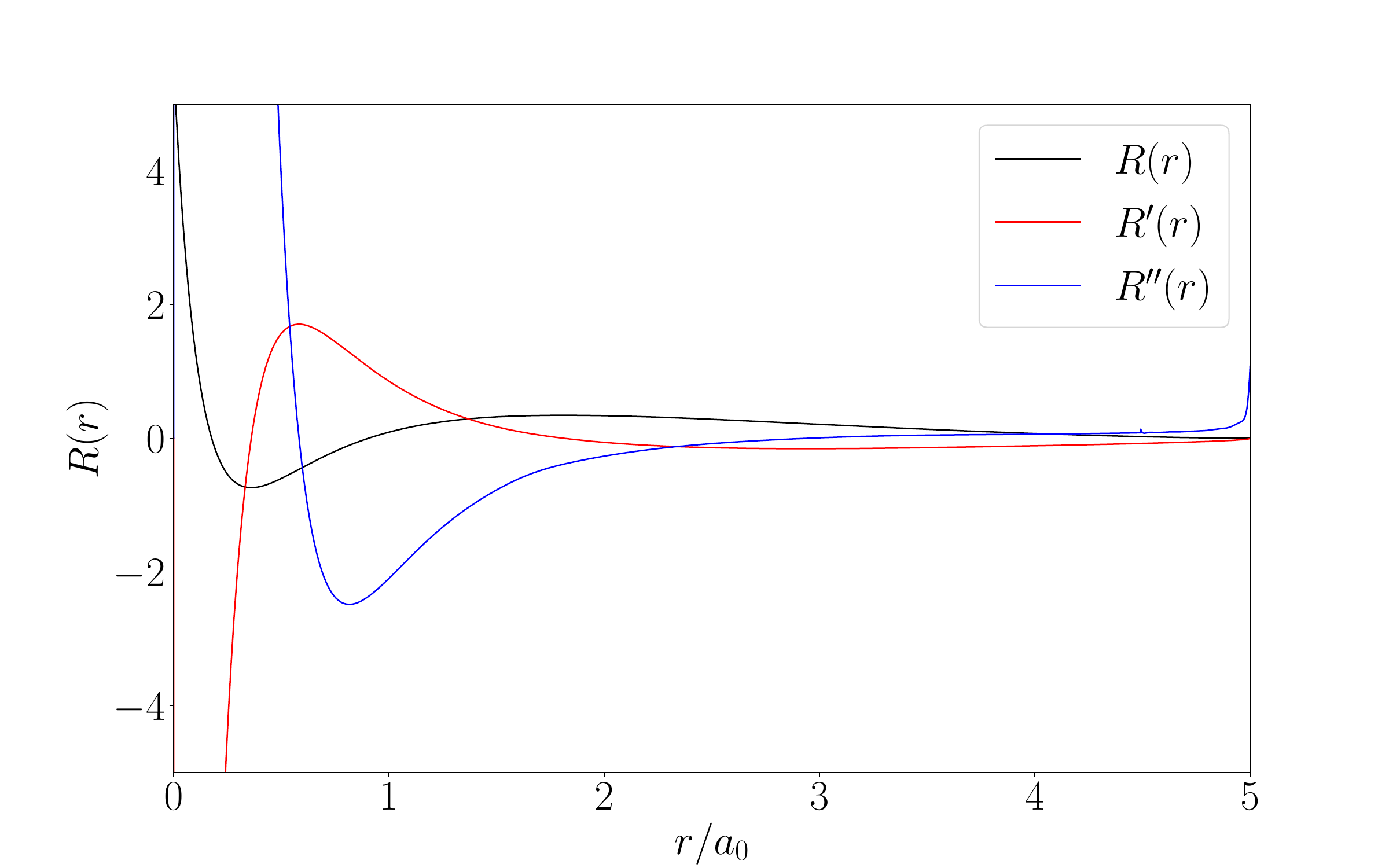}
\caption{$V_0=1.0$ \Eh.}
\label{fig:der_1}
\end{subfigure}
\begin{subfigure}[b]{.49\textwidth}
\includegraphics[width=\textwidth]{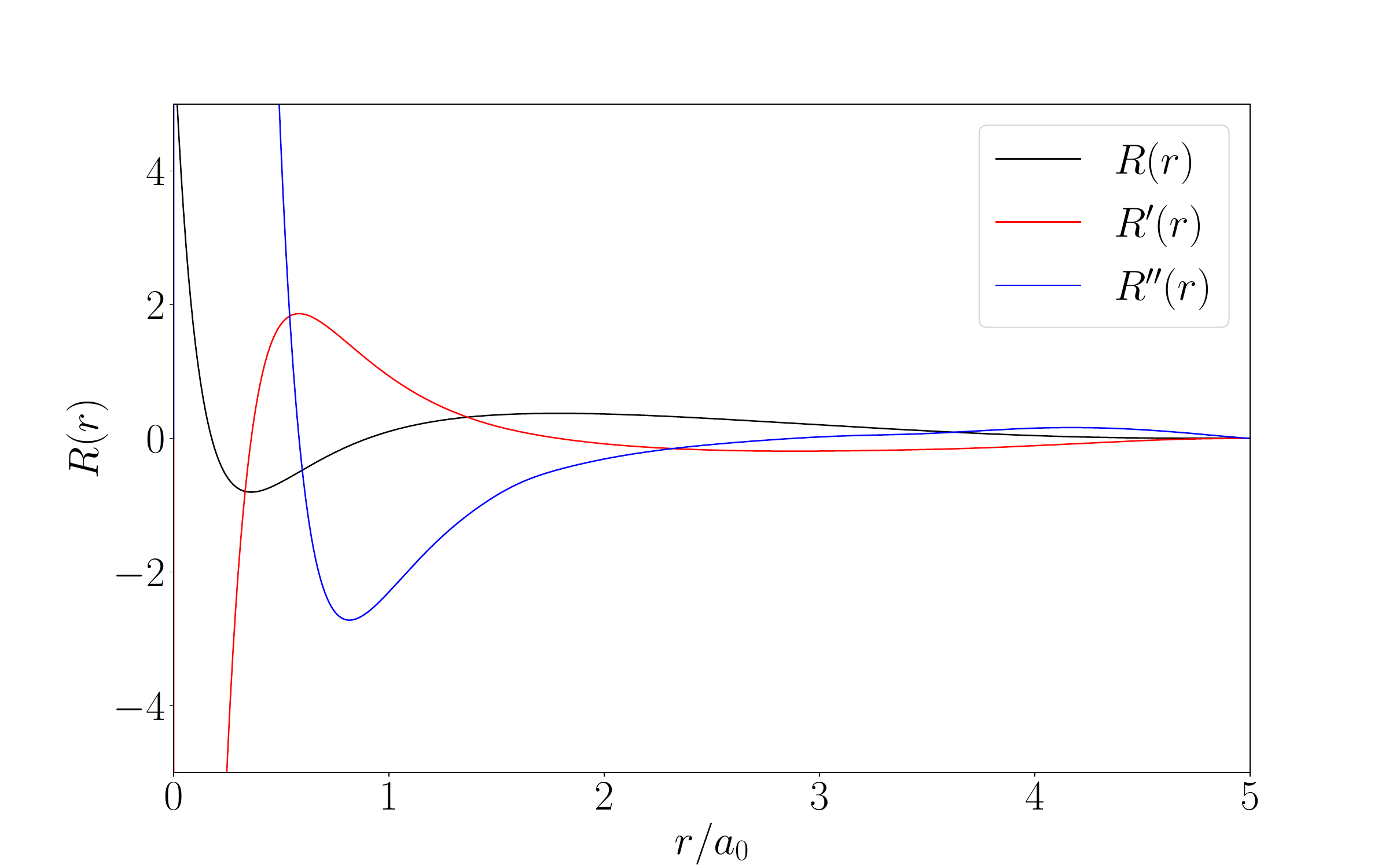}
\caption{$V_0=10.0$ \Eh.}
\label{fig:der_2}
\end{subfigure}
\begin{subfigure}[b]{.49\textwidth}
\includegraphics[width=\textwidth]{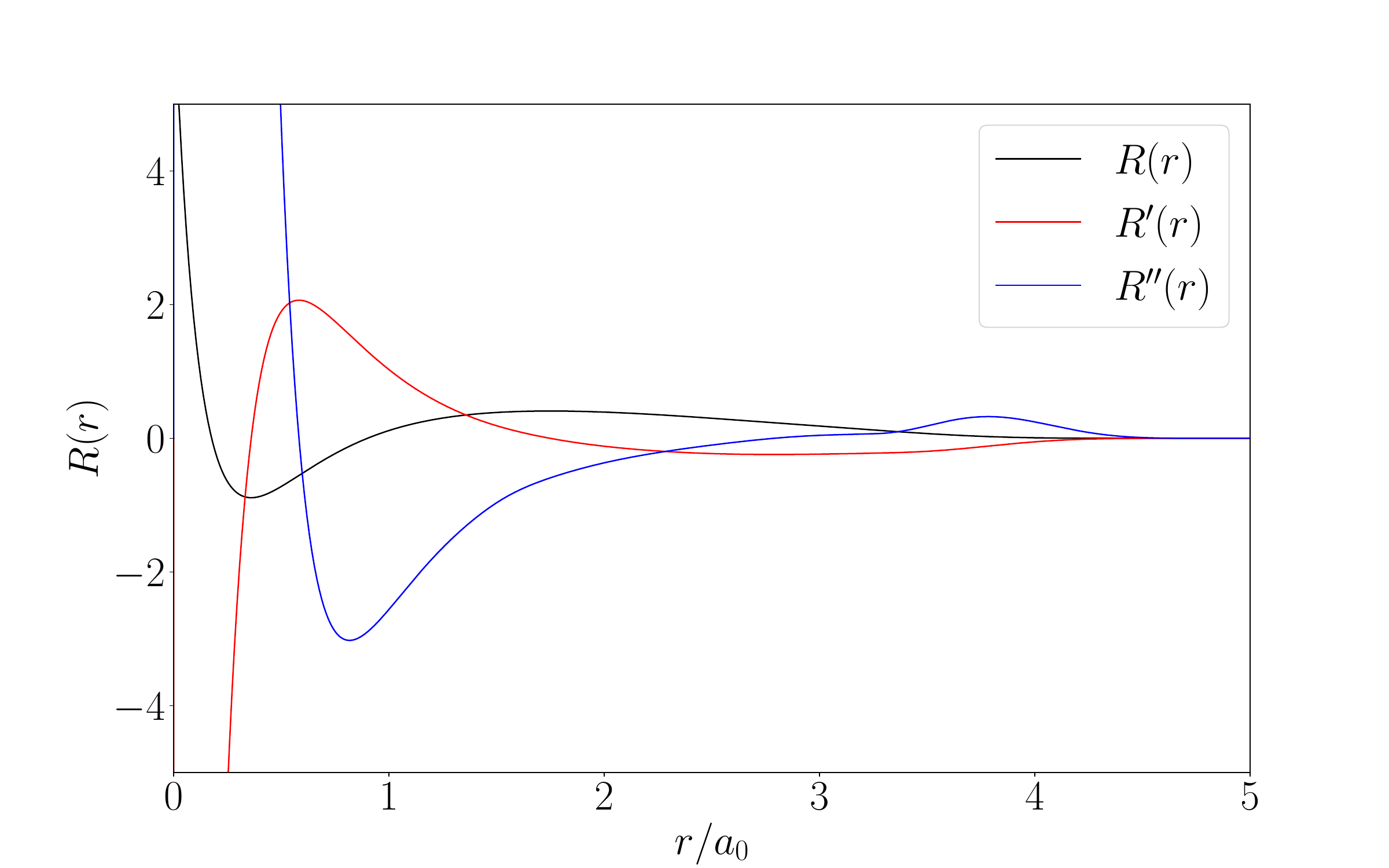}
\caption{$V_0=100.0$ \Eh.}
\label{fig:der_3}
\end{subfigure}
\caption{The radial part of the 3s orbital of the Mg atom as well as its first and second derivatives in the singular potential of \cref{eqn:genblum-pot} with $r_i=3\ a_0$, $r_c=5\ a_0$, $n=2$, and various $V_0$.}
\label{fig:blum-derivative}
\end{figure}

Moreover, as the potentials in \cref{eqn:junq-pot,eqn:blum-pot,eqn:genblum-pot} are continuous at $r=r_i$, the first-derivative discontinuity in the resulting radial functions is also avoided; however, the confinement potential will lead to a similar kink in the second derivative as for the polynomial potential as discussed by \citeitcomma{Blum2009_CPC_2175} for example.
This behavior is illustrated in \cref{fig:blum-derivative}.

\subsubsection{Approaching the hard-wall limit \label{sec:singular-hw}}

\begin{figure}
\centering
\includegraphics[width=.5\textwidth]{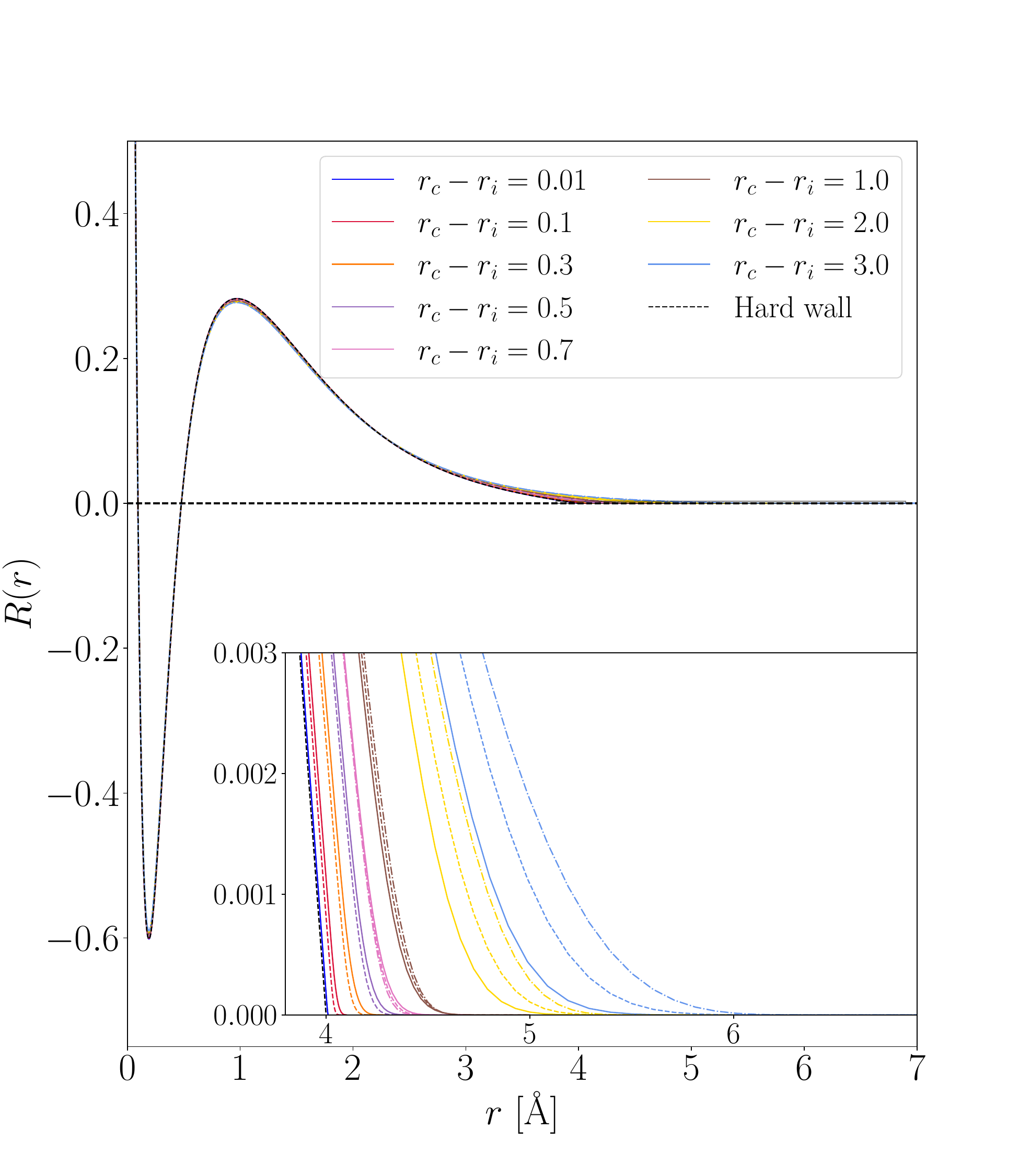}
\caption{The radial part of the 3s orbital of Mg confined by the singular potential for $n=1$ resulting in \cref{eqn:junq-pot} (solid lines), $n=2$ resulting in \cref{eqn:blum-pot} (dashed lines), and $n=3$ (dash-dotted lines) and various values of $r_c-r_i$ in Å as well as the hard-wall at $r_\infty=r_i$.
Note that the unit of $r$ is \AA{} and not $a_0$ as in the other figures.}
\label{fig:singular-hw}
\end{figure}

We go on to study the Mg 3s orbital in increasingly steep singular potentials.
We do this by fixing $V_0=250$ \Eh{}, and decreasing the difference $r_c-r_i$ for $n\in\{1,2,3\}$.
The results are shown in \cref{fig:singular-hw}.
We see that also the singular potentials approach the hard-wall limit in a systematic and smooth manner and when $r_c-r_i=0.01$ Å, the orbital is practically indistinguishable from the hard-wall confined orbital.
However, we note that we were not able to converge the calculations for $n=3$ when $r_c-r_i\leq0.5$ Å.

\subsubsection{Basis-set truncation errors \label{sec:bste}}

As the final part of this study, we examine the basis-set truncation errors for the H--Xe atoms arising from the singular potentials.
This part of the study thus measures how well the NAOs generated with the various potentials reproduce the exact solution.
As in \cref{sec:singular-hw}, we take the parameters from the FHI-aims 2020 species defaults for all atoms.

As  the NAO generator in \textsc{FHI-aims} does not support the use of meta-GGA functionals to the best of our knowledge, we prepare for the use of meta-GGAs in fully self-consistent NAO calculations---where also the NAO basis is generated with the same functional---by studying how the truncation error behaves for three levels of functionals, in analogy to our previous work in \citeref{Aastroem2025_JPCA_2791}: the Perdew--Wang local-density approximation,\cite{Perdew1992_PRB_13244} the Perdew--Burke--Ernzerhof generalized-gradient approximation (GGA),\cite{Perdew1996_PRL_3865, Perdew1997_PRL_1396} and the $r^2$SCAN meta-GGA.\cite{Furness2020_JPCL_8208, Furness2020_JPCL_9248}

\begin{figure}
\centering
\includegraphics[width=\linewidth]{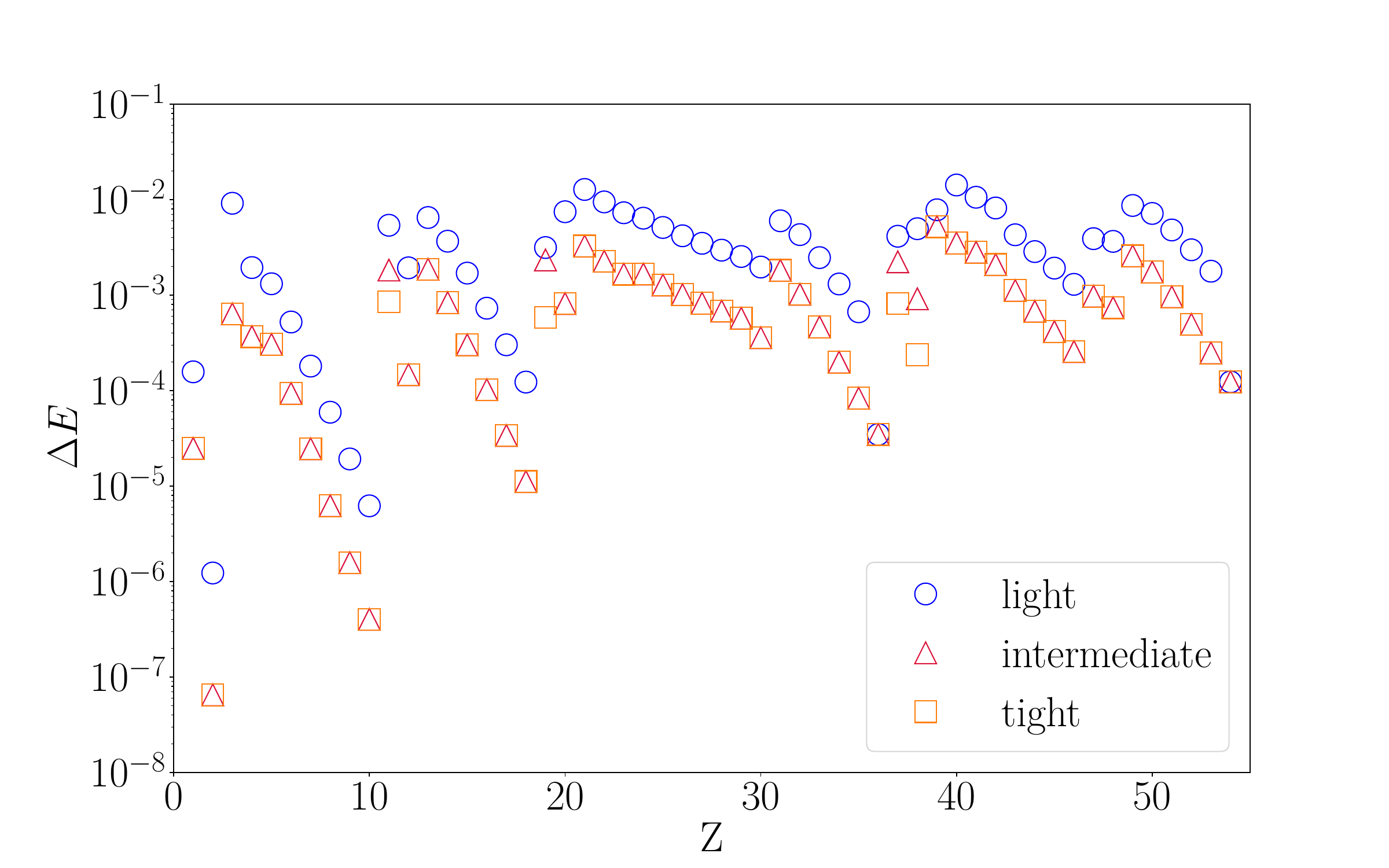}
\caption{Truncation errors of atoms with the PBE functional with the ``light'', ``intermediate'', ``tight'' and ``really tight'' defaults in FHI-aims.}
\label{fig:bste-pbe}
\end{figure}

For each functional, we compute the truncation error that would arise in an atomic calculation with the generated NAO basis functions as
\begin{equation} \label{eqn:bste}
\Delta E(r_i)=E_\mathrm{confined}(r_i)-E_\mathrm{confinement}(r_i)-E_\mathrm{unconfined}
\end{equation}
where $E_\mathrm{confined}(r_i)$ is the self-consistent total energy of the atom in confinement, $E_\mathrm{confinement}(r_i)$ is the confinement energy included in the previous term, and $E_\mathrm{unconfined}$ is the energy of the unconfined atom.
These results are shown in \cref{fig:bste-pbe} for the PBE functional.
The PW92 and $r^2$SCAN results are left to the SI, since as expected, the results are effectively independent of the employed functional.

We can see in \cref{fig:bste-pbe} that the BSTE for the 2020 species defaults fluctuates significantly across the periodic table.
The smallest BSTEs are found for the He atom with errors around 1 $\upmu \Eh$, while the largest error is around 10 m\Eh{} in the ``light'' setting.
The ``intermediate'' and ``tight'' settings appear to yield BSTEs that are systematically around one order of magnitude smaller than the ``light'' setting.

\begin{figure}
\centering
\begin{subfigure}[b]{.5\textwidth}
\includegraphics[width=\textwidth]{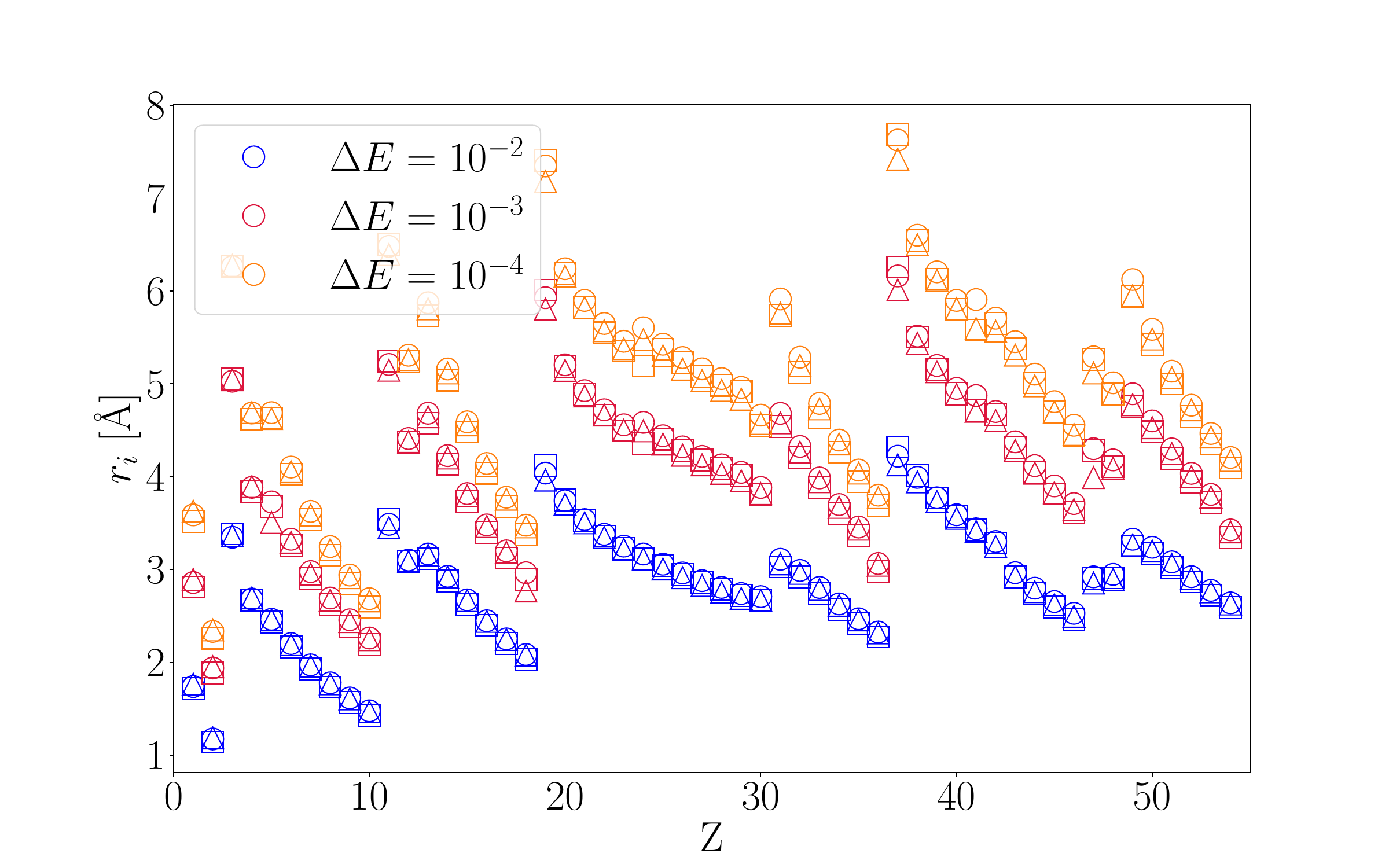}
\caption{Spherical symmetric DFT.}
\label{fig:bste-dft}
\end{subfigure}
\begin{subfigure}[b]{.5\textwidth}
\includegraphics[width=\textwidth]{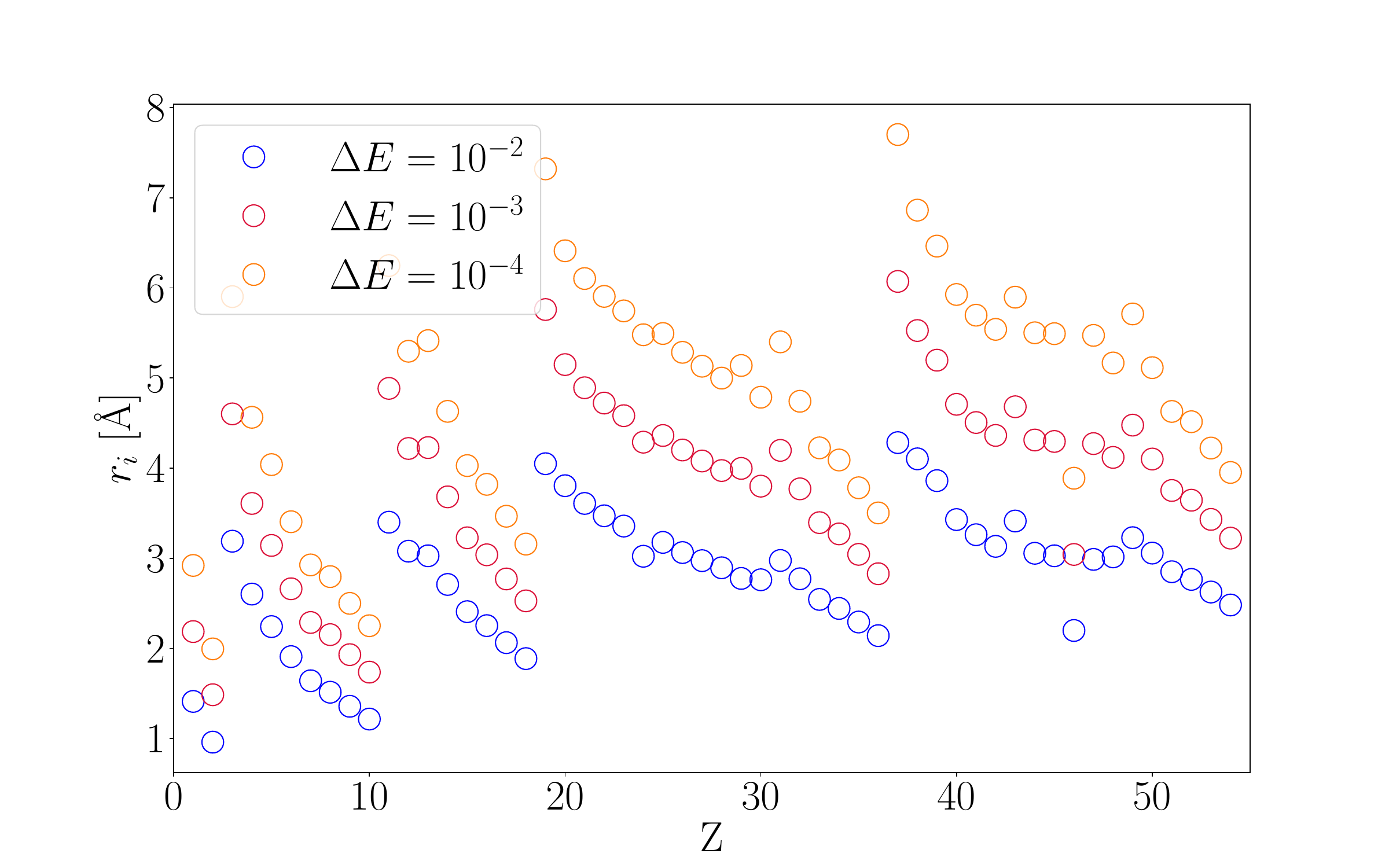}
\caption{Hartree--Fock.}
\label{fig:bste-hf}
\end{subfigure}
\caption{The parameter $r_i$ of \cref{eqn:blum-pot} corresponding to fixed BSTE (\cref{eqn:bste}). The parameter $r_c = r_i +2.0$ Å in all calculations.
PBE values are indicated with circles, PW92 values with triangles, and $r^2$SCAN values with squares.}
\label{fig:fixed-bste}
\end{figure}

The \textsc{FHI-aims} manual points out that the time to set up the Hamiltonian for a densely packed solid scales as $r_c^6$.
This means that being able to employ smaller values of $r_c$ can lead to huge computational savings.
We now explore an alternative scheme, where we fix the BSTE to a certain value, and instead calculate the parameter $r_i$ that yields this BSTE, when the width of the transition $r_c-r_i$ is kept at the \textsc{FHI-aims} default.
The resulting values of $r_i$ for the H--Xe atoms are illustrated in \cref{fig:fixed-bste} for BSTEs fixed to $10^{-2}$, $10^{-3}$, and $10^{-4}$ \Eh.
The results in \cref{fig:bste-dft} are obtained with PW92, PBE, and $r^2$SCAN with spherically symmetric densities.
We also include results for non-symmetric atoms computed with unrestricted Hartree--Fock\cite{Lehtola2019_IJQC_25945} in \cref{fig:bste-hf}.
As expected, we now observe periodic fluctuations for $r_i$ similarly to the fluctuations in BSTEs observed in \cref{fig:bste-pbe}.
The largest differences between the radii corresponding to BSTEs of $10^{-4}$ and $10^{-2}$ are 4 Å for the alkali elements.
Furthermore, we observe no significant differences between the two methods in \cref{fig:bste-dft} and Hartree--Fock in \cref{fig:bste-hf}.
The only exception is the Pd atom with a particularly stable ground state electron configuration which is not affected by confinement.\cite{Aastroem2025_JPCA_2791}
In the HF calculations the radius of Pd atom stands out compared to neighbouring atoms in contrast to the spherically symmetric DFT calculations.
We hope to further explore these types of systematic ways to determine the confinement potentials in future work.

\begin{figure}
\includegraphics[width=\linewidth]{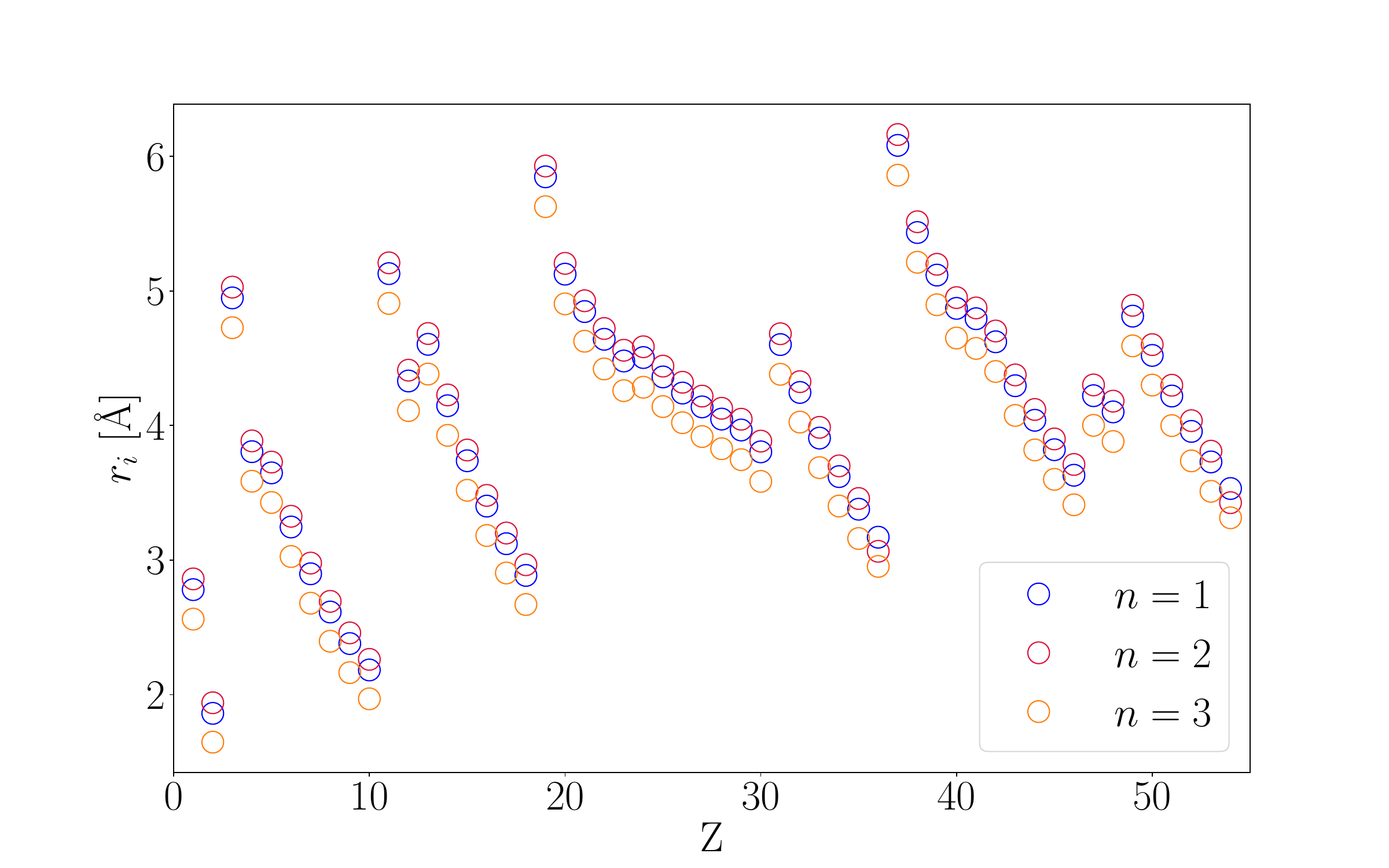}
\caption{The parameter $r_i$ obtained from the singular potentials with various exponents and by setting the truncation error of \cref{eqn:bste} to $10^{-3}$ \Eh{} for the PBE functional.}
\label{fig:bste-n}
\end{figure}

Finalising the analysis for the singular potentials, we set the target BSTE to $10^{-3}$ \Eh{} and calculate the values of $r_i$ for $n\in\{1,2,3\}$ using the PBE functional.
The obtained radii are depicted in \cref{fig:bste-n}.
Surprisingly, $n=2$ gives rise to the largest radii, while $n=1$ gives slightly smaller radii, except for the Xe atom.
$n=3$ gives the smallest radii for all atoms.
However, all differences appear to be $\lesssim0.5$ Å.

\section{Summary and conclusion \label{sec:summary}}

We have discussed the widespread use of various confinement potentials in several contexts, such as the generation of numerical atomic orbital\cite{Averill1973_JCP_6412} (NAO) basis functions and the study of confined atoms, molecules, and quantum dots.
We pointed out that the dissimilar fields of use of confinement potentials do not appear to be fully aware of each other.
We carried out a thorough review of the various confinement potentials used in these dissimilar fields in \cref{sec:theory}.

For the calculations of this work, we considered (i) the well-established finite-barrier (\cref{eqn:soft-wall}) and (ii) polynomial soft confinement (\cref{eqn:rn-pot}) potentials that have been used in many NAO and confinement studies, (iii) the exponential soft confinement potential suggested in \cref{eqn:exp-pot} of this work, as well as (iv) the family (\cref{eqn:genblum-pot}) of singular potentials of \cref{eqn:junq-pot,eqn:blum-pot} also familiar from the NAO literature, which we generalized to various exponents in the denominator in \cref{eqn:genblum-pot}.
Although the soft potential of \cref{eqn:rn-pot} has been widely used in earlier literature,\cite{Eschrig1978_PSSB_621, Porezag1995_PRB_12947, Koepernik1997_PRB_5717, Horsfield1997_PRB_6594, Delley2000_JCP_7756} it does not appear to have been recently employed for NAO basis set generation, as it does not enforce that the orbitals are identically zero beyond a certain radius, and it also affects the core electrons.
However, these issues can be adressed by combining the potential with a hard-wall potential,\cite{Delley2000_JCP_7756} and by shifting its turn-on point, respectively.

To illustrate the employed confinement potentials, we examined the behavior of the ground state orbitals of Mg and Ca with various parameters for each potential.
We observed that the form of the valence orbital is qualitatively independent of the employed soft confinement potential.
As discussed by \citeitcomma{Delley2000_JCP_7756} we demonstrated that finite support of the NAO basis is achievable even for soft confinement potentials when an additional hard-wall boundary is placed suitably far away.

We observed that the orbitals disappear rapidly when the finite barrier is made sufficiently high or when the soft confinement potential is made sufficiently steep, allowing us to truncate the orbitals to finite support with a hard-wall boundary at $r_\infty$ only slightly larger than the employed value of $r_0$ that describes the onset of the potential.
We also observed that the exponential soft confinement potential proposed in \cref{eqn:exp-pot} of this work leads to even faster decay of the orbitals than that observed in polynomial confinement with the same $N$ parameter.

We investigated how the soft potentials approach the hard-wall boundary and saw that all potentials approach the hard-wall potential in a smooth and systematic manner.
This allows us to employ a steep potential that is strictly zero in the core region and forces the valence orbitals to decay arbitrarily quickly.

Finally, we studied the use of the orbitals generated with the singular potentials through basis set truncation errors (BSTEs) for the H--Xe atoms at three levels of density-functional theory.
We observed large fluctuations in the BSTEs when employing the 2020 species defaults of \textsc{FHI-aims}.
We suggested instead determining consistent sets of confinement potential parameters for the periodic table by fixing the BSTEs to a certain value for all atoms.
This method was found to lead to large periodic variations in the truncation radius, suggesting that considerable computational savings may be achievable by further exploration of this scenario.

Our study of atoms under confinement is the first step towards a reusable library for electronic structure calculations with NAO basis sets.
The project of reusable software for electronic structure\cite{Lehtola2023_JCP_180901} is simultaneously proceeding on other fronts.\cite{Lehtola2025__,Lehtola2025_JPCA_5651}
In future work, we wish to address the issues of numerical quadrature in polyatomic NAO calculations, and the generation of optimal NAO basis sets.

\begin{acknowledgments}
We thank Volker Blum and Eric Canc\`{e}s for discussions.
H.\AA{}. thanks the Finnish Society for Sciences and Letters for
funding. S.L. thanks the Academy of Finland for financial support
under project numbers 350282 and 353749.
\end{acknowledgments}
  
\section*{Appendix: Asymptotic Orbital Behavior}\label{sec:appendix}
\setcounter{equation}{0}
\renewcommand\theequation{A\arabic{equation}}

\begin{figure}
  \centering
  \includegraphics[width=\linewidth]{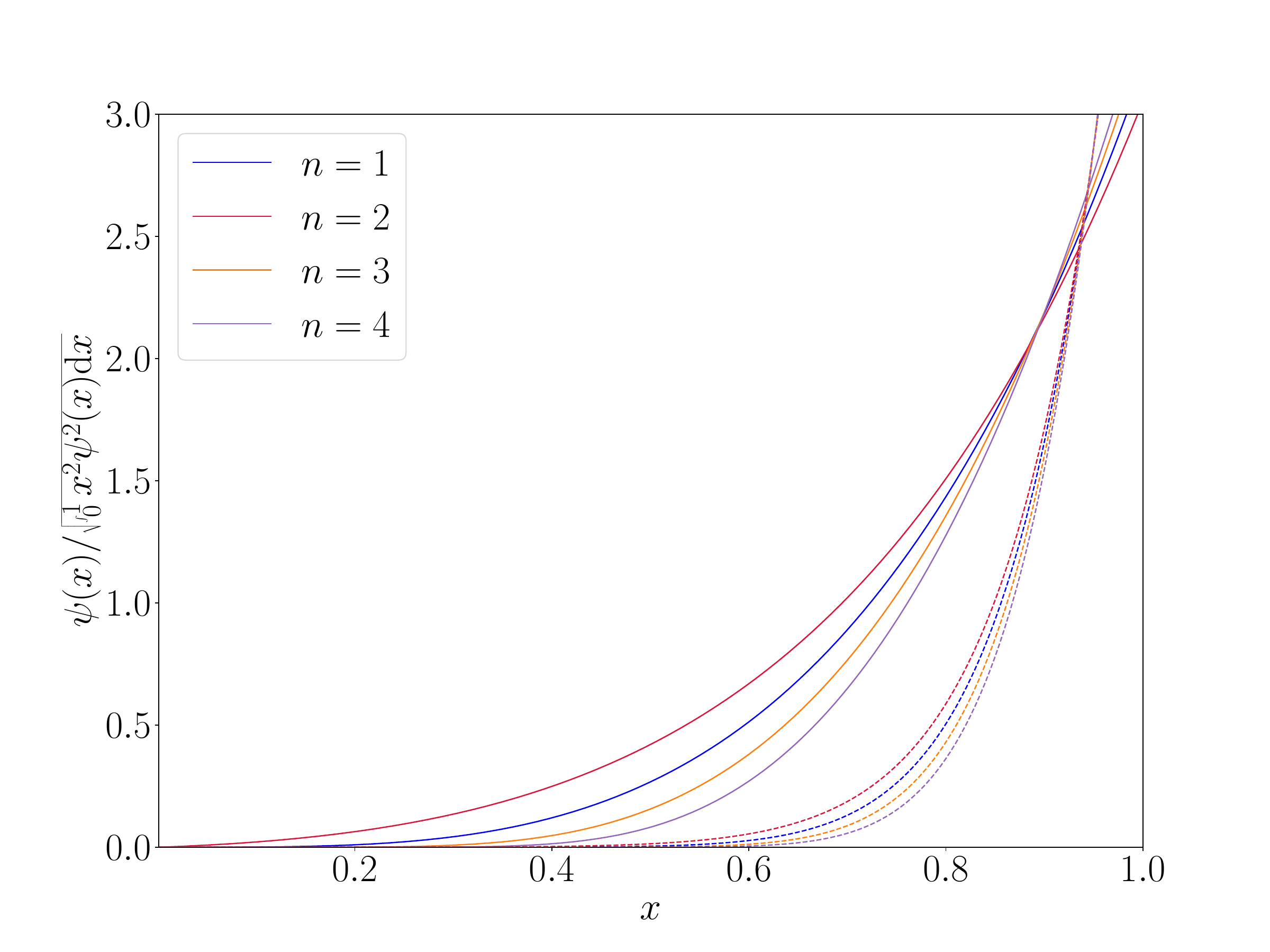}
  \caption{Asymptotic behavior of confined orbitals as shown by
    \cref{eqn:junq-asymptote,eqn:blum-asymptote,eqn:n3-asymptote,eqn:n4-asymptote}
    corresponding to choices $n=1$, $n=2$, $n=3$, and $n=4$ in
    \cref{eqn:simplerad}, respectively. The barrier heights are $V_0=25$
    \Eh{} (solid lines), and $V_0=250$ \Eh{} (dashed lines).}
  \label{fig:asymptote}
\end{figure}

We analyze herein the asymptotic behavior of the orbitals with the various confinement potentials.
The radial Schr\"odinger equation is given by
\begin{align}
  & \frac {\partial^2 \psi} {\partial r^2} - \frac {2} {r} \frac {\partial \psi} {\partial r} - \frac {l(l+1)} {r^2} \psi(r) \nonumber \\
  + & [V_\textrm{Hxc}(r) + V_\textrm{conf}(r)] \psi(r) = E \psi(r) \label{eqn:radschr}
\end{align}
where $V_\textrm{Hxc}(r)$ contains the Coulomb and exchange-correlation potentials.
\Cref{eqn:genblum-pot} diverges when $r \to r_c$ for all $n$.
The confinement potential thus dominates, and we can study the asymptotic behavior with a simplified equation.
Switching variables as $x = r_c - r$, so that $r \to r_c$ corresponds to the case $x\to 0$, the behavior of \cref{eqn:genblum-pot} is now given at this limit by the simplified equation
\begin{equation}
 - \frac {\partial^2 \psi} {\partial x^2} +  V_1 x^{-n} \psi(x) = 0. \label{eqn:simplerad}
\end{equation}
where $V_1 = V_0/e$, and $n=1$ for \cref{eqn:junq-pot} and $n=2$ for \cref{eqn:blum-pot}.
The solution of this differential equation is readily obtained with \textsc{Maple}, yielding the solution
\begin{equation}
  \label{eqn:blum-asymptote}
  \psi(x) \propto x^{(1 + \sqrt{1 + 4 V_1})/2}
\end{equation}
for \cref{eqn:blum-pot} and
\begin{equation}
  \label{eqn:junq-asymptote}
  \psi(x) \propto \sqrt{x} I_{1}(2\sqrt{V_1 x})
\end{equation}
for \cref{eqn:junq-pot}, where $I_{1}(x)$ is a modified Bessel
function of the first kind; the other solution is excluded in each
case as it diverges in the limit $x\to 0$, the asymptotic solutions in
\cref{eqn:blum-asymptote,eqn:junq-asymptote} remaining regular.
For $n=3$ one obtains two solutions, one of which diverges as $x\to 0+$ but the other has the asymptotic behavior
\begin{equation}
\label{eqn:n3-asymptote}
\psi(x)\propto \sqrt{x}K_1\left(2\sqrt{\frac{V_1}{x}}\right),
\end{equation}
where $K_1(x)$ is a modified Bessel function of the second kind.
Repeating the analysis for $n=4$ one gets the solution
\begin{equation}
  \label{eqn:n4-asymptote}
  \psi(x) \propto x\exp(-\sqrt{V_1}/x),
\end{equation}
which looks especially promising for its expected fast decay, but
calculations on atoms in confinement with $n=4$ failed to converge.
The four choices are compared in \cref{fig:asymptote} for various
values of $V_0$. For an ambiguous comparison, all functions have been
normalized to the same number of electrons in the region $x\in[0,1]$.
We see that the solutions are qualitatively independent of $n$, but
are heavily influenced by the choice of $V_0$, a larger $V_0$ leading
to faster decay as expected.

\section*{Supporting Information}
The following data is included in the supporting information PDF file
\begin{enumerate}
\item Plots of the radial parts of the core orbitals in the finite-barrier, polynomial, and exponential potentials.
\item Plots of the radial density of the valence orbitals in the finite-barrier, polynomial, and exponential potentials for various $r_0$ both with a large converged radial grid, as well as a truncated radial grid.
\item Plots of the radial parts of the valence orbitals in the shifted polynomial and exponential potentials for various $N$.
\item Basis set truncation error for the H--Xe atoms in the singular potentials with the PW92 and $r^2$SCAN functionals and the \textsc{FHI-aims} 2020 default values for $r_i$ and $r_c$.
\end{enumerate}
All of the data for 1--3 above is provided both the Mg and Ca atoms.
In addition, all analogous plots and tables to the discussion in the main text for the Mg atom are included in the SI for the Ca atom.

\FloatBarrier
\bibliography{citations,solid}

\end{document}

%% file: tables/shifttabmg.tex
\newcommand*\shifttabmgpolynomial{
\begin{tabular}{l|cc|cc|cc|cc}
$N$ & \multicolumn{2}{|c}{\small{2}} & \multicolumn{2}{|c}{\small{4}} & \multicolumn{2}{|c}{\small{6}} & \multicolumn{2}{|c}{\small{10}}\\ 
\hline
\hline
$(r_0,\delta)$& $r_\infty$ & $||\Delta||$& $r_\infty$ & $||\Delta||$& $r_\infty$ & $||\Delta||$& $r_\infty$ & $||\Delta||$\\ 
\hline
\hline
$(0.1,3.9)$ & $4.96$ & $1.82\times10^{-4}$ & $4.52$ & $2.83\times10^{-5}$ & $4.34$ & $6.61\times10^{-6}$ & $4.16$ & $7.75\times10^{-6}$\\ 
$(0.2,3.8)$ & $5.16$ & $2.97\times10^{-4}$ & $4.68$ & $6.59\times10^{-5}$ & $4.48$ & $2.06\times10^{-5}$ & $4.30$ & $4.08\times10^{-6}$\\ 
$(0.3,3.7)$ & $5.26$ & $3.96\times10^{-4}$ & $4.78$ & $1.07\times10^{-4}$ & $4.57$ & $3.88\times10^{-5}$ & $4.37$ & $9.09\times10^{-6}$\\ 
$(0.5,3.5)$ & $5.37$ & $5.83\times10^{-4}$ & $4.91$ & $1.98\times10^{-4}$ & $4.69$ & $8.35\times10^{-5}$ & $4.46$ & $2.12\times10^{-5}$\\ 
$(0.7,3.3)$ & $5.40$ & $7.76\times10^{-4}$ & $4.98$ & $3.00\times10^{-4}$ & $4.76$ & $1.39\times10^{-4}$ & $4.53$ & $4.25\times10^{-5}$\\ 
$(1.0,3.0)$ & $5.37$ & $1.11\times10^{-3}$ & $5.01$ & $4.80\times10^{-4}$ & $4.81$ & $2.42\times10^{-4}$ & $4.59$ & $8.45\times10^{-5}$\\ 
$(4.0,0.0)$ & $4.07$ & $1.05\times10^{-2}$ & $4.11$ & $6.23\times10^{-3}$ & $4.20$ & $3.85\times10^{-3}$ & $4.31$ & $1.71\times10^{-3}$\\ 
\hline
\end{tabular}
}
\newcommand*\shifttabmgexponential{
\begin{tabular}{l|cc|cc|cc|cc}
$N$ & \multicolumn{2}{|c}{\small{2}} & \multicolumn{2}{|c}{\small{4}} & \multicolumn{2}{|c}{\small{6}} & \multicolumn{2}{|c}{\small{10}}\\ 
\hline
\hline
$(r_0,\delta)$& $r_\infty$ & $||\Delta||$& $r_\infty$ & $||\Delta||$& $r_\infty$ & $||\Delta||$& $r_\infty$ & $||\Delta||$\\ 
\hline
\hline
$(0.1,3.9)$ & $4.22$ & $1.88\times10^{-5}$ & $4.16$ & $5.26\times10^{-6}$ & $4.12$ & $4.68\times10^{-6}$ & $4.07$ & $2.11\times10^{-6}$\\ 
$(0.2,3.8)$ & $4.29$ & $7.45\times10^{-5}$ & $4.23$ & $2.48\times10^{-5}$ & $4.19$ & $9.44\times10^{-6}$ & $4.12$ & $3.64\times10^{-6}$\\ 
$(0.3,3.7)$ & $4.32$ & $1.50\times10^{-4}$ & $4.27$ & $5.29\times10^{-5}$ & $4.23$ & $2.30\times10^{-5}$ & $4.17$ & $5.80\times10^{-6}$\\ 
$(0.5,3.5)$ & $4.33$ & $3.56\times10^{-4}$ & $4.32$ & $1.39\times10^{-4}$ & $4.28$ & $6.27\times10^{-5}$ & $4.22$ & $2.05\times10^{-5}$\\ 
$(0.7,3.3)$ & $4.31$ & $6.33\times10^{-4}$ & $4.34$ & $2.62\times10^{-4}$ & $4.31$ & $1.24\times10^{-4}$ & $4.26$ & $3.71\times10^{-5}$\\ 
$(1.0,3.0)$ & $4.25$ & $1.19\times10^{-3}$ & $4.33$ & $5.14\times10^{-4}$ & $4.34$ & $2.57\times10^{-4}$ & $4.30$ & $8.77\times10^{-5}$\\ 
$(4.0,0.0)$ & $3.80$ & $1.04\times10^{-2}$ & $4.00$ & $6.20\times10^{-3}$ & $4.13$ & $3.83\times10^{-3}$ & $4.27$ & $1.70\times10^{-3}$\\ 
\hline
\end{tabular}
}
\newcommand*\shifttabmgbarrier{
\begin{tabular}{l|cc}
$V_0$ & $r_\infty$ & $||\Delta||$\\ 
\hline
\hline
3.0 & $4.43$ & $4.42\times10^{-4}$\\ 
10.0 & $4.23$ & $9.36\times10^{-5}$\\ 
30.0 & $4.13$ & $2.06\times10^{-5}$\\ 
300.0 & $4.04$ & $7.47\times10^{-7}$\\ 
3000.0 & $4.01$ & $4.84\times10^{-8}$\\ 
10000.0 & $4.01$ & $1.54\times10^{-8}$\\ 
\hline
\end{tabular}
}

%% file: tables/rmax_table-barrier-Mg.tex
\begin{table}
\centering
\begin{tabular}{l|ccc|ccc|ccc}
\hline
 & \multicolumn{3}{c|}{\small{LIP}} & \multicolumn{3}{c|}{\small{HIP}} & \multicolumn{3}{c}{\small{HIP}'} \\ 
\hline
\hline
\small{$V_0$} & \scriptsize{2.0} & \scriptsize{3.0} & \scriptsize{4.0} & \scriptsize{2.0} & \scriptsize{3.0} & \scriptsize{4.0} & \scriptsize{2.0} & \scriptsize{3.0} & \scriptsize{4.0} \\ 
\hline
\small{3} & \tiny{6.00} & \tiny{5.89} & \tiny{6.50} & \tiny{6.00} & \tiny{5.89} & \tiny{6.50} & \tiny{6.01} & \tiny{5.90} & \tiny{6.51} \\ 
\small{10} & \tiny{3.81} & \tiny{4.49} & \tiny{5.32} & \tiny{3.81} & \tiny{4.49} & \tiny{5.32} & \tiny{3.81} & \tiny{4.50} & \tiny{5.33} \\ 
\small{30} & \tiny{2.96} & \tiny{3.82} & \tiny{4.73} & \tiny{2.96} & \tiny{3.82} & \tiny{4.73} & \tiny{2.97} & \tiny{3.83} & \tiny{4.74} \\ 
\small{300} & \tiny{2.28} & \tiny{3.24} & \tiny{4.21} & \tiny{2.28} & \tiny{3.24} & \tiny{4.21} & \tiny{2.28} & \tiny{3.24} & \tiny{4.21} \\ 
\hline
\end{tabular}
\caption{Values of practical infinity ($r_\infty$) in $a_0$ resulting in a 1 \textmu \Eh{} energy increase from a converged radial grid for the Mg atom in finite-barrier confinement with $r_0\in\{2.0,3.0,4.0\}a_0$. LIP: Lagrange interpolating polynomial, HIP: Hermite interpolating polynomial, HIP': HIP with zero derivative at $r_\infty$.}
\label{tab:rmax-table-barrier-Mg}
\end{table}

%% file: tables/rmax_table-polynomial-Mg.tex
\begin{table}
\centering
\begin{tabular}{l|ccc|ccc|ccc}
\hline
 & \multicolumn{3}{c|}{\small{LIP}} & \multicolumn{3}{c|}{\small{HIP}} & \multicolumn{3}{c}{\small{HIP}'} \\ 
\hline
\hline
\small{$N$} & \scriptsize{2.0} & \scriptsize{3.0} & \scriptsize{4.0} & \scriptsize{2.0} & \scriptsize{3.0} & \scriptsize{4.0} & \scriptsize{2.0} & \scriptsize{3.0} & \scriptsize{4.0} \\ 
\hline
\small{1} & \tiny{7.09} & \tiny{7.80} & \tiny{8.32} & \tiny{7.09} & \tiny{7.80} & \tiny{8.32} & \tiny{7.11} & \tiny{7.82} & \tiny{8.34} \\ 
\small{2} & \tiny{5.49} & \tiny{6.47} & \tiny{7.24} & \tiny{5.49} & \tiny{6.47} & \tiny{7.24} & \tiny{5.50} & \tiny{6.48} & \tiny{7.26} \\ 
\small{4} & \tiny{4.27} & \tiny{5.43} & \tiny{6.43} & \tiny{4.27} & \tiny{5.43} & \tiny{6.43} & \tiny{4.28} & \tiny{5.44} & \tiny{6.44} \\ 
\small{6} & \tiny{3.73} & \tiny{4.94} & \tiny{6.02} & \tiny{3.73} & \tiny{4.94} & \tiny{6.02} & \tiny{3.74} & \tiny{4.95} & \tiny{6.03} \\ 
\small{8} & \tiny{3.42} & \tiny{4.64} & \tiny{5.75} & \tiny{3.42} & \tiny{4.64} & \tiny{5.75} & \tiny{3.43} & \tiny{4.65} & \tiny{5.77} \\ 
\small{10} & \tiny{3.21} & \tiny{4.43} & \tiny{5.56} & \tiny{3.21} & \tiny{4.43} & \tiny{5.56} & \tiny{3.22} & \tiny{4.44} & \tiny{5.57} \\ 
\hline
\end{tabular}
\caption{Values of practical infinity ($r_\infty$) in $a_0$ resulting in a 1 \textmu \Eh{} energy increase from a converged radial grid for the Mg atom in polynomial confinement with $r_0\in\{2.0,3.0,4.0\}a_0$. LIP: Lagrange interpolating polynomial, HIP: Hermite interpolating polynomial, HIP': HIP with zero derivative at $r_\infty$.}
\label{tab:rmax-table-polynomial-Mg}
\end{table}

%% file: tables/rmax_table-exponential-Mg.tex
\begin{table}
\centering
\begin{tabular}{l|ccc|ccc|ccc}
\hline
 & \multicolumn{3}{c|}{\small{LIP}} & \multicolumn{3}{c|}{\small{HIP}} & \multicolumn{3}{c}{\small{HIP}'} \\ 
\hline
\hline
\small{$N$} & \scriptsize{2.0} & \scriptsize{3.0} & \scriptsize{4.0} & \scriptsize{2.0} & \scriptsize{3.0} & \scriptsize{4.0} & \scriptsize{2.0} & \scriptsize{3.0} & \scriptsize{4.0} \\ 
\hline
\small{1} & \tiny{5.01} & \tiny{6.01} & \tiny{6.75} & \tiny{5.01} & \tiny{6.01} & \tiny{6.75} & \tiny{5.02} & \tiny{6.02} & \tiny{6.76} \\ 
\small{2} & \tiny{4.67} & \tiny{5.71} & \tiny{6.54} & \tiny{4.67} & \tiny{5.71} & \tiny{6.54} & \tiny{4.68} & \tiny{5.72} & \tiny{6.55} \\ 
\small{4} & \tiny{4.03} & \tiny{5.18} & \tiny{6.17} & \tiny{4.03} & \tiny{5.18} & \tiny{6.17} & \tiny{4.04} & \tiny{5.20} & \tiny{6.18} \\ 
\small{6} & \tiny{3.63} & \tiny{4.83} & \tiny{5.89} & \tiny{3.63} & \tiny{4.83} & \tiny{5.89} & \tiny{3.64} & \tiny{4.84} & \tiny{5.90} \\ 
\small{8} & \tiny{3.36} & \tiny{4.57} & \tiny{5.68} & \tiny{3.36} & \tiny{4.57} & \tiny{5.68} & \tiny{3.37} & \tiny{4.58} & \tiny{5.69} \\ 
\small{10} & \tiny{3.17} & \tiny{4.38} & \tiny{5.51} & \tiny{3.18} & \tiny{4.39} & \tiny{5.51} & \tiny{3.18} & \tiny{4.39} & \tiny{5.52} \\ 
\hline
\end{tabular}
\caption{Values of practical infinity ($r_\infty$) in $a_0$ resulting in a 1 \textmu \Eh{} energy increase from a converged radial grid for the Mg atom in exponential confinement with $r_0\in\{2.0,3.0,4.0\}a_0$. LIP: Lagrange interpolating polynomial, HIP: Hermite interpolating polynomial, HIP': HIP with zero derivative at $r_\infty$.}
\label{tab:rmax-table-exponential-Mg}
\end{table}